\documentclass[12pt]{article}
\usepackage{amsmath}
\usepackage{amssymb}
\usepackage{graphicx}

\input{epsf}

\begin{document}

\begin{center}
{\Large \bf  Computing Spin Networks }
\end{center}
\vspace{6pt}

\begin{center}
{\large
{\sl Annalisa Marzuoli}}\\

\vspace{4pt}
{\small{Dipartimento di Fisica Nucleare e Teorica,
Universit\`a degli Studi di Pavia and 
Istituto Nazionale di Fisica Nucleare, Sezione di Pavia, \\
via A. Bassi 6, 27100 Pavia (Italy) \\
annalisa.marzuoli@pv.infn.it}} \\
\end{center}

\begin{center}
and
\end{center}

\begin{center}
{\large
{\sl Mario Rasetti}}\\

\vspace{4pt}
{\small{Dipartimento di Fisica and Istituto Nazionale di Fisica della Materia,\\
Politecnico di Torino,\\
corso Duca degli Abruzzi 24, 10129 Torino (Italy) \\
rasetti@isi36a.isi.it}} \\
\end{center}

\vspace{6pt}

\begin{center}
{\bf Abstract}
\end{center}

{\small{We expand a set of notions recently introduced providing the general setting for a 
universal representation of the quantum structure on which quantum information stands. 
The dynamical evolution process associated with generic quantum information manipulation 
is based on the (re)coupling theory of $SU(2)$ angular momenta. Such scheme automatically 
incorporates all the essential features that make quantum information encoding much more 
efficient than classical: it is fully discrete; it deals with inherently entangled states, 
naturally endowed with a tensor product structure; it allows for generic 
encoding patterns. The model proposed can be thought of as the non--Boolean generalization 
of the {\em quantum circuit} model, with unitary gates expressed in terms of $3nj$ 
coefficients connecting inequivalent binary coupling schemes of $n+1$ angular momentum 
variables, as well as Wigner rotations in the eigenspace of the 
total angular momentum. A crucial role is played by elementary $j$--gates ($6j$ symbols) 
which satisfy algebraic identities that make the structure of the model similar to 
$''$state sum models$\, ''$  employed in discretizing Topological Quantum Field Theories and 
quantum gravity. The spin network simulator can thus be viewed also as a Combinatorial QFT 
model for computation. The semiclassical limit (large $j$) is discussed.}}  

\vspace{12pt}


\vfill
\newpage

\section{Introduction}

It is by--now a generally accepted fact that the laws 
of quantum theory provide in principle a radically novel,
and more powerful way to process information than any 
classically operated device \cite{Rev}.
In the past few years a big deal of activity has been 
devoted to devise and to implement schemes for taking 
actual advantage from such quantum extra power. In 
particular in Quantum Computation (QC) the states of a quantum 
system ${\cal S}$ are used for encoding information in such a 
way that the final state, obtained by the appropriate unitary 
time evolution of ${\cal S}$, encodes the solution to a given 
computational problem. A system ${\cal S}$ with state--space 
${\cal H}$ (the {\em Quantum Computer}) supports universal QC 
if any unitary transformation $U\in{\cal U}({\cal H})$ can be 
approximated with arbitrarily high accuracy by a sequence (the 
{\em network}) of simple unitaries (the {\em gates}) that the 
experimenter is supposed to be able to implement. The case in 
which ${\cal S}$ is a multi--partite system is the most relevant, 
as it allows for entanglement, a unique quantum feature that 
is generally believed to be one of the crucial elements from 
which quantum speed--up (polynomial or possibly exponential) is 
generated \cite{Jo}.\\
In the above picture of QC the realization of the quantum 
network is achieved at the physical level by turning on and 
off external fields coupled to ${\cal S}$ as well as local 
interactions among the subsystems of ${\cal S}$. In other words
the experimenter $''$owns$\, ''$ a basic set of time--dependent 
Hamiltonians that she/he activates at will to perform the 
necessary sequences of quantum logic gates.

At variance with 
such a standard dynamical view of QC, more recently several 
authors considered geometrical and topological approaches 
\cite{Kit}--\cite{PreMe}. The peculiarity of these proposals is 
far reaching: over the manifold ${\cal C}$ of quantum codewords 
one can have a trivial Hamiltonian, for example $H \bigr 
|_{\cal C} =0$, yet obtain nevertheless a non--trivial quantum 
evolution due to the existence of an underlying 
geometrical/topological global structure.
The quantum gates -- or parts thereof -- in this latter case 
are realized in terms of operations having a purely 
geometrical/topological nature. Besides being conceptually 
intriguing on their own, these schemes have some built--in 
fault--tolerant features. This latter attractive characteristic  
stems out of the fact that often certain topological as well 
as geometrical quantities are inherently stable against local 
perturbations. This in turn allows for quantum information 
processing inherently stable against special classes of 
computational errors.

There has been a number of proposals suggesting general conceptual schemes 
of interpretation of quantum computation. Most of them are indeed based on 
topological notions, even though this is not always explicitly stated. 
Among these, anyonic quantum computation \cite{Kit}, fermionic quantum 
computation \cite{BrKi}, localised modular functor quantum field computation 
\cite{FrLaWa}, holonomic quantum computation \cite{ZaRa}, \cite{PaZaRa} have mostly 
attracted attention. However, such models appear to be simply different realizations 
of a unique conceptual scheme that incorporates all of them as particular 
instances (once one focuses on their suitable $''$discretized$\,''$ counterparts).

We propose here, expanding a set of notions introduced in \cite{MaRa}, 
a general setting for a universal representation of the very quantum 
structure on which quantum information stands. The associated dynamical evolution 
process, giving rise to information manipulation, is based on the (re)coupling theory of 
$SU(2)$ angular momenta (see \cite{Lituani}, \cite{BiLo9} Topic 12, \cite{Russi}).
The scheme automatically incorporates all the essential features that make quantum 
information encoding so much more efficient than classical: it is fully discrete 
(both for space--like and time--like variables); it deals with inherently 
entangled states, naturally endowed with a (non associative) 
tensor product structure; it allows for generic encoding patterns. 
The minimal set of requirements listed by Feynman \cite{Fey} as essential for 
the proper characterisation of an efficient quantum simulator is 
automatically satisfied: i) locality of interactions; ii) number of 
$''$computer$\,'' $ elements proportional to a function which is at most polynomial 
in the space--time volume of the physical system; iii) time discreteness (time 
is itself $\,''$simulated$\, ''$ in the computer by computational steps).
 
Key element of our argument is the fact that all such basic features are typical of spin networks.
It should be enphasized that by {\em spin networks} we mean here -- contrary to 
what happens in solid state physics, but somewhat in the spirit of combinatorial 
approach to quantum space--time representation \cite{Pen} -- graphs the node and edge 
sets of which can be labelled by quantum numbers associated with $SU(2)$ irreducible 
representations and by $SU(2)$ recoupling coefficients, respectively. For this reason 
spin networks can be thought of as an ideal candidate conceptual framework for dealing 
with tensorial transformations and topological effects in groups of observables. The 
idea is to exploit to their full extent the discreteness hypotheses ii) 
and iii), by modelling the computational space in terms of a set of combinatorial and 
topological rules that mimic space--time features in a way that automatically includes 
quantum mechanics. 

The model proposed can be thought of as a non--Boolean generalization of the 
{\em quantum circuit} model, with unitary gates expressed in terms of: 
a) recoupling coefficients ($3nj$ symbols) between inequivalent binary
coupling schemes of $N=n+1$ $SU(2)$--angular momentum variables ($j$--gates);
b) Wigner rotations in the eigenspace of the total angular momentum ($M$--gates).
These basic ingredients of the spin network simulator, namely computational
Hilbert spaces and admissible elementary gates, are discussed in details in 
Sections 2 and 3, respectively. 
The picture does  contain the Boolean case  as the particular case when 
all $N$ angular momenta are spin $\frac{1}{2}$.

In Section 4 both the architecture and the computational capabilities of 
the simulator are described in full extent. On the kinematical side (Section 4.1),
the computational space is shown to be modeled as an $SU(2)$--decorated graph
or, more precisely, as a fiber space structure over a discrete base space --the
Rotation graph -- which encodes all possible computational Hilbert spaces as well
as  gates for any fixed number $N$ of incoming angular momenta. A crucial role is played
by elementary $j$--gates (Racah transforms, related to $6j$ symbols of
$SU(2)$) which satisfy suitable algebraic identities \cite{Russi} and make the structure of
the model similar to $''$state sum models$''$  employed in discretizing Topological Quantum
Field Theories (TQFT) and quantum gravity (see Section 5).
In Section 4.2, after a discussion of the hypotheses of Feynman \cite{Fey}, 
general circuit--type 
computation processes on the spin network are described and classified into
{\em computing classes}. Virtual, polylocal Hamiltonians are generated by the simulator 
which evolves in an intrinsic discrete time variable.
Section 4.3 deals with questions in (quantum) computational complexity which turn out
to be closely related to graph combinatorics. We argue that our new framework, 
when implemented on the basis of explicit encoding schemes \cite{GaMaRa}, 
could be suitable to handle $''$combinatorially hard $''$ problems more 
efficiently than any classical machine.

The key ingredient of Section 5 is the Ponzano--Regge asymptotic formula \cite{PoRe}
for the $6j$ symbol which plays a twofold role. On the one hand, it gives a precise
meaning to the semiclassical analog of a $j$-gate providing, together with the
asymptotics of Wigner rotation matrices \cite{BiLo9} (Topic 9), the
notion of $''$approaching a classical simulator$\,''$ out of a quantum one. 
Moreover, as stated in \cite{MaRa}, 
the whole conceptual scheme of spin network computing can be reformulated in 
terms of density matrix formalism, namely resorting no longer to sharp eigenstates of 
angular momenta but rather to generalised multipole moments (see {\em e.g.} \cite{BiLo8}, Ch. 7.7). 
Such generalisations can be summarised in the following diagram 

\vfill
\newpage

\bigskip 
\begin{center}
\fbox{ SEMICLASSICAL SIMULATOR}
\end{center}

\begin{center}
{\large $\Uparrow\;\;$}(large quantum numbers){\large $\;\;\Uparrow$}
\end{center}

\begin{center}
\fbox{\fbox{{\bf SPIN NETWORK QU-SIMULATOR}}}
\end{center}

\begin{center}
{\large $\Downarrow\;\;$}(extension to){\large $\;\;\Downarrow$}
\end{center}

\begin{center}
\fbox{MIXED STATES COMPUTING MACHINE}
\end{center}

\noindent On the other hand, the Ponzano--Regge asymptotic formula opens
the intriguing possibility of bridging the quantum theory of angular momenta
to Euclidean gravity in
dimension three. More precisely, a {\em state sum functional} for  triangulated
3--dimensional space--time manifolds, built up by associating a $6j$ symbol with each tetrahedron, is shown to
correspond, in the asymptotic limit, to the semiclassical partition function of gravity \cite{PoRe}
with a classical action representing the discretized counterpart of the Einstein--Hilbert action
of general relativity
\cite{Reg}. Since quantum gravity in dimension three is strictly related to
a TQFT with an $SU(2)$ Chern--Simons--type action (see references quoted at 
the end of Section 5 and
at the beginning of Section 6) we present in this part of the paper some
known results concerning spin networks viewed as $''$Combinatorial Quantum
Field Theories$\,''$, to be interpreted as discretized versions of TQFTs 
based on $SU(2)$--decorated triangulations. 
At the end of Section 5 we compare the discrete partition functions of Ponzano--Regge gravity
with the partition functions introduced in Section 4.2 in connection with the simulator's dynamics.
We conclude that a spin network simulator working by switching on $j$--gates acting on
states with $N$ incoming spins is able
to simulate some subclasses of triangulated surfaces and possibly 
some subclasses of triangulated 3--manifolds but, since partition functions for 
quantum field theories must be $''$sums over all configurations$\,''$ (apart from regularization),
we cannot infer the possibility of fully simulating Combinatorial QFT  
(unless we take some sort of termodynamical limit
for $N \rightarrow \infty$ which does not sound good when dealing with 
quantum circuit schemes for computation).

We start Section 6 (Spin network and topological quantum computation) 
by reviewing some basic definitions on TQFTs.
Section 6.1 addresses holonomic quantum computation    
and we give indications that the discrete setting developed in Section 4 could support
also such kind of computational processes.
In Section 6.2 we compare the spin network approach with the approach of Freedman and
collaborators \cite{FrLaWa}. We provide a (not unique) mapping between the spin network
and the modular functor approach by introducing combinatorial marked 2--disks which
display localised interactions between spins. 
The algebraic structures of the two approaches, summarised in the Yang--Baxter
identity for the standard topological one and in the (hexagon + pentagon) identities
for the spin network, suggest that the partition functions of the two models are
related to each other in the same way as the regularized version of Ponzano--Regge
functional corresponds to a double Chern--Simons partition function.
 
In Appendix A we present the graph--theoretical rationale underlying spin network
combinatorics by collecting  results spread over a number of
references in discrete mathematics, binary couplings and
recoupling theory of angular momenta, complexity theory. 
Binary coupling trees are defined in Appendix A1, Twist--Rotation and Rotation graphs
in A2 and some results in (classical) combinatorial complexity theory are summarised in
A3.  Appendix B contains (standard) technical results  concerning the composition 
of Wigner rotation matrices (B1)
and U--rotation matrices (B2) which are employed in Section 3.2 ($M$--gates).

\bigskip 

The twofold possible interpretation of the deliberately ambiguous title we have chosen
for the paper should have become clear at this point:\\
\noindent - on the one hand, spin networks are computing devices supporting simulations
of the dynamical behaviour of composite quantum systems described in terms of pure angular
momentum eigenstates;\\
\noindent - such computing devices, on the other hand, are able to simulate classes of 
extended geometrical objects modeled as spin networks.

We may summarise the content of the paper in the following diagram, where the
spin network simulator may be viewed both as a generalised quantum circuit and
as a Combinatorial QFT model for computation.
The standard Boolean quantum
circuit (shown to be equivalent to the topological approach \cite{FrLaWa}) 
is a particular case of this general scheme for computation. To complete the picture,
the combinatorial approach can be suitably mapped into
the purely topological one as discussed in Section 6.2. 

We plan to develop in the next future the upper connection which points toward 
$''$quantum automata$''$ since our framework seems quite promising to address
such issues like quantum languages and grammars,
quantum encoding \cite{GaMaRa} and quantum complexity classes of algorithms, naturally 
related here to enumerative combinatorics of graphs.

\bigskip\bigskip  
\begin{center} 
\thicklines  
\begin{picture}(250,280)(-94,-135)  
\put(-140,37){\framebox(130,32){BOOLEAN CIRCUIT}}  
\put(-140,-32){\framebox(130,32){TOPOLOGICAL QFT}}  
\put(-78,28){\vector(0,-1){25}}
\put(-77,3){\vector(0,1){25}}
\put(-77,28){\vector(0,-1){25}}
\put(-78,3){\vector(0,1){25}}
\put(108,87){\vector(0,1){25}}
\put(109,87){\vector(0,1){25}}
\put(108,29){\vector(0,-1){25}}
\put(109,4){\vector(0,1){25}}
\put(109,29){\vector(0,-1){25}}
\put(108,4){\vector(0,1){25}}
\put(108,-51){\vector(0,-1){25}}
\put(109,-76){\vector(0,1){25}}
\put(109,-51){\vector(0,-1){25}}
\put(108,-76){\vector(0,1){25}}
\put(0,51){\vector(1,0){40}} 
\put(0,50){\vector(1,0){40}} 
\put(35,-105){\vector(-3,4){49}}
\put(35,-106){\vector(-3,4){49}}
\put(55,116){\framebox(120,46){QU-AUTOMATA}} 
\put(55,37){\framebox(120,46){}} 
\put(115,70){\makebox(0,0){GENERALISED}}
\put(115,50){\makebox(0,0){QU-CIRCUIT}}
\put(55,-46){\framebox(120,46){}}
\put(53,-48){\framebox(124,50){}}
\put(67,-16){\makebox{SPIN NETWORK}} 
\put(67,-37){\makebox{QU-SIMULATOR}}
\put(55,-125){\framebox(120,46){}} 
\put(115,-92){\makebox(0,0){COMBINATORIAL}} 
\put(105,-112){\makebox(0,0){QFT}} 
\end{picture}  
\end{center} 

\bigskip 

\section{Computational Hilbert spaces}

Following \cite{BiLo9} (Topic 12)
let us consider $N=n+1$ mutually commuting angular
momentum operators of the algebra of $SU(2)$

\begin{equation*}
{\bf J}_1,\;{\bf J}_2,\;{\bf J}_3,\ldots,{\bf J}_{n+1}\,\equiv \,\{{\bf J}_i\}
\end{equation*}

\noindent and the corresponding components 

\begin{equation*}
\{J_{i\,(z)}\}_{i=1,2,\ldots,n+1}
\end{equation*}

\noindent along the quantization axis.
For each $i=1,2,\ldots,n+1$ the simultaneous eigenstates of the complete sets 
${\bf J}^2_i$ and $J_{i(z)}$ are

\begin{align}\label{Jist}
{\bf J}^2_i \,|j_i\,m_i \rangle & = j_i\,(j_i+1)\;|j_i\,m_i \rangle \nonumber\\
J_{i(z)}\,|j_i\,m_i \rangle & = m_i\;|j_i\,m_i \rangle,
\end{align}

\noindent where we set $\hbar=1$ and the eigenvalues range over

\begin{align}\label{Jieigen}
j_i & =\;0,\frac{1}{2},1, \frac{3}{2},\ldots \ldots;\nonumber\\
-j_i & \leq\,m_i\,\leq\,j_i\;\;(\text{integer steps}).
\end{align}

\noindent Denoting by

\begin{equation*}
{\cal H}^{j_i}\;\doteq\;\text{span}\;\{\;\,|j_i\;m_i \rangle\;\}
\end{equation*}

\noindent the $(2j_i+1)$--dimensional Hilbert space supporting the $j_i$--th irreducible 
representation of $SU(2)$, the tensor product

\begin{align}\label{fact}
{\cal H}^{j_1}\;\otimes\;{\cal H}^{j_2}\;\otimes\;{\cal H}^{j_3}\,\otimes\ldots \ldots 
\otimes \; {\cal H}^{j_{n}}\;\otimes\;
{\cal H}^{j_{n+1}}\nonumber\\
\doteq \;\text{span}\; \{\, |\,j_1\,m_1\rangle\, \otimes\, \ldots \otimes\,
 |\,j_{n+1}\,m_{n+1}\rangle \;\}
\end{align}

\noindent represents the simultaneous eigenspace of the $2(n+1)$ operators 
$\{{\bf J}^2_i\,;\,J_{i\,(z)}\}$ and may be used {\em e.g.} to describe the state of
$N=(n+1)$ kinematically independent particles. By setting $j_1=j_2=\ldots=j_{n+1}$ $=1/2$ 
in \eqref{fact} we would get $\otimes^{N}$ $\mathbb{C}^2$, namely the $N$--qubits space of 
the Boolean quantum circuit model.

To address the interacting case, we make explicit the basic assumptions on
which the model of simulator discussed in this paper relies: 

\begin{enumerate}
\item The simulator computational states are suitable pure $(n+1)$--angular momenta states 
(to be described below) in Wigner--coupled Hilbert spaces of the total angular momentum
operator

\begin{equation}\label{jtot}
{\bf J}_1\,+\,{\bf J}_2\,+\,{\bf J}_3\,+\ldots+{\bf J}_{n+1}\;\doteq\;{\bf J}
\end{equation}

\noindent and of its projection $J_{z}$. The corresponding quantum numbers, $J$ and $M
\equiv m_1+m_2+\ldots+m_{n+1}$ ($-J\leq M\leq J$ in integer steps),
label the resulting representation spaces (which are referred to as $JM$--representations). 

\item Interactions are not fixed {\em a priori} but rather are generated by
the simulator itself as polylocal virtual Hamiltonian operators (see Section 4.2). 

\item The admissible interactions are modeled on (a finite number of) combinations 
of the following basic types:

\begin{itemize}
\item binary couplings of the computational states which involve
only the spin quantum numbers;
\item actions of rotation operators over states in $JM$--representations involving
the total magnetic quantum number and depending on continuous sets of parameters 
({\em e.g.} Euler angles).
\end{itemize}

\end{enumerate}

Given ${\bf J}$ as in \eqref{jtot}, the simultaneous 
eigenspace of the operators ${\bf J}^2$ and $J_z$ (namely a $JM$--representation 
space for any fixed $J$) 

\begin{align}\label{JMst}
{\bf J}^2\,|J\,M \rangle & = J\,(J+1)\;|J\,M \rangle \nonumber\\
J_z\,|J\,M \rangle & = M\;|J\,M \rangle
\end{align}

\noindent turns out to be degenerate. 
The degeneracy is 
partially removed by noticing that ${\bf J}_1^2$, ${\bf J}_2^2$,
$\ldots$, ${\bf J}_{n+1}^2$ commute with ${\bf J}^2$ and $J_z$ and thus
$j_1,\,j_2,\,\ldots,j_{n+1}$ still are good quantum numbers (while the individual
$m_1,m_2,\ldots,m_{n+1}$ are not). The ket vectors $|J\,M\rangle$ in \eqref{JMst}
can be rewritten for the moment as $|j_1,j_2,\ldots,j_{n+1};\,J\,M \rangle$, 
namely in terms of $(n+1)+2$ quantum numbers. 

The complete removal of the degeneracy would be achieved by introducing a new set of
$(n-1)$ Hermitean operators -- commuting with each other and with the previous ones --
in order to get a total amount of $2(n+1)$ quantum numbers (this number equals
the number of operators needed to specify the eigenstates in the
factorized Hilbert space \eqref{fact}). The most effective way to reach the goal is 
to consider {\em equivalently}:

\begin{itemize}
\item binary coupling schemes in the sequence

\begin{equation}\label{bcou}
{\bf J}_1\,+\,{\bf J}_2\,+\,{\bf J}_3\,+\ldots+{\bf J}_{n+1}\;=\; {\bf J}
\end{equation}

\item binary bracketings on the factorized Hilbert space

\begin{equation}\label{bbra}
{\cal H}^{j_1}\;\otimes\;{\cal H}^{j_2}\;\otimes\;{\cal H}^{j_3}\,\otimes\ldots \ldots \otimes \;
{\cal H}^{j_{n}}\;\otimes\;
{\cal H}^{j_{n+1}}.
\end{equation}

\end{itemize}

As a simple example, consider the case $(n+1)=3$: the binary coupling schemes are
$({\bf J}_1\,+\,{\bf J}_2)\,+\,{\bf J}_3\,=\,{\bf J}$;
${\bf J}_1\,+\,({\bf J}_2\,+\,{\bf J}_3)\,=\,{\bf J}$; $({\bf J}_1\,+\,{\bf J}_3)\,+\,{\bf J}_2
\,=\,{\bf J}$,
and the corresponding binary bracketings are 
$({\cal H}^{j_1}\otimes{\cal H}^{j_2})\otimes{\cal H}^{j_3};$
${\cal H}^{j_1}\otimes({\cal H}^{j_2}\otimes{\cal H}^{j_3});$
${(\cal H}^{j_1}\otimes{\cal H}^{j_3})\otimes{\cal H}^{j_2}$, respectively.\\
The crucial point is to realize that each binary coupling in \eqref{bcou} generates -- 
by using the Clebsch--Gordan series of $SU(2)$ -- an
intermediate angular momentum operator whose quantum number will be added to the
set $\{j_1,\,j_2,\,\ldots,j_{n+1};J\,M\}$. In the case $(n+1)=3$ 
the first coupling scheme $({\bf J}_1\,+\,{\bf J}_2)\,+\,{\bf J}_3$ $=\, {\bf J}$ splits into

\begin{align}\label{j12}
({\bf J}_1\,+\,{\bf J}_2) & = {\bf J}_{12}\nonumber\\ 
{\bf J}_{12}\,+\,{\bf J}_3 & = {\bf J}
\end{align}

\noindent with $|j_1-j_2|\leq j_{12}\leq j_1+j_2$ and $J\,=\,j_{12}+j_3$, while the second
coupling scheme 
${\bf J}_1\,+\,({\bf J}_2\,+\,{\bf J}_3)$ $={\bf J}$ splits into 

\begin{align}\label{j23}
({\bf J}_2\,+\,{\bf J}_3) & = {\bf J}_{23}\nonumber\\ 
{\bf J_1}\,+\,{\bf J}_{23} & = {\bf J}
\end{align}

\noindent with $|j_2-j_3|\leq j_{23}\leq j_2+j_3$ and $J\,=\,j_1+j_{23}$.\\
The reformulation of the same example in terms of binary bracketings on \eqref{bbra}
leads to the expressions

\begin{align}\label{H123}
(\,({\cal H}^{j_1} & \otimes{\cal H}^{j_2})_{j_{12}}\otimes{\cal H}^{j_3}\,)_J\nonumber\\
(\,{\cal H}^{j_1} & \otimes({\cal H}^{j_2}\otimes{\cal H}^{j_3})_{j_{23}}\,)_J
\end{align}

\noindent where the inner brackets have been labeled by the q--numbers associated with the
corresponding intermediate angular momentum operators and we have added an overall 
bracket labeled by the total $J$. Note that these Hilbert spaces, although isomorphic,
are not identical since they actually correspond 
to (partially) different complete sets of physical observables,
namely
$\{{\bf J}^2_1,\,{\bf J}^2_2,\,{\bf J}^2_{12},\,{\bf J}^2_3,\,{\bf J}^2,\,J_z\}$ and 
$\{{\bf J}^2_1,\,{\bf J}^2_2,\,{\bf J}^2_3,\,{\bf J}^2_{23},\,{\bf J}^2,\,J_z\}$
respectively (in particular, ${\bf J}^2_{12}$ and
${\bf J}^2_{23}$ cannot be measured simultaneously). On the mathematical side this 
remark reflects the fact that the tensor product $\otimes$ is not an associative operation.

Coming to the general case, a counting argument explained 
in Appendix A1 (based on combinatorics of rooted labeled binary trees)
shows that the number of binary bracketings one can accommodate on
the $(n+1)$--fold tensor product in \eqref{bbra} is $(n-1)$, plus the external
bracket $(\ldots )_J$. Thus we get exactly the number of intermediate angular momenta 
operators we need to remove completely the degeneracy in the $JM$ space \eqref{JMst}.
More precisely, given a particular binary bracketing structure (for the moment we can think of
an ordered sequence of incoming angular momenta $\{{\bf J}_1$, ${\bf J}_2$, ${\bf J}_3$,
$\ldots$, ${\bf J}_{n+1}\}$),
we get a unique set of (ordered) mutually commuting operators denoted by

\begin{equation}\label{Ks}
{\bf K}_1,\,{\bf K}_2,\,{\bf K}_3,\,\ldots,\,{\bf K}_{n-1}
\end{equation}

\noindent with quantum numbers $k_1,\;k_2,\;k_3,\ldots,k_{n-1}$, each running over
a suitable finite range ({\em cfr.} \eqref{j12} and \eqref{j23}).
 
An explicit example of one possible bracketing structure is given by

\begin{equation}\label{seq}
(\ldots(((({\cal H}^{j_1}\otimes{\cal H}^{j_2})_{k_1}\otimes{\cal H}^{j_3})_{k_2}\otimes \ldots \otimes
{\cal H}^{j_{n}})_{k_{n-2}}\otimes
{\cal H}^{j_{n+1}})_{k_{n-1}})_J
\end{equation}

\noindent where the incoming angular momenta are coupled sequentially.  To denote the basis
vectors belonging to such a space we could write either

\begin{equation}\label{altbin1}
\{\,|j_1,\,j_2,\,j_3,\ldots,j_{n+1};k_1,\ldots,k_{n-1};\,JM\, \rangle,\;
-J\leq M\leq J \}
\end{equation}

\noindent (where both the sequences of quantum numbers $j$'s and $k$'s are ordered)
or

\begin{equation}\label{altbin2}
\{\,|\,(\ldots((((j_1,\,j_2)_{k_1}\,,j_3)_{k_2},\ldots),j_{n+1})_{k_{n-1}})_J\,;\,JM\, \rangle,\;
-J\leq M\leq J \},
\end{equation}

\noindent where the binary bracketing structure $(\,\centerdot \otimes\,\centerdot)_k\,$ of \eqref{seq} has been 
exactly transferred 
on the string of quantum numbers inside the symbol $|\ldots \rangle$ as $(\,\centerdot\,,\centerdot)_k\,$.
 
However, when dealing with other types of binary 
bracketing structures none of the above notations turns out to be well suited. First, we would like to 
consider any permutation of the incoming angular momenta quantum numbers
$\{j_1,j_2,j_3,\ldots,j_{n+1}\}$ and not just a sequence with a fixed ordering. 
It is clear that the notation adopted in
\eqref{altbin1} is not flexible in this respect. Secondly, we have to get rid of all possible
binary arrangements of the incoming variables and at the same time of the 
resulting (partially ordered) set of intermediate
$k$'s which appears in the example \eqref{altbin2} as subscripts under brackets.

According to the above remarks we shall denote from now on a binary coupled basis of $(n+1)$ angular
momenta in the $JM$--representation 
(and the corresponding Hilbert space) as

\begin{equation*}
\{\,|\,[j_1,\,j_2,\,j_3,\ldots,j_{n+1}]^{\mathfrak{b}}\, ;k_1^{\mathfrak{b}\,},\,k_2^{\mathfrak{b}\,}
,\ldots,k_{n-1}^{\mathfrak{b}}\, ;\,JM\, \rangle,\;
-J\leq M\leq J \}
\end{equation*}
\begin{equation}\label{genba}
=\;{\cal H}^{J}_{\,n}\;(\mathfrak{b})\;\doteq\;\text{span}\;\{\;|\,\mathfrak{b}\,;JM\,\rangle_n\,\}\;,
\end{equation}

\noindent where  the string inside $[j_1,\,j_2,\,j_3,\ldots,j_{n+1}]^{\mathfrak{b}\,}$ is not necessarily
an ordered one, $\mathfrak{b}$ indicates the current binary bracketing structure and 
the $k$'s are uniquely associated with the chain of pairwise couplings given by $\mathfrak{b}$.

As explained in details in Appendix A1, the combinatorial structure underlying the above 
computational Hilbert spaces (for fixed $n$ and for any $J$)  is provided by 
{\em rooted labeled binary trees}
with:

\begin{itemize}
\item $(n+1)$ terminal nodes (or leaves) labeled by $\{j_1,\,j_2,\,j_3,\ldots,j_{n+1}\}$;
\item $(n-1)$ internal nodes labeled by $\{k_1,\,k_2,\,k_3,\ldots,k_{n-1}\}$;
\item the root, labeled by $J$.
\end{itemize}

For instance, the binary coupling tree structure corresponding to
\eqref{seq} and \eqref{altbin2} is depicted in Fig. 12 of Appendix A1.

As shown in the following Section, processing information through changes
of binary coupling schemes of the Hilbert spaces \eqref{genba} involves only spin 
quantum numbers 
({\em cfr.} the first assumption in point 3.) and
can be modeled on other types of combinatorial structures, Twist--Rotation and Rotation
graphs as discussed in Section 4.1.

On the other hand, the quantum number $M$ can be activated by noticing that 
a vector in ${\cal H}^{J}_{\,n}$ $(\mathfrak{b})$ is expressed in terms of the
basis \eqref{genba} as

\begin{equation}\label{binvec}
|\mathfrak{b}\;;J\,\rangle_n
\;=\;\sum_{M=-J}^{J}\;^{(\mathfrak{b})}\,a^J_M\;
|\,\mathfrak{b}\,;JM\,\rangle_n,
\end{equation}

\noindent where $^{(\mathfrak{b}\,)}a^J_M$ $(M=-J,-J+1,\ldots,J-1,J)$ 
represent the (complex) components of the vector.
The natural unitary transformations acting on the $M$--dependence of basis vectors and components 
(Wigner rotation operators
$D^J_{\,M\,M'}(\alpha \beta \gamma)$ introduced in Section 3.2) are not discrete, 
but rather depend on continuous parameters as claimed in the second assumption of 3..

{\bf Remark 2.1.} A more fundamental question arising in connection with assumptions 1., 2., 
and 3. is the following: we could have included the factorized states \eqref{fact}
too in the set of computational states of the simulator, {\em e.g.} as non interacting
input states. Then any binary coupled state \eqref{genba} could be generated by means of a suitable
sequence of Clebsch--Gordan coefficients representing a unitary transformation depending on the whole
set of quantum numbers $\{j_1,j_2,\ldots,j_{n+1};$ $ m_1,m_2,\ldots,m_{n+1};$ 
$k_1,k_2,\ldots,k_{n-1}; J,M\}$ (see \cite{BiLo9}, Topic 12). 
For instance, a vector in the binary basis associated with
\eqref{j12} is obtained  from the factorized basis through

\begin{equation*}
|\,(\,(j_1,j_2)_{j_{12}}\,,\,j_3)_{J}\,;\,JM\,\rangle 
\end{equation*}
\begin{equation}\label{CG}
=\,\sum_{m_1\,m_2\,m_3} \;C^{\,JM}_{j_{12}m_{12}\,j_3\,m_3}\;
C^{\,j_{12\,}m_{12}}_{j_{1}m_{1}\,j_2\,m_2}
\;\;|j_1\,m_1\,\rangle \otimes |j_2\,m_2\,\rangle \otimes |j_3\,m_3\,\rangle,
\end{equation}
   
\noindent where there appear two Clebsch--Gordan coefficients. This expression
can obviously be inverted, providing us with one factorized basis vector in terms of
a combination of binary coupled states in the $JM$--representation.

However, such kind of procedure -- namely going through the factorized basis at each step
of $''$computation$\, ''$ -- is not satisfactory in many respects and certainly very cumbersome
when dealing with many spins. Even if we could assume that the input
state of the simulator is indeed a factorized, non--interacting one (and thus 
a transformation of the type \eqref{CG} is certainly allowed), any further unitary computing 
step would force us to deal with (pure angular momenta) {\em coupled states}. Now, quantum theory
of angular momentum gives us a powerful framework to manipulate the {\em binary} coupled 
states of the type we are considering: other kinds 
of couplings ({\em e.g.} symmetrical) cannot be efficiently manipulated since the 
degeneracy in the total $JM$--representation space cannot be completely removed for 
$N \equiv n+1>3$ (see the remark at the end of \cite{BiLo9}, Topic 12).

Summing up, a model of simulator whose computational states are binary coupled ones 
actually relies on the most exhaustive theory we have at
our disposal in order to deal 
with many--angular momenta quantum systems whose interactions are
modeled (or can be well approximated) by polylocal two--body--type interactions.
It is worth noting that two--body interactions were shown to be sufficient
to implement universal quantum computation in the decoherence--free subspaces approach
according to \cite{KeLi} (see other references therein). 
$\blacktriangle$

\section{Gates}
\subsection{$j$--gates}

By $j$--{\em gates} we mean unitary transformations on the computational Hilbert spaces
\eqref{genba} which act on the set of the spin variables 
$\{j_1,j_2,j_3,\ldots,j_{n+1},$ $k_1,k_2,k_3,\ldots,k_{n-1}\}$ of the eigenstates
without changing the quantum numbers $J$ and $M$.

According to the recoupling theory of angular momenta \cite{Lituani}, \cite{BiLo9} 
(Topic 12), \cite{Russi} (see also \cite{Belgi99}, \cite{Belgi02})
the most general unitary transformation between two computational states
characterised by different binary coupling schemes $\mathfrak{b}$ and $\mathfrak{b}'$

\begin{equation*}
|\,[j_1,\,j_2,\,j_3,\ldots,j_{n+1}]^{\mathfrak{b}\,};k_1^{\mathfrak{b}\,},\,k_2^{\mathfrak{b}\,}
,\ldots,k_{n-1}^{\mathfrak{b}\,}\,;\,JM\, \rangle\; \longmapsto
\end{equation*}
\begin{equation}\label{btrans}
\longmapsto\;
|\,[j_1,\,j_2,\,j_3,\ldots,j_{n+1}]^{\mathfrak{b}'\,};k_1^{\mathfrak{b}'\,},\,k_2^{\mathfrak{b}'\,}
,\ldots,k_{n-1}^{\mathfrak{b}'\,}\,;\,JM\, \rangle
\end{equation}

\noindent is a recoupling coefficient of $SU(2)$ (or $3nj$--symbol) denoted by

\begin{equation}\label{formal3nj}
{\mathcal U}_{\,3nj}\;
\left[\,\begin{array}{ccc}
k_1^{\mathfrak{b}} & \ldots & k_{n-1}^{\mathfrak{b}}\\
k_1^{\mathfrak{b}\,'} & \ldots & k_{n-1}^{\mathfrak{b}\,'}\\
\end{array}\,\right]\;\;\doteq\;\;{\mathcal U}_{\,3nj}\;[\mathfrak{b};\mathfrak{b}\,'],
\end{equation}

\noindent where the variables $\{j\}$, $\{k^{\mathfrak{b}}\}$,
$\{k^{\mathfrak{b}\,'}\}$,
$J,\,M$ appearing in 
states \eqref{btrans} have been partially or totally dropped ({\em cfr.} the shorthand notation
for Hilbert spaces and state vectors introduced in \eqref{genba}). According to this definition,
$|{\mathcal U}_{\,3nj}\;[\mathfrak{b};\mathfrak{b}\,']|^2$
represents the probability that a quantum system prepared in the state  
$|\mathfrak{b};JM\rangle_n$ is measured in the state $|\mathfrak{b}\,';\,JM\rangle_n$. 

Note that \eqref{formal3nj}
is a reduced tensor operator, namely the magnetic quantum numbers do not appear at all in
its expression in view of the Wigner--Eckart theorem \cite{BiLo8}. More explicitly, by looking at 
the {\em re}--coupling between two states 
of the type \eqref{btrans} --but with different $M, M'$-- we would get

\begin{equation*}
\langle\,[j_1,\ldots,j_{n+1}]^{\mathfrak{b}'\,};k_1^{\mathfrak{b}'\,},
\ldots,k_{n-1}^{\mathfrak{b}'\,}\,;\,JM'\,{\bf |}
\,[j_1,\ldots,j_{n+1}]^{\mathfrak{b}\,};k_1^{\mathfrak{b}\,}
,\ldots,k_{n-1}^{\mathfrak{b}\,}\,;\,JM\, \rangle
\end{equation*}
\begin{equation}\label{recou3nj}
=\; \delta_{M\,M'}\;
\langle\,[\ldots]^{\mathfrak{b}'\,};k_1^{\mathfrak{b}'\,},
\ldots,k_{n-1}^{\mathfrak{b}'\,}\,;\,JM'\,{\bf ||}
\,[\ldots]^{\mathfrak{b}\,};k_1^{\mathfrak{b}\,}
,\ldots,k_{n-1}^{\mathfrak{b}\,}\,;\,JM\, \rangle,
\end{equation}

\noindent where $\delta$ is the Kronecker delta and the symbol $\langle \ldots ||\ldots\rangle$
denotes the reduced operator (which coincides with 
${\mathcal U}_{\,3nj}\;[\mathfrak{b};\mathfrak{b}\,']$
in \eqref{formal3nj} for $M=M'$).

For each fixed $n$ there exist inequivalent types of $3nj$ symbols
(up to degenerate cases and phase transforms, see below): one $6j$ symbol,
one $9j$, two $12j$ symbols, five $15j$'s, eighteen $18j$'s etc. \cite{Lituani}.
The recoupling coefficients classified as {\em type I} and {\em type II} 
may be expressed through
single sums of products of $6j$ symbols, while {\em types III, IV, V,} $\ldots$
(appearing for $n\geq 5$) may be represented either by single sums of more complex products
of $6j$ and $9j$ symbols or by multiple sums of products of symbols of lower orders.
In the present context we do not really need any such complicated expressions 
since we take advantage of the results illustrated in details 
in \cite{BiLo9} (Topic 12) and collected in the following:

\vspace{.5cm}
{\bf Theorem} (Biedenharn--Louck).\\
For each $n$ any
${\mathcal U}_{\,3nj}\;[\mathfrak{b};\mathfrak{b}\,']$
  is the composition of (a finite number of) two
elementary unitary transformations, namely\\
\begin{itemize}
\item Racah transform
\begin{equation}\label{racah}
{\cal R}\;:\,| \dots (\,( a\,b)_d \,c)_f \dots;JM \rangle\; \mapsto \;
\,|\dots( a\,(b\,c)_e\,)_f \dots;JM \rangle, 
 \end{equation}

\item Phase transform
\begin{equation}\label{phase}
\Phi\; :\,|\dots (a\,b)_c \dots;JM \rangle \mapsto \;|\dots (b\, a)_c \dots;JM\rangle,
 \end{equation}
\end{itemize}

\noindent where we are using Latin  letters $a,b,c,\ldots$ to denote both incoming ($j'$s
in the previous notation) 
and intermediate ($k'$s) spin quantum numbers.

The explicit expression of \eqref{racah} reads

\begin{equation*}
|(a\,(b\,c)_e\,)_f\,;M\rangle\,=
\end{equation*}
\begin{equation}\label{6j} 
\sum_{d}\,(-1)^{a+b+c+f}\; [(2d+1)
(2e+1)]^{1/2}
\left\{ \begin{array}{ccc}
a & b & d\\
c & f & e
\end{array}\right\}\;|(\,(a\,b)_d \,c)_f \,;M\rangle,
\end{equation}
\noindent where there appears the Racah--Wigner $6j$ symbol of $SU(2)$ and the weights $(2d+1)$,
$(2e+1)$ are the dimensions of the irreps labeled by $d$ and $e$, respectively.\\
Recall that the $6j$ symbol may be expressed in turn as a sum over magnetic
quantum numbers of the product of
four Clebsch--Gordan coefficients with entries in the set
$\{a,b,c,d,e,f;$ $m_a,m_b,m_c,m_d,m_e,m_f\}$, where $m_a$ ($-a \leq m_a\leq a$ in
integer steps) is the magnetic quantum number associated with the spin variable $a$ 
(and similarly for the others) \cite{Russi}.
The numerical value of the $6j$ symbol depends on normalization: we are tacitly assuming
through the whole paper the standard Condon--Shortley conventions.

Finally, the phase transform \eqref{phase} reads 
\begin{equation}\label{exphase}
 |\dots (a\,b)_c \dots;JM \rangle \,=\,(-1)^{a+b-c}\;|\dots (b\, a)_c \dots; JM \rangle.
\;\;\;\;\;\;\blacktriangle
\end{equation}

Summing up, the unitary transformations 
$\{\,{\mathcal U}_{\,3nj}\;[\mathfrak{b};\mathfrak{b}\,']\,\}$
connecting any pair of computational states $|\mathfrak{b}\,';\,JM\rangle_n$, 
$|\mathfrak{b};\,JM\rangle_n$ are identified as $j$--gates in the present 
computational context and the theorem stated above naturally single 
out two types of universal {\em elementary} $j$--gates.
In particular, the Racah transform switches the current coupling
scheme to a physical inequivalent one ({\em cfr.} the explicit expression \eqref{6j}).

As a further remark note that the elementary unitary $j$--gates \eqref{racah}
and \eqref{phase} are in one--to--one correspondence with elementary topological 
operations between pairs of rooted labeled binary trees, Rotations and Twists
respectively ({\em cfr.} Appendix A2 and pay attention to the fact that $''$rotation$\, ''$
will be used in the following part of the present Section in a totally different way).

\subsection{$M$--gates}

As anticipated in Section 1, $M$--gates are implemented --inside each 
computational Hilbert space ${\cal H}^{J}_{\,n}\;(\mathfrak{b})$
introduced in \eqref{genba}-- by the Wigner $D$--functions $D^{\,J}_{M\,M'}$, namely
the matrix elements of the (unitary) rotation operator
$\mathfrak{D}^{\,J}_{M\,M'}$ in the $JM$ representation.
By using Euler angles $\alpha$, $\beta$, $\gamma$ to specify the rotation, 
the Wigner $D$--functions --or W--rotation matrices--
are defined by

\begin{equation}\label{genW}
\langle\,JM\,|\mathfrak{D}^{\,J}_{M\,M'}\;(\alpha \beta \gamma)\,|\,J'M'\rangle
\;\doteq\; \delta_{JJ'}\;D^{\,J}_{M\,M'}\;(\alpha \beta \gamma),
\end{equation}

\noindent where $0\leq \alpha < 2\pi$; 
$0\leq \beta \leq \pi$ or $2\pi\leq \beta \leq 3\pi$; 
$0\leq \gamma < 2\pi$ and we adopt in this Section the notations of \cite{Russi} (see
also \cite{BiLo8} for a general treatment of $SU(2)$ representation theory).
A W--matrix can  be always expressed as

\begin{equation}\label{redD}
D^{\,J}_{\,M\,M'}\;(\alpha \beta \gamma)\;=\;e^{-iM\alpha}\;d^{\,J}_{\,M\,M'}\;(\beta)\;
e^{-iM'\gamma},
\end{equation}

\noindent where $d^{\,J}_{\,M\,M'}\;(\beta)$ is the {\em reduced} W--matrix.

Generally speaking, W--rotation matrices appears when we look at 
transformations of the wave function  of a quantum mechanical 
system under a coordinate rotation. Given a basis vector 
of ${\cal H}^{J}_{\,n}\;(\mathfrak{b})$
as in \eqref{genba}, its dependence on polar coordinates $(\theta,\phi)$ is better described
if we change our previous notation according to

\begin{equation}\label{psiba}
|\mathfrak{b}\;;JM\,\rangle_n\;\;\;\rightsquigarrow\;\;\;
^{(\mathfrak{b})}\,\psi^J_M(\theta,\phi) 
\end{equation}

\noindent where we dropped the subscript $n$ for simplicity.
Then the action of a W--matrix \eqref{genW} on any such basis vector is expressed by

\begin{equation}\label{rotwave}
^{(\mathfrak{b})}\,\psi^J_{M'}\;(\theta',\phi')\;=\;
\sum_{M=-J}^{+J}\;^{(\mathfrak{b})}\,\psi^J_M\;(\theta, \phi)
\;\;D^{\,J}_{M\,M'}\;(\alpha \beta \gamma),
\end{equation}

\noindent where $(\theta,\phi)$ and $(\theta',\phi')$ are polar angles in the initial and
rotated coordinate systems, respectively.
The action of a rotation on the components of vectors ({\em cfr.} the decomposition written in \eqref{binvec})
involves the Hermitean conjugate $\tilde{D\,}^J_{M'M}$ of the corresponding $D^J_{MM'}$, namely
$^{(\mathfrak{b})}\,a'^{\,J}_{M'}$ $=\sum \tilde{D\,}^J_{M'M}$ $^{(\mathfrak{b})}\,a^J_M$, where the sum
is over $M$ as in \eqref{rotwave}. 

\vspace{0.5cm}

It is worth to recall  that in the general case ($N$ any integer and 
$j_1,j_2,\ldots$, $j_{n+1}$ chosen in $\{0,1/2,1,3/2,\ldots\}$)
the {\em reducible} $(2J+1)\times$ $(2J+1)$ W--rotation matrix
$D^J_{M M'}$ 
will admit a block diagonal decomposition into irreducible rotation matrices of 
lower ranks. On the other hand, by changing $N$ and/or the values of the
incoming spins, the resulting $D^{\bar{J}}_{\bar{M} \bar{M'}}$ 
will decompose into different elementary blocks. 
From the computational
viewpoint this provides a more general notion of {\em universal} set
of elementary $M$--gates than that currently adopted in (Boolean) quantum information 
schemes, typically given in terms of $2\times2$ and/or $4\times4$ unitary matrices 
\cite{BaBe}. 
  
However, as described in details in Appendix B1, each matrix element of any $D^J_{M M'}$ 
can be {\em factorized} in a well--defined way, and this procedure 
is independent of the binary bracketing structure of ${\cal H}^{J}_{\,n}\;(\mathfrak{b})$.
The explicit expression of such factorisation
given in \eqref{NWcsser1} can be written symbolically as in \eqref{NWcsser2}, namely 

\begin{equation*} 
\boldsymbol{D}^{\,J}\;(\alpha \beta \gamma)\;=
\end{equation*} 
\begin{equation}\label{NWcsser} 
\sum_{\{m,\,m'\}}\;\,
\prod_{i=1}^{N}\;\left(\,
\boldsymbol{C}^{\,\kappa_i}_{\kappa_{i-1}\;j_i}\;\,
\boldsymbol{D}^{\,j_i}\;(\alpha \beta \gamma)\;\,
\boldsymbol{C}^{\,\kappa_i\,}_{\kappa_{i-1}\;j_i}\;\right),
\end{equation} 

\noindent where we dropped the matrix indices 
$M,M'$,$m_i,m'_i$
on the W-matrices $\boldsymbol{D}^{\,J}$,
$\boldsymbol{D}^{\,j_i}$ $(i=1,2,\ldots,N\equiv n+1)$ and similarly
$\boldsymbol{C^{\bullet}_{\bullet \bullet}}$'s are Clebsch--Gordan coefficients 
with $m$--type entries omitted.
The summation is over all magnetic quantum numbers of the angular momentum operators
$\{{\bf J}_i\}$, while $\{\kappa_i\}$ are spin quantum numbers associated with the 
intermediate operators defined in \eqref{keys}.

In \eqref{NWcsser} there appear $N$ ($\equiv$ $\#$  of incoming spins)
factors, each containing a W--matrix in the irreducible
$j_i$--th representation of dimension $(2j_i+1)$, and a total amount 
of $2N$ Clebsch--Gordan coefficients. As explained at the end of Appendix B1, each
matrix element of
$\boldsymbol{D}^{\,j_i}$ may be further factorized into the (sum of) product of
$2j_i$ W--matrices in the fundamental $j=1/2$ representation of $SU(2)$. The explicit form 
of $D^{\frac{1}{2}}_{m\,m'}$ $(m,m'\in \{\frac{1}{2},\,-\frac{1}{2}\})$
reads

\begin{equation}\label{d12}
D^{\frac{1}{2}}_{m\,m'}\;(\alpha\,\beta\,\gamma)\;=\;
\begin{pmatrix}
e^{-i\alpha/2}\cos (\beta/2)\;e^{-i\gamma/2} &  -e^{-i\alpha/2}\sin (\beta/2)\;e^{i\gamma/2}\\
e^{i\alpha/2}\sin (\beta/2)\;e^{-i\gamma/2} &  e^{i\alpha/2}\cos (\beta/2)\;e^{i\gamma/2}
\end{pmatrix}.
\end{equation}

\noindent As a consequence of the above remarks we conclude that the $''$elementary factors$\, ''$
appearing in the right hand side of \eqref{NWcsser} needed to determine one matrix element
of $\boldsymbol{D}^{\,J}$, namely $D^{\,J}_{M\,M'}\;(\alpha\,\beta\,\gamma)$ for some $M M'$,
are  

\begin{align}\label{couM1}
2N &\;\;\mbox{C--G coefficients}\nonumber\\
2J & \equiv\,2\sum_{i=1}^N\;j_i\;\;\mbox{W--matrices}\;\,D^{\frac{1}{2}}_{m\,m'}
\end{align}

\noindent and the number of factors one needs in order to evaluate the whole 
$\boldsymbol{D}^{\,J}$ amounts to $[2(J+N)]^{(2J+1)^2}$. These estimates represent in fact 
upper bounds on the number of factors, since we
may reduce the number of elementary W--matrices by employing some $3\times3$ matrices 
$\boldsymbol{D}^{\,1}$ of the $j=1$ irrep.
Moreover, by considering a purely fermionic (bosonic) symmetric $N$--multiplet, the
expression \eqref{NWcsser}
does not contain C--G coefficients anymore and the number of elementary factors 
to be taken into account in \eqref{couM1} is simply $2J\equiv N$
($J\equiv N$, respectively), as discussed in Appendix B1.
\vspace{.5cm}

{\bf Remark 3.1.} Our framework turns out to be much richer than the Boolean case,
but contains the usual Boolean gates as particular examples. Note also that
$D^{\frac{1}{2}}_{m\,m'}$ is  an {\em elementary} gate in any situation; it is also {\em 
universal} for the two particular cases discussed above. In particular, the $N$ 
spin--$\frac{1}{2}$ case is compatible with the scheme proposed in \cite{KeLi}, Sect.VII. 

{\bf Remark 3.2.} In each representation labeled by $J$ the W--matrices form a group under 
multiplication, namely

\begin{equation}\label{Wgrou}
\boldsymbol{D}^J\,(\alpha_1 \beta_1 \gamma_1)\;\boldsymbol{D}^J\,(\alpha_2 \beta_2 \gamma_2)\;
\;=\;\boldsymbol{D}^J\,(\alpha \beta \gamma),
\end{equation}

\noindent where
$(\alpha \beta \gamma)$ are related to $(\alpha_1 \beta_1 \gamma_1)$
and $(\alpha_2 \beta_2 \gamma_2)$ by quite involved expressions  (see \cite{Russi} Ch. 1.4.7).

{\bf Remark 3.3.}
Instead of W--matrices we 
could have used other parametrisations for rotations, for instance the matrices $U^J_{M M'}$
defined in terms of rotation axis and rotation angle. For completeness we collect
in Appendix B2 some standard formulas
relating these two types of transformations together with some explicit examples \cite{Russi}.
$\blacktriangle$ \\ 

\section{Spin Network Quantum Circuit}

By exploiting the basic ingredients introduced in the previous
 sections (computational Hilbert spaces, $j$--gates and $M$--gates) we present here the structural
 setting of a quantum simulator $\mathfrak{M}$ --the spin network 
simulator-- modeled as a generalised 
({\em i.e.} not Boolean) quantum circuit model. In a broader sense such a computing machine could be reinterpreted 
as a concrete realization of what should be a Quantum Automaton (see {\em e.g.} \cite{BeVa}, \cite{MoCr}),
namely a theoretical 
framework able to deal consistently with quantum languages and grammars \cite{GaMaRa}.

\subsection{Combinatorial kinematics}

 The computational space of the simulator $\mathfrak{M}$ turns 
out to be modeled on an $SU(2)$--decorated graph (or {\em spin network}, according
to Penrose's similar structures introduced  in \cite{Pen}). 

For each fixed $n$ 
the underlying network structure is denoted for the moment by $\mathbf{G}_n(V,E)$, where $V$ and 
$E$ are the vertex and edge sets of the graph, respectively. Both the vertices and the 
edges (arcs connecting pairs of vertices) of $\mathbf{G}_n(V, E)$ are decorated by algebraic
 objects 
from $SU(2)$--representation theory we introduced previously. The vertices are in one-to-one  
correspondence with the set of computational Hilbert spaces
$\mathcal{H}^J_n (j_1, j_2,\ldots, j_{n+1}; k_1, k_2,\ldots, k_{n-1};\,J M)$ $\doteq
\mathcal{H}^J_n (\,\mathfrak{b}$) introduced in \eqref{genba}

\begin{equation}\label{vert1}
\text{vertex set}\,V\;\longleftrightarrow\; \{\mathcal{H}^J_n \,(\mathfrak{b}) \}
\end{equation}

\noindent and since each $\mathcal{H}^J_n (\mathfrak{b})$ has dimension $(2J + 1)$ over 
$\mathbb{C}$ there exists one isomorphism

\begin{equation}\label{Hniso}
\mathcal{H}^J_n (\mathfrak{b})\;\;\; \cong _{\,\mathfrak{b}}\;\;\; \mathbb{C}^{2J+1}
\end{equation}

\noindent
for each admissible binary coupling scheme $\mathfrak{b}$ of $(n + 1)$ incoming spins. Thus each 
vertex in the spin network is decorated with a copy of $\mathbb{C}^{2J+1}$

\begin{equation}\label{vert2}
\mathbb{C}^{2J+1}\;\;\;\rightsquigarrow\;\;\;  v(\mathfrak{b})\;\in \,V, 
\end{equation}

\noindent
where each $v \in V$ is labeled by the unique corresponding $\mathfrak{b}$. 

The construction of the graph proceeds by establishing the connections between vertices. 
The edge set $E = \{e\}$ of $\mathbf{G}_n(V, E)$ is a subset of the Cartesian
product $(V \times V )$ selected by the action of elementary $j$--gates. More precisely, 
an (undirected) arc between two vertices $v(\mathfrak{b})$ and $v(\mathfrak{b}')$

\begin{equation}\label{edge1}
e\,(\mathfrak{b},\mathfrak{b}')\;\doteq \;(v(\mathfrak{b}),\, v(\mathfrak{b}')) 
\;\in \;(V \times V)
\end{equation}

\noindent
exists if, and only if, the underlying Hilbert spaces are related to each other by 
one of elementary unitary operations defined in Biedenharn--Louck Theorem of Section 3.1. 
Note that at the kinematical level the resulting decorated edges in the net are to be considered 
undirected since any such operation is invertible.

This general combinatorial structure can be concretized in two different ways, 
depending on what types of binary bracketing schemes --associated with the vertex 
set of $\mathbf{G}_n(V, E)$ as stated in \eqref{vert1}-- we are going to consider as inequivalent
 (we refer to Appendix A2 for more details on the combinatorics of the resulting graphs). 
If we distinguish pairs $\mathfrak{b}$, $\mathfrak{b}'$ which differ either by a Racah transform 
$\mathcal{R}$ \eqref{racah} or by a Phase 
transform $\Phi$ \eqref{phase}, then the structure we get is a Twist--Rotation graph

\begin{equation}\label{TR1}
\mathbf{G}^{TR}_n \,(V, E)\;\doteq\; {\hat{\mathfrak{G}}}_n \,(V, E) 
\end{equation}

\noindent with

\begin{align}\label{TR2}
\hat{\mathfrak{G}}_n \;\supset\;\; &  V \,= \,\{v(\mathfrak{b}) ,\,\text{all admissible}\,\mathfrak{b}\}\nonumber\\ 
\hat{\mathfrak{G}}_n \;\supset\;\; & E \,=\,\{e(\mathfrak{b},\,\mathfrak{b}')\;
\leftrightarrow\;\mathcal{R},\,\Phi\,\}.
\end{align}

\noindent For each $n$ the Twist--Rotation graph is a regular, cubic (trivalent) graph 
representing pictorially all possible types of $3nj$ symbols introduced in Section 3.1 
and extensively analysed in literature ({\em cfr.} \cite{Belgi99}, \cite{Belgi02}, \cite{BiLo9}, \cite{AqCo}).
In Fig. 21 of 
Appendix A2 the Twist--Rotation graph $\hat{\mathfrak{G}}_3$ (binary coupling schemes of four angular momenta 
and associated $9j$ symbols) is depicted.

 Since quantum states which differ by phase 
transformations give rise to the same physical probabilities of observables, it seems 
quite natural to consider a combinatorial structure in which $\mathfrak{b}$, $\mathfrak{b}'$ 
correspond to different 
vertices only if the states are related by a Racah transform (recall that in this case there 
is a change of one intermediate spin variable). This reduction operation on the 
Twist--Rotation graph \eqref{TR1} involves an equivalence relation to be imposed on
the vertex set. Moreover, each of the surviving edges turns out to be associated with a 
Racah transform $\mathcal{R}$, but there could appear some additional weights and/or phase factors 
in its explicit expression \eqref{6j}. With these premises we define the Rotation graph

\begin{equation}\label{R1}
 \mathbf{G}^{R}_n \,(V, E)\;\doteq\;\mathfrak{G}_n \,(V, E) 
\end{equation}

\noindent with

\begin{align}\label{R2}
\mathfrak{G}_n \;\supset\;\; &  V \,= \,\{\text{equivalence classes of}\,
v(\mathfrak{b})\}\nonumber\\ 
\mathfrak{G}_n \;\supset\;\; & E \,=\,\{e(\mathfrak{b},\,\mathfrak{b}')\;
\leftrightarrow\;\mathcal{R},\,\text{mod \,weights/phases}\}.
\end{align}

\begin{center}
{\bf Figure 1} 
\end{center}

\noindent A picture of $\mathfrak{G}_3$ is given in Fig. 1, while more details 
on the above construction are 
collected in Appendix A2. Here we just recall that the Rotation graph 
$\mathfrak{G}_n$ is a regular 
(not planar) $2(n - 1)$--valent graph with a number of vertices given by

\begin{equation}\label{R3}
|V|\,\equiv\,\text{card}\, V \;=\;(2n - 1)!! 
\end{equation}

\noindent
where $(\,)$!! denotes the double factorial number. In the following we shall deal 
mainly with the Rotation graph structure, but we point out in advance that applications 
of such combinatorial machinery in the computing context work on the Twist--Rotation graph 
as well.

On the spin network the action of a $j$--gate defined by a recoupling coefficient
$\mathcal{U}_{3nj} \,[\mathfrak{b}; \mathfrak{b}']$ introduced in \eqref{formal3nj} 
is represented formally by a {\em piecewise path} in $\mathfrak{G}_n$ connecting the vertices 
$v(\mathfrak{b}$) and $v(\mathfrak{b}')$

\begin{equation}\label{3njpath}
v(\mathfrak{b})\;\xrightarrow{\mathcal{U}_{3nj}\; [\mathfrak{b}; \mathfrak{b}']}\;
v(\mathfrak{b}'). 
\end{equation}

\noindent
Two state vectors ({\em e.g.} basis vectors, for simplicity, with the same value of the 
q--number $M$) in the Hilbert spaces \eqref{vert1} attached to the vertices 
labeled by $\mathfrak{b}$ (the initial point) and $\mathfrak{b}'$ (the terminal point) 
are denoted by

\begin{equation}\label{3njst}
|\mathfrak{b}; \,JM \rangle_n \doteq\, |\text{in}\,(\mathfrak{b});\,JM\rangle_n\;\;\;\text{and}
\;\;\; 
|\mathfrak{b}'; \,JM \rangle_n \doteq\, |\text{out}\,(\mathfrak{b}');\,JM\rangle_n \, , 
\end{equation}
and can be interpreted consistently $''$in$\, ''$ and $''$out$\, ''$ as referring 
to an initial state (input) and a final state (output), respectively.\\ 
To deal with the formal expression \eqref{3njpath} we need some results from angular momenta 
recoupling theory summarised in the following

\vspace{.5cm}
{\bf Theorem} (see \cite{Lituani}, \cite{BiLo9} Topic 12, and the original references therein).
 
Consider all possible paths in $\mathfrak{G}_n$ connecting two states as in \eqref{3njst}. 
Then the transition 
probability amplitudes for {\em any} pair of paths with given endpoints, say 
$\gamma_1$ and $\gamma_2$, are equal

\begin{equation}\label{3njampl}
\langle \text{in}(\mathfrak{b});JM\,|\, \text{out}(\mathfrak{b}'); JM\rangle_{\gamma_1}\,=\,
\langle \text{in}(\mathfrak{b});JM\,|\, \text{out}(\mathfrak{b}'); JM\rangle_{\gamma_2}
\end{equation}

\noindent since we can freely deform such paths one into each other. As a consequence of that, 
probabilities too turn out to be equal owing to the fact that

\begin{equation}\label{3njpro}
|\,\mathcal{U}_{3nj}\, [\mathfrak{b};\mathfrak{b}']\,|^2\;\equiv\;
|\,\langle \text{in}(\mathfrak{b});JM\,|\, \text{out}(\mathfrak{b}'); JM\rangle \,|^2 
\end{equation}

\noindent is actually the probability that a quantum system prepared in the state $''$in$\, ''$ 
will be measured in the state $''$out$\, ''$. The proof of \eqref{3njampl} 
relies on the existence 
of fundamental algebraic identities involving $6j$ symbols, namely\\

$\bullet$ the Biedenharn--Elliott identity

\begin{align}\label{BEid}
\sum_{x}(-)^{R+x}\,(2x+1)&\begin{Bmatrix}
a & b & x\\
c & d & p
\end{Bmatrix}
\begin{Bmatrix}
c & d & x\\
e & f & q
\end{Bmatrix}
\begin{Bmatrix}
e & f & x\\
b & a & r
\end{Bmatrix}\nonumber\\
& =\;
\begin{Bmatrix}
p & q & r\\
e & a & d
\end{Bmatrix}
\begin{Bmatrix}
p & q & r\\
f & b & c
\end{Bmatrix};
\end{align}

$\bullet$ the Racah identity

\begin{equation}\label{Rid}
\sum_{x}(-)^{p+q+x}\,(2x+1)\,
\begin{Bmatrix}
a & b & x\\
c & d & p
\end{Bmatrix}
\begin{Bmatrix}
a & b & x\\
d & c & q
\end{Bmatrix}\,=\,
\begin{Bmatrix}
a & c & q\\
b & d & p
\end{Bmatrix},
\end{equation}

\noindent where the spin variables $\{a, b, c,\ldots, x\}$ run over 
$\{0,\frac{1}{2},1,\frac{3}{2},\ldots\}$ 
and must satisfy suitable triangular inequalities inside each $6j$ symbol (otherwise the 
symbol itself would vanish). The weight $(2x + 1)$ is the dimension of the representation
 labeled by the quantum number $x$, the sum over $x$ is constrained only by the triangular 
conditions quoted above and $R$ in the phase factor of
the first identity is the combination $(a + b + c + d + e + f + p + q + r)$. 
Note that these identities, together with the orthogonality relation

\begin{equation}\label{ort6j}
\sum_{x}\,(2x+1)\,
\begin{Bmatrix}
a & b & x\\
c & d & p
\end{Bmatrix}
\begin{Bmatrix}
c & d & x\\
a & b & q
\end{Bmatrix}\,=\,
\frac{\delta_{pq}}{(2p+1)},
\end{equation}

\noindent define uniquely the Racah--Wigner $6j$ symbol (considered as the hypergeometrical
polynomial which generates the Askey hierarchy \cite{Ask}).

Without entering into details about the proof of the above theorem,
we collect below some remarks which should make the rationale of its proof as clear as possible.

\begin{itemize}
\item The spin network $\mathfrak{G}_2$ (see Fig. 22 in Appendix A2) is a 
closed loop of triangular shape: 
the vertices correspond to the three inequivalent binary couplings of $(n + 1) = 
3$ spins and each edge is associated with a Racah transform. Then Racah identity \eqref{Rid} 
ensures that we may implement the transition from one vertex ($''$in$\, ''$) 
to another ($''$out$\, ''$) 
traversing either the edge connecting them directly or the other two.
\item The spin network $\mathfrak{G}_3$ shown in Fig. 1 is characterised by triangular
and pentagonal closed loops (bounding triangular and pentagonal pla\-quet\-tes). 
Triangles are associated with Racah identity \eqref{Rid} (as happens for $\mathfrak{G}_2$), while 
pentagonal plaquettes turns out to correspond to the Biedenharn--Elliott identity \eqref{BEid}: 
if the chosen path embraces two edges of a pentagon, one can freely deform it  
traversing the other three edges (and viceversa). By using both \eqref{BEid} and 
orthogonality relation 
\eqref{ort6j} we can also deform a piece of path connecting 
two contiguous vertices into a path which 
touches the other four edges of the pentagon (and viceversa). Thus we conclude that the
 improvement of \eqref{3njampl} for $\mathfrak{G}_3$ relies on all the three algebraic 
identities written above.
\item The spin networks $\mathfrak{G}_n$ $(n > 3)$ display plaquettes with other types
of polygonal boundaries. 
Each type of plaquette can be associated with a suitable algebraic 
identity which can be derived from the fundamental ones by making use of the explicit 
expression for the $3nj$ symbol involved. The procedure for improving \eqref{3njampl} 
goes on as in the previous cases.
\end{itemize}
\noindent As a final comment on the equi--amplitudes 
of paths in $\mathfrak{G}_n$ under the action of pure $j$--gates, notice that such result holds 
as far as we are dealing with
the computational space of the simulator at the kinematical level. When we shall 
ask the spin network to perform a computation by means of a sequence of $j$-gates such 
an invariance will be broken (as we are going to explain in the following section). 
$\blacktriangle$ 

\vspace{.5cm}

In order to include in the combinatorial setting the action of $M$--gates (see Section 3.2) 
we may employ either the Wigner rotation matrices \eqref{genW} parametrised by Euler angles or 
$U$--matrices written in terms of rotation axis and rotation angle introduced in Appendix B2: 
here we agree to making use of the former ones as in the rest of the main text. By 
analogy with the formal expression \eqref{3njpath} representing a $j$--gate on the spin network, 
we write formally the action of an $M$--gate on $\mathfrak{G}_n$ as

\begin{equation}\label{Mgate}
v(\mathfrak{b})\;\xrightarrow{\mathcal{D}^J\,(\alpha \beta \gamma)}\;
v(\mathfrak{b}),
\end{equation}

\noindent where we drop the matrix indices $M,M'$ by using the operatorial 
notation as in \eqref{genW}. Since rotations do not alter 
the binary bracketing structure of the computational Hilbert space 
$\mathcal{H}^J_n\,(\mathfrak{b})$
we may activate an M--gate independently at any vertex $v(\mathfrak{b})$ $\subset \mathfrak{G}_n$.
To recover 
the explicit expression of the action of an $M$--gate we have to pick up a vector in 
$\mathcal{H}^J_n (\mathfrak{b})$, change back our notation as in \eqref{psiba} 
(to make the angular dependence explicit) 
and finally recover the expression given in \eqref{rotwave}. However, in order to have at our disposal 
a unified notation for states to be considered as $''$input$\, ''$ and $''$output$\, ''$ in a 
quantum circuit framework, we are forced here to use a hybrid notation by setting

\begin{equation}\label{ibdri}
\mathcal{H}^J_n (\mathfrak{b})\;\doteq\; \{\text{span}\,\; |\mathfrak{b};\theta,\, \phi;\,
JM\rangle_n\,\}. 
\end{equation}

\noindent With this convention we write down the action of an $M$--gate 
(for a given choice of the parameters $\alpha,\beta,\gamma$) on an input (basis) state as

\begin{equation}\label{Winput1}
\mathcal{D}^J \,(\alpha \beta \gamma)\;:\;
|\mathfrak{b};\text{in}(\theta,\, \phi;\,M)J\,\rangle_n\,\rightarrow\,
|\mathfrak{b};\text{out}(\theta',\, \phi';\,M')J\,\rangle_n 
\end{equation}

\noindent where the output state is

\begin{equation}\label{Winput2}
|\mathfrak{b};\text{out}(\theta',\, \phi';\,M')J\,\rangle_n\;=\;
\sum_{M=-J}^{J}\;D^J_{MM'}\,(\alpha \beta \gamma)\,
|\mathfrak{b};\theta,\, \phi;\,JM\,\rangle_n
\end{equation}

\noindent and the input basis state appears in the combination on the right--hand side 
with its particular M--label.

The actions of both types of gates on the spin network can be visualised
by looking at Fig. 2: we can move from one vertex to a different one along an edge
as in \eqref{3njpath} 
(without changing $M$ ) or choose to perform a rotation \eqref{Mgate} inside the computational 
Hilbert space associated with a vertex.

\begin{center}
{\bf Figure 2} 
\end{center}

\noindent According to \eqref{ibdri} and \eqref{binvec}, a suitable 
unified notation for generic state vectors to be used in actual computations should be

\begin{equation}\label{genstate1}
|\mathfrak{b};\theta,\,\phi;\,J\,\rangle_n
\end{equation}

\noindent and consequently input/output states in the particular cases \eqref{3njst}
and \eqref{Winput1} have to be set in the form

\begin{equation}\label{genstate2}
|\text{in}(\mathfrak{b};\,\theta,\, \phi);\,J\,\rangle_n\,;\;
|\text{out}(\mathfrak{b};\theta',\, \phi');\,J\,\rangle_n
\end{equation}

\noindent possibly with additional $M$--labels 
if basis vectors are considered.
 
Summing up, the kinematical ingredients of the spin 
network simulator based on the Rotation graph $\mathfrak{G}_n(V, E)$ are

\begin{align}\label{kinem}
V\;=\;\{v(\mathfrak{b})\} & \leftrightarrow \{\mathcal{H}^J_n\,(\mathfrak{b}\}\nonumber\\
E\;=\;\{e(\mathfrak{b},\mathfrak{b}')\} & \leftrightarrow
\text{elementary}\,j\text{--gates}\nonumber\\
\{\mathcal{D}^J\;:\;\mathcal{H}^J_n(\mathfrak{b}) \;\rightarrow
\mathcal{H}^J_n(\mathfrak{b})\} & \leftrightarrow
M\text{--gates}
\end{align}

\noindent where the discrete structure encoded in $(V, E)$ is endowed with transformations 
$\mathcal{D}^J$ depending on both discrete and continuous parameters ({\em cfr.} 
the points discussed in Section 2).

\vspace{.5cm}

{\bf Remark 4.1.} The combinatorial setting described above can be interpreted as a fiber 
space structure $(V,\mathbb{C}^{2J+1},SU(2)^J)$ where
\begin{itemize}
\item $V = \{v(\mathfrak{b})\}$ is the (discrete) base space;
\item $\mathbb{C}^{2J+1}$ is the typical fiber, a copy of which is attached to each 
$v(\mathfrak{b})$
through the isomorphisms given in \eqref{Hniso};
\item $SU(2)^J$ is the automorphism group of the fiber realized by the $(2J +1) \times
(2J +1)$ W--matrices which form a group under the composition law \eqref{Wgrou};
\item $E = \{e (\mathfrak{b}, \mathfrak{b}')\}$ are arcs connecting pairs of 
contiguous vertices in the
base space ({\em cfr.} \eqref{edge1}), but they may be also considered as mappings

\begin{align}\label{etrans}
V\,\times\,\mathbb{C}^{2J+1}\;& \rightarrow\,
V\,\times\,\mathbb{C}^{2J+1}\nonumber\\
(v(\mathfrak{b}),\,\mathcal{H}^J_n (\mathfrak{b})\,)\,& \mapsto\,
(v(\mathfrak{b}'),\,\mathcal{H}^J_n (\mathfrak{b}')\,)
\end{align}

connecting each given decorated vertex to one of its nearest $2(n-1)$ 
vertices.
\end{itemize}

In such a vector bundle framework one could take advantage of the above 
transport prescriptions (W--matrices along the fiber and maps \eqref{etrans} 
along horizontal sections) to get a notion of $\,''$connection$\, ''$ 
in the total fiber space $V \times \mathbb{C}^{2J+1}$ as illustrated in more details in 
\cite{MaRa2}. This remark opens the possibility 
of discussing relations between the spin network scheme and the holonomic 
q--computation approach. We shall come back on this point in Section 6.1.
$\blacktriangle$ \\ 

\subsection{Dynamics and computing}

The kinematical structure of the spin network $\mathfrak{M}$ 
complies with all the requisites of an universal q--simulator as defined by 
Feynman \cite{Fey}, namely
\begin{itemize}
\item {\em locality}, reflected in the binary bracketing structure of the computational 
Hilbert spaces, which --together with the action of W--rotations -- bears on 
the existence of local interactions;
\item  {\em discreteness of the computational space}, reflected in the combinatorial
structure of $\mathfrak{G}_n$;
\item {\em discreteness of time}, to be discussed below; 
\item {\em universality}, guaranteed 
by the properties of gates we described in Section 3: any unitary transformation 
operating on computational Hilbert spaces can be reconstructed by taking a finite 
sequence of Racah transforms (and possibly phases) followed by the application of 
a finite number of W--rotations.
\end{itemize}

Thus we have explicitly defined the class of $''$exact imitators$\, ''$ of any finite, 
discrete quantum system (described by pure angular momentum states) with no need 
of resorting to the notions coming from the (inherently classical) Boolean circuit 
theory.

In order to describe the dynamical behavior of the spin network, we notice preliminarily 
that the rule to $''$move$\, ''$ from a state (say a vector \eqref{genstate1}) to a nearest one 
have been already established: apply either one $j$--gate or one particular $M$--gate 
(for fixed $\alpha,\beta,\gamma$ ). 
Thus a natural discrete time unit, denoted by $\tau$, can be associated with one elementary 
step in such a cellular automaton scheme. 
However, as pointed out by Feynman himself, this naive assumption can at best make the 
simulator to $''$imitate$\, ''$ time. If we pretend the spin network 
$\mathfrak{M}$ to $''$simulate$\, ''$ time, we 
have to go through a genuine space--time dynamics providing Hamiltonians and intrinsical 
evolution in actual time intervals.

Generally speaking, the basic 
data to implement computation in a circuit model are an input state and a program 
$\mathcal{P}$ giving instructions to manipulate information stored in the machine states: output 
states must belong to the set of $''$accepted$\, ''$ states (if computation halts, as we are 
tacitally assuming). In the spin network $\mathfrak{G}_n$ the choice of a particular program 
$\mathcal{P}$ is 
interpreted as the selection of a subset of unitary transformations

\begin{equation}\label{pro1}
\mathcal{P}\;\,\longleftrightarrow\;\,\{\mathfrak{U}_{\mathcal{P}}\}\;
\subset\;\{\mathfrak{U}\}
\end{equation}

\noindent among all the kinematically allowed $\{\mathfrak{U}\}$. Since we are going to 
deal with sequences of states we are forced to change again our last notation \eqref{genstate1} 
into a simplified one, namely

\begin{equation}\label{genstate3}
|\mathfrak{b};\,\theta,\,\phi;\,J\rangle_n\;\;\rightsquigarrow\;|\mathfrak{v}\rangle_n.
\end{equation}

\noindent Then a computation based on the program $\mathcal{P}$, represented formally as

\begin{equation}\label{progr1}
|\mathfrak{v}_{\text{in}}\,\rangle_n\;
\xrightarrow{\mathfrak{U}_{\mathcal{P}}}\;
|\mathfrak{v}_{\text{out}}\,\rangle_n
\end{equation}

\noindent is a collection of directed paths in $\mathfrak{G}_n$, all starting from the input 
state $|\mathfrak{v}_{\text{in}}\,\rangle_n$ and ending in some accepted
$|\mathfrak{v}_{\text{out}}\,\rangle_n$.
By a directed path 
we mean a (time) ordered sequence of states

\begin{equation}\label{dirpath1}
|\mathfrak{v}_{\text{in}}\,\rangle_n\equiv
|\mathfrak{v}_{0}\,\rangle_n\rightarrow
|\mathfrak{v}_{1}\,\rangle_n\rightarrow\cdots\rightarrow
|\mathfrak{v}_{s}\,\rangle_n\rightarrow\cdots\rightarrow
|\mathfrak{v}_{L}\,\rangle_n\equiv
|\mathfrak{v}_{\text{out}}\,\rangle_n
\end{equation}

\noindent where $s=0,1,2,\ldots ,L$ is the lexicographical labelling of the states along the 
given path and $L$ is the length of the path, which turns out to be proportional to the 
time duration of the computation process $L \cdot \tau \doteq T$ in units of the 
discrete time step
$\tau$. The integer $L$ characterising the particular directed path 
in \eqref{dirpath1} represents the
 number of time--ordered elementary operations (computational steps) needed to 
get $|\mathfrak{v}_{\text{out}}\,\rangle_n$ from
$|\mathfrak{v}_{\text{in}}\,\rangle_n$
by employing the program $\mathcal{P}$. It should be clear that 
from the same input 
$|\mathfrak{v}_{\text{in}}\,\rangle_n$
the program $\mathcal{P}$
could select different paths to get (possibly) different output states. 
For instance we may represent by

\begin{equation}\label{dirpath2}
|\mathfrak{v}_{\text{in}}\,\rangle_n\equiv
|\mathfrak{v}_{0}\,\rangle_n\rightarrow
|\mathfrak{v}'_{1}\,\rangle_n\rightarrow\cdots\rightarrow
|\mathfrak{v}'_{s}\,\rangle_n\rightarrow\cdots\rightarrow
|\mathfrak{v}'_{L'}\,\rangle_n\equiv
|\mathfrak{v}'_{\text{out}}\,\rangle_n
\end{equation}

\noindent another path of length $L'$ in the collection \eqref{progr1} ending in an accepted 
$|\mathfrak{v}'_{\text{out}}\,\rangle_n$. 
Each arrow in \eqref{dirpath1} or \eqref{dirpath2} stands for one of the elementary 
operations (Racah transforms, 
Wigner rotations) described in the previous sections ({\em cfr.} also Fig. 2). 
Using our current notation an elementary computational step is

\begin{align}\label{elemstep}
\text{either}\;& 
|\mathfrak{v}_{s}\,\rangle_n\;\xrightarrow{\mathcal{R}}
|\mathfrak{v}_{s+1}\,\rangle_n
\nonumber\\
\text{or}\;&
|\mathfrak{v}_{s}\,\rangle_n\;\xrightarrow{\mathcal{D}(\alpha \beta \gamma)}\;
|\mathfrak{v}_{s+1}\,\rangle_n
\end{align}

\noindent and the expression of a particular unitary transformation 
$\mathfrak{U}_{\mathcal{P}}$ (of length $L$) in \eqref{progr1} turns out to be a well 
defined time--ordered composition

\begin{equation}\label{progr2}
\mathfrak{U}_{\mathcal{P}}\;=\;
\mathcal{U}_{L}\,\circ \,
\mathcal{U}_{L-1}\,\circ \,\cdots \circ
\mathcal{U}_{2}\,\circ \,
\mathcal{U}_{1} \, , 
\end{equation}

\noindent where each $\mathcal{U}$ is given by one of the operations \eqref{elemstep}. 

The framework developed so far 
is an extremely flexible and powerful circuit modelization of quantum computing and 
we shall examine later on in this section the range of different types of computations 
that can be actually carried out. However, to complete the dynamical setting based 
on the identification made in \eqref{pro1} and \eqref{progr1} of a program 
$\mathcal{P}$ with a collection of 
directed paths in $\mathfrak{G}_n$, we have to call into play algorithms. Then the 
program $\mathcal{P}(\mathcal{A})$ 
to perform a particular algorithm $\mathcal{A}$ is the specification of a suitable directed 
path in the collection \eqref{progr1}, starting from a given $|\mathfrak{v}_{\text{in}}\rangle_n$
and ending in one particular $|\mathfrak{v}_{\text{out}}\rangle_n$.
The associated unitary transformation is denoted by

\begin{equation}\label{proalg1}
\mathcal{P}(\mathcal{A})\;\;\longleftrightarrow\;\; \mathfrak{U}_{\mathcal{P}(\mathcal{A})}\;
\subset\;\{ \mathfrak{U}_{\mathcal{P}}\}
\end{equation}

\noindent and for the path itself we may use the same notation as in \eqref{dirpath1}. 
Alternatively, by broadening the meaning of the symbol $\mathcal{P}(\mathcal{A})$, 
we agree that it represents also an ordered sequence of labelings

\begin{equation}\label{proalg2}
\mathcal{P}(\mathcal{A})\;\longleftrightarrow\; \{s=0,1,2,\ldots, L(\mathcal{P}(\mathcal{A})) \}  
\end{equation}

\noindent which turns out to be in one--to--one correspondence 
with the states $\{|\mathfrak{v}_{s}\rangle_n$,
$s=0,1,2,\ldots , L(\mathcal{P}(\mathcal{A})) \}$. The time lapse required to get the 
output is nothing but the length of the path in units of the discrete time step $\tau$ , namely

\begin{equation}\label{timealg}
T\,(\mathcal{P}(\mathcal{A}))\,=\, L(\mathcal{P}(\mathcal{A}))\cdot \tau.
\end{equation}

A circuit--type computation process in $\mathfrak{G}_n$ based on a program 
$\mathcal{P}$ performing the 
algorithm $\mathcal{A}$ (which could be formally written as in \eqref{progr1}) 
is actually represented 
by the expectation value of the unitary operator \eqref{proalg1} evaluated between the given 
input and output

\begin{equation}\label{expect1}
\langle \mathfrak{v}_{\text{out}}\,|\,\mathfrak{U}_{\mathcal{P}(\mathcal{A})}\,|\,
\mathfrak{v}_{\text{in}}\,\rangle_n.
\end{equation}

\noindent This quantity gives the physical transition probability amplitude to 
get $|\mathfrak{v}_{\text{out}}\,\rangle_n$ by acting with 
$\mathfrak{U}_{\mathcal{P}(\mathcal{A})}$
on $|\mathfrak{v}_{\text{in}}\,\rangle_n$ and obviously its square  
modulus is the quantum probability to be assigned to the corresponding 
computation. 
By taking advantage of the possibility of decomposing 
$\mathfrak{U}_{\mathcal{P}(\mathcal{A})}$
 uniquely into an ordered sequence of elementary operators (gates), \eqref{expect1} becomes

\begin{equation}\label{expect2}
\langle \mathfrak{v}_{\text{out}}\,|\,\mathfrak{U}_{\mathcal{P}(\mathcal{A})}\,|\,
\mathfrak{v}_{\text{in}}\,\rangle_n\;=\;
\lfloor\,
\prod_{s=0}^{L-1}\,
\langle \mathfrak{v}_{s+1}\,|\,\mathcal{U}_{s,s+1}\,|\,
\mathfrak{v}_{s}\,\rangle_n\;\rfloor_{\mathcal{P}(\mathcal{A})}
\end{equation}

\noindent with $L\equiv L(\mathcal{P}(\mathcal{A}))$ for short. The symbol  
$\lfloor \; \rfloor_{\mathcal{P}(\mathcal{A})}$ denotes the ordered 
product along the path $\mathcal{P}(\mathcal{A})$ and provides a sort of superselection rule 
which induces destructive interference of the forbidden ({\em i.e.}, not leading to the correct 
result) paths in $\mathfrak{G}_n$. Each elementary operation introduced in the generic expression 
\eqref{progr2} 
is now better denoted by $\mathcal{U}_{s, s+1}$ to stress its $''$one--step$\, ''$ 
character with respect to computation.
 Consequently, each elementary transfer matrix in \eqref{expect2} 
turns out to be associated with a
 local Hamiltonian operator arising from

\begin{equation}\label{expect3}
\langle \mathfrak{v}_{s+1}\,|\,\mathcal{U}_{s,s+1}\,|\,
\mathfrak{v}_{s}\,\rangle_n\;=\;\exp\,\{i\, \mathbf{H}_n\,(s,s+1)\cdot \tau\}
\end{equation}

\noindent  and representing the unitary evolution of the simulator in one unit of its 
intrinsic time variable $(s =0,1,2, \ldots L(\mathcal{P}(\mathcal{A})))$. We indicate with the 
shorthand notation $(s,s+1)$ the dependence of $\mathbf{H}_n$ on its variables to make clear 
the local nature of this operator with respect to the computational space $\mathfrak{G}_n$ (a more 
detailed description should involve the quantum numbers characterising both states and operation). 
When \eqref{expect3} is inserted in \eqref{expect2}, such {\em virtual} 
Hamiltonians generally do not commute with each 
other but nonetheless the whole computational process may be identified with a well defined 
unitary evolution of the simulator in the internal time interval given in \eqref{timealg}.

The above remarks justify the statement made at the beginning of this section, 
namely that the spin network simulates intrinsically time evolution (without resorting 
to {\em ad hoc} external Hamiltonians as happens for instance in
the Universal Quantum Simulator model proposed in \cite{Ll2}). 
Moreover, by changing 
the program $\mathcal{P}$ (and the algorithm $\mathcal{A}$) 
the machine is able to select different types of
 dynamical behaviors, and thus to simulate complex poly--local many angular momenta
 interactions modeled as binary couplings and Wigner rotations (see the assumptions
 discussed in Section 2). What we mean in particular is that 
different types of evolutions 
in $\mathfrak{G}_n$ can be grouped into $''$computing classes$\, ''$ based on the choice of 
gates that each program has to employ. Then a program $\mathcal{P}$ (defined in 
\eqref{progr1} and associated with a collection of directed
 paths as in \eqref{dirpath1} or, equivalently, with decompositions into a number of gates
 as in \eqref{progr2}) may be based on either $M$--gates alone, 
or $j$--gates alone, or some fixed sequence of $M$ and $j$--gates.\\

$\blacklozenge$ An $M$--{\em computing class} contains programs which employ only $M$--gates 
at each step in their associated directed paths. The binary bracketing structure of the
computational Hilbert spaces described in Section 2 is not involved, and it is 
not difficult to realize that such kind of computation, when applied to $N$ 
$\frac{1}{2}$--spins,
reproduces the usual Boolean quantum circuit ({\em cfr.} the end of Section 3.2
and Appendices B1 and B2).$\blacktriangle$ \\

$\blacklozenge$  A $j$--{\em computing class} includes programs which employ only $j$-gates at
 each computational step. This class is particularly interesting since it shares 
many features with suitable types of discretized field theories (the so--called
 state sum models, to be discussed in Section 6) as we already noticed in \cite{MaRa}.
 Now the combinatorial structure of Rotation graphs becomes prominent owing to the existence
 of an one--to--one correspondence between allowed elementary operations and the edge set $E$ of 
$\mathfrak{G}_n$, for each $n$ ({\em cfr.} \eqref{R2} and more generally Appendix A2).
 
In the present framework it is
convenient to switch back to notations used in the first part of Section 4.1. Then states
will be labeled again by $|\mathfrak{b}\rangle_n$ as in \eqref{3njst} (dropping $JM$ ) 
and a program $\mathcal{P}$ is represented formally as

\begin{equation}\label{pro3nj1}
\mathcal{U}_{3nj}\, [\mathfrak{b}_{\text{in}}\; \xrightarrow{\mathcal{P}}\; 
\mathfrak{b}_{\text{out}}]\;:\; 
|\mathfrak{b}_{\text{in}}\rangle_n\;\longrightarrow\;|\mathfrak{b}_{\text{out}}\rangle_n\;, 
\end{equation}

\noindent where, as before, $|\mathfrak{b}_{\text{in}}\rangle_n$
is fixed and $|\mathfrak{b}_{\text{out}}\rangle_n$ is an accepted state. The set of operators 
$\mathfrak{U}_{\mathcal{P}}$ in \eqref{progr1} has been replaced by the unitary 
operators $\mathcal{U}_{3nj}$ introduced in \eqref{formal3nj} and already used in \eqref{3njpath}. 
The collection of directed paths associated with \eqref{pro3nj1}
is defined as in \eqref{dirpath1} and a particular path of length $L$ corresponding 
to a time--ordered sequence of states is represented as

\begin{equation}\label{pro3nj2}
|\mathfrak{b}_{\text{in}}\,\rangle_n\equiv
|\mathfrak{b}_{0}\,\rangle_n\rightarrow
|\mathfrak{b}_{1}\,\rangle_n\rightarrow\cdots\rightarrow
|\mathfrak{b}_{s}\,\rangle_n\rightarrow\cdots\rightarrow
|\mathfrak{b}_{L}\,\rangle_n\equiv
|\mathfrak{b}_{\text{out}}\,\rangle_n
\end{equation}

\noindent Each arrow corresponds now to the first type of operation in \eqref{elemstep}, 
namely a Racah
 transform (possibly up to weights/phases). When one particular path is chosen we 
would recover expressions similar to \eqref{proalg1}--\eqref{expect3}, referring 
to a computation process 
based on a program $\mathcal{P}$ performing an algorithm $\mathcal{A}$.
 
However, in the $j$--computing class one
may address other types of problems, namely: selected two states in $\mathfrak{G}_n$, say
$|\mathfrak{b}_{\text{in}}\rangle_n$ and
$|\mathfrak{b}_{\text{out}}\rangle_n$, 
consider all possible $\mathcal{P}(\mathcal{A})$ that compute
$|\mathfrak{b}_{\text{out}}\rangle_n$
as the result of the application of some

\begin{equation}\label{pro3nj3}
\mathcal{U}_{3nj}\,[\mathfrak{b}_{\text{in}}\;\xrightarrow{\mathcal{P}(\mathcal{A})}\;
\mathfrak{b}_{\text{out}}] 
\end{equation}

\noindent to $|\mathfrak{b}_{\text{in}}\rangle_n$. 
The functional on $\mathfrak{G}_n$ which takes care 
of such multiple choices is a $''$path sum$\, ''$ (a discretized Feynman's path integral) 
which may be written as

\begin{equation}\label{pathsum1}
\mathbf{Z}[\mathfrak{b}_{\text{in}},\,\mathfrak{b}_{\text{out}}]\;=\;
\sum_{\mathcal{P}(\mathcal{A})}\,
W_{\mathcal{P}(\mathcal{A})}\,
\langle\mathfrak{b}_{\text{out}}\,|\,
\mathcal{U}_{3nj}[\mathfrak{b}_{\text{in}}\;\xrightarrow{\mathcal{P}(\mathcal{A})}\;
\mathfrak{b}_{\text{out}}]\,|\,\mathfrak{b}_{\text{in}}\,\rangle_n\,,
\end{equation}

\noindent where the summation is over all paths with fixed endpoints and $W_{\mathcal{P}(\mathcal{A})}$ is a 
weight to be assigned to each path.
 
Notice that if we should give the same 
weight, say $W_{\mathcal{P}(\mathcal{A})} = 1$ to each path, then the results on 
equi--probability amplitudes
 collected in the theorem of Section 4.1 ensure us that the functional \eqref{pathsum1}
is a combinatorial invariant, namely it is actually independent of the particular 
path connecting 
$\mathfrak{b}_{\text{in}}$ and $\mathfrak{b}_{\text{out}}$. 
On the other hand, if we insert non trivial weights in
\eqref{pathsum1}, we may naturally address questions about most efficient 
algorithms and time complexity.
 For instance we could weigh paths with the inverse of their lengths $L({\cal P}({\cal A}))$; 
then the minimum--length path (the optimal algorithm) will be dynamically singled out in 
the path sum. As a matter of fact, even such simple example turns out to be highly 
non trivial owing to the combinatorial complexity of ${\mathfrak{G}}_n$. We shall come back 
on such issues in the following Section 4.3. $\blacktriangle$ \\ 

$\blacklozenge$  An {\em altered} $j$--{\em computing class} is a modification of 
the $j$--class obtained 
on applying just one Wigner rotation ($M$--gate) to the input state. 
A typical directed path in this class can be represented by mixing 
our previous notations (see \eqref{genstate1}, \eqref{genstate3}, \eqref{dirpath1} 
and \eqref{pro3nj2}) to get

\begin{align}\label{dirpath3}
|\mathfrak{v}_{\text{in}}\,\rangle_n\equiv\;&
|\mathfrak{b}_{0};\theta,\phi;J\,\rangle_n\rightarrow
|\mathfrak{b}_{0};\theta',\phi';J\,\rangle_n\rightarrow
|\mathfrak{b}_{1}\,\rangle_n\rightarrow
|\mathfrak{b}_{2}\,\rangle_n\rightarrow\cdots\rightarrow\nonumber\\
&
\rightarrow |\mathfrak{b}_{s}\,\rangle_n\rightarrow\cdots\rightarrow
|\mathfrak{b}_{L+1};\theta',\phi';J\,\rangle_n
\equiv
|\mathfrak{v}_{\text{out}}\,\rangle_n
\end{align}

\noindent where the first arrow represents a Wigner rotation 
$\mathfrak{D}^J(\alpha \beta \gamma)$ 
and all the others are alterations of the binary bracketing structure 
while keeping the angular dependence of the states fixed. Such computing 
class seems sufficiently general and shares some features with the $j$--class 
for what concerns the path sum interpretation. The counterpart of the path 
sum functional \eqref{pathsum1} reads

\begin{align}\label{pathsum2}
\mathbf{Z}[\mathfrak{v}_{\text{in}},\,\mathfrak{v}_{\text{out}}]\;=&
\sum_{\mathcal{P}(\mathcal{A})}\,
W_{\mathcal{P}(\mathcal{A})}\,
\langle\mathfrak{b}_{L+1}\,|\,
\mathcal{U}_{3nj}[\mathfrak{b}_{0}\;\xrightarrow{\mathcal{P}(\mathcal{A})}\;
\mathfrak{b}_{L+1}]\,|\,\mathfrak{b}_{0}\,\rangle_n\nonumber\\
&
\langle \mathfrak{b}_0;\theta',\phi';J\,|\,\mathcal{D}^J\,(\alpha \beta \gamma)\,|\,
\mathfrak{b}_0;\theta,\phi;J\,\rangle_n\,,
\end{align}

\noindent from which we see in particular that combinatorial invariance is broken even
 if we assign to each path the same weight.$\blacktriangle$ \\ 

$\blacklozenge$ An {\em alternating computing class} includes programs which 
alternates $M$ and $j$--gates and the length of each of the associated
directed paths is an even number $L = 2\ell$. This class is quite general 
with respect to capability of simulating real physical systems and includes 
all the former computing categories since anyone of the elementary gates could
 eventually be realized by an identity transformation. 
$\blacktriangle$ \\ 

\subsection{Computational complexity}

In view of the role that binary coupling trees play in our model for quantum computation
we define an encoding map

\begin{equation}\label{A1}
\mathcal{H}^{\,J}_n\, (\mathfrak{b}) \;\longrightarrow \;T (\mathfrak{b})
\end{equation}

\noindent where, as in (15), the shorthand notation 
$\mathfrak{b}$ stands for the string of quantum numbers 
$( [j_1, j_2, \ldots j_{n+1}]^{\mathfrak{b}};\, k^{\mathfrak{b}}_1, k^{\mathfrak{b}}_2,
\ldots, k^{\mathfrak{b}}_{n-1})$,
and $T (\mathfrak{b})$ is the coupling tree uniquely 
associated with the computational Hilbert space $\mathcal{H}^{\,J}_n$ 
for given $J$ and $n$ (see Appendix A1). This coding is 
intrinsically quantum, namely $T(\mathfrak{b})$ in (96) is not a simple
 device in which classical 
information can be stored (as happens for instance when  
search trees are considered). 
The quantum behaviour of $T(\mathfrak{b})$ emerges in particular when 
we look at the nature of the internal labelings. 
Recall from Section 2 that an intermediate angular momentum, say 
$\mathbf{K}_1 = \mathbf{J}_1 + \mathbf{J}_2$, has eigenvalue $k_1$ ranging 
between $|j_1 - j_2|$ and $j_1 + j_2$ and thus such {\em quantum trees} -- 
even when equipped with definite values 
of the incoming quantum numbers chosen in $\{0,\frac{1}{2}, 1, \frac{2}{3},\ldots \}$
-- take care consistently of the 
range of different values that may be assigned to the internal nodes. 

It is worth noting that the binary bracketing notations 
introduced in (12) and (14) of Section 2 can be interpreted as the quantum counterpart of the 
$''$word construction$\, ''$ outlined in Remark A.1 of Appendix A1 for a generic binary operation. 
To formalise this observation we 
introduce a new map -- equivalent to \eqref{A1} from a quantum mechanical point of view
-- which encodes 
information carried by the Hilbert space $\mathcal{H}^{\,J}_n$ 
into a {\em quantum word}, namely a string of quantum numbers 
plus parenthesization. Denote formally this map by

\begin{equation}\label{A3}
\mathcal{H}^{\,J}_n\, (\mathfrak{b}) \;\longrightarrow \;
( [j_1, j_2, \ldots j_{n+1}]^{\mathfrak{b}};\, k^{\mathfrak{b}}_1, k^{\mathfrak{b}}_2,
\ldots, k^{\mathfrak{b}}_{n-1}),
\end{equation}

\noindent since a more explicit 
form as in (14) would force us to write down just one particular type of binary coupling. 
In our opinion a closer inspection of the encoding map \eqref{A3} (instead 
of \eqref{A1} which will be 
exploited in the following) could represent a promising starting point to establish 
a truly quantum Formal Theory including languages, grammars, G\"odel numberings 
and related automaton 
models \cite{GaMaRa}.

Coming back to the encoding map \eqref{A1}, and referring to the topological transformations
on binary coupling trees discussed in Appendix A2, 
we easily recognise that a Racah 
transform $\mathcal{R}$ defined in (21) is encoded in a rotation 
(more precisely, the explicit expression given in (23) 
is encoded into the operation depicted at the bottom of Fig. 19). On the other hand, 
a phase transform 
$\Phi$ defined in (22) turns out to be encoded into a twist (compare {\em e.g.} the explicit 
expression (24) with the twist depicted at the bottom of Fig. 20). Consequently,

\begin{align}\label{A8}
\text{Racah transform}\;\; \mathcal{R}\, & \longrightarrow \;\text{Rotation}\nonumber\\
\text{Phase transform} \;\;\Phi\, & \longrightarrow \;\text{Twist}
\end{align}

\noindent represent encoding maps associating the two types of unitary elementary 
$j$--gates introduced in 
Section 3.1 with basic topological moves on quantum trees. The role of these 
two sets of operations 
is specular also for what concerns composition, since the Biedenharn--Louck 
Theorem (Section 3.1) has its counterpart 
in the fact that any pair of binary coupling trees can be connected by a sequence 
of rotations and twists
(see Appendix A2). 

The quantum encoding maps \eqref{A1}, \eqref{A8}
make manifest that combinatorics of (Twist)--Rotation graphs (Appendix A2) 
and of the computational space of the simulator (Section 4.1) share identical features, 
at least as far as $j$--computing classes (Section 4.2) are implemented. This crucial remark 
justifies the fact that we may speak about combinatorial and computational complexity questions 
by employing a common language and concepts. Note however that these similarities can be exploited 
only to some extent since combinatorial complexity of graphs is usually addressed in a classical 
information theory context (see Appendix A3). In the remaining part of this section we shall 
illustrate in brief the computational potentialities of the spin network simulator. 
Going beyond questions in number theory, we argue that our model is suitable to deal 
with $\# {\bf P}$ ($''$hard enumerative/combinatorial$\, ''$) problems more efficiently 
than any classical machine.

For what concerns space complexity capacity of the spin network simulator in 
the sense of capability 
of storing information, we realize that it behaves as the cardinality of the 
(Twist)--Rotation graphs 
({\em cfr.} \eqref{A4}, \eqref{A5},
\eqref{A6}, Table 2 and Table 3 of Appendix A1). When the number $n$ of incoming angular 
momenta grows, the number of states which becomes accessible for computation increase at least 
exponentially. To quantify these asymptotic growth, consider first the Catalan numbers (96) which 
represent some sort of lower bound for the various enumerations of rooted labeled binary trees
shown in Table 3 of Appendix A1. They have the following asymptotic expansion for 
$n \rightarrow \infty $ \cite{Sta} 

\begin{equation}\label{asyCat}
C_n \,\thickapprox\,
\frac{e^{n \ln 4}}{\sqrt {\pi n(n + 1)^2}}\;\; 
\left\{
1 - \frac{1}{8n} +
\frac{1}{128n^2} + \ldots \right\}
\end{equation}

\noindent where $\ln = \log_e$. On the other hand, we may estimate approximatively 
rates of grows of the double 
factorial $D_n$ in \eqref{A6} and of the quadruple factorial $\hat{C}_n$ in \eqref{A5}
by using Stirling formula. We get

\begin{equation}\label{asyDou}
D_n \,\thickapprox n^{\,n} \,\exp \,\{n \ln 2 - n\}
\end{equation}

\begin{equation}\label{asyQua}
\hat{C}_n \,\thickapprox n^{\,n} \,\exp \,\{2n \ln 2 - n\}
\end{equation}

\noindent where the subleading terms are decreasing (increasing) exponentials, respectively. 
Thus the case considered in Section 4.1, namely the computational space modeled on the Rotation 
graph $\mathfrak{G}_n$, turns out to exhibit for large n space complexity of factorial class 
$\thickapprox n!$ as in \eqref{asyDou}.

To address analogously {\em time complexity}, we need first the notion of $''$input 
length$\, ''$ (for an instance of some given problem) 
which turns out to be related with the encoding scheme employed. By using the map
\eqref{A1} we may say that a typical input length is the number of symbols required
to specify a (quantum) labeled tree, namely $(2n +1)$ (terminal nodes, intermediate
nodes and the root). Thus it seems natural to assume such a number as a typical measure of the
size of the input (note however that in order to specify one particular quantum state
we should choice also a value for the total magnetic number $M$). 
Although the input length is linear in the number of symbols, the quantum nature
of the computational space is reflected by the fact that the size of the configuration  space
accessible for computation 
grows factorially with $n$ as discussed in the previous remark.

With these premises and by exploiting the estimate on the Diameter of the rotation graph
$\mathfrak{G}_n$ given in \eqref{A19} of Appendix A3, we assert that the (time) 
complexity function for any possible algorithm running over the spin network simulator 
can be expected to be polynomially bounded as a function of the input size $n$.
An effective discussion of both space and time complexity, however, requires of course 
reference to a specific algorithm, which in turn can be formulated only provided the 
necessary encoding scheme is defined. Work is in progress along these lines \cite{GaMaRa}. \\

\section{Semiclassical simulator\\ 
and $SU(2)$ state sum models}

According to the Bohr correspondence principle, classical concepts
become increasingly valid in the regime where quantum numbers are large.
In handling with angular momenta variables measured in units of $\hbar$,
the classical limit $\hbar \rightarrow 0$ implies that, for finite angular 
momenta, both the $j$--quantum numbers and the magnetic ones are 
much bigger than one.
For what concerns pure angular momentum states -- and in particular the
computational Hilbert spaces introduced in Section 2 and involved in 
dynamical processing as illustrated in Section 4.2 -- when approaching 
classical limit all the components of the operators 
$\{\mathbf{J}_i (i=1,2,\ldots, n+1), 
\mathbf{J}\}$ are confined to narrower ranges around specific values.
Thus geometrical concepts typical of the semiclassical vector model 
arise naturally and the corresponding physical quantities 
have to be thought as averaged out.
As we shall see below, angular momentum functions such as Racah 
transforms and Wigner rotation matrices admit well defined {\em asymptotic limits},
whose absolute squares (probabilities) correspond to {\em classical limits} of the related
physical quantities. 

With these preliminary remarks, and on the basis of \cite{PoRe} and \cite{BiLo9}, Topic 9 (in which a
self contained discussion of the various asymptotics is given, together with 
the list of original references),
we are going to set up a semiclassical counterpart of the spin network simulator
which represents, to our knowledge, the first explicit example of a quantum circuit
model mapped onto a (classical) probabilistic automaton scheme. 

Let us focus for the moment on probabilities, namely on asymptotic expansions
of absolute squares of transition amplitudes. Consider an elementary $j$--gate,
namely a Racah transform expressed in terms of a $6j$ symbol as in 
\eqref{6j}. When all the six angular momenta in the $6j$ become $\gg 1$ in $\hbar$ units,
the square of the symbol has the limiting value given by the Wigner formula

\begin{equation}\label{6jsquare}
\begin{Bmatrix}
a & b & d\\
c & f & e
\end{Bmatrix}^{\,2}\;\;\sim\;\; \frac{1}{12 \pi\,V}
\end{equation}

\noindent where $V$ is the Euclidean volume of the tetrahedron  formed by the six angular
momentum vectors whose lengths are the arguments of the coefficient ($V^2$ can be computed 
from $(a,b,c,d,e,f)$ by using the Cayley determinant).
This result can be exploited to find the probability of measuring a coupling scheme
$(a(bc)_e )_f$ having prepared the system in the scheme $((ab)_d c)_f$ . 
Denoting by $P(e)$ this probability and using 
\eqref{6j}
and \eqref{6jsquare} we get

\begin{equation}\label{prob6j}
P(e)\;\;=\;\;\frac{(2e+1)(2d+1)}{12 \pi\,V}.
\end{equation}

Coming to elementary $M$-gates, and in particular to the expression 
\eqref{redD}
for a W--rotation matrix in terms of Euler angles, we see that

\begin{equation}\label{probWmatrix}
|D^{\,J}_{M'M}\,(\alpha \beta \gamma)\,|^2\;\;=\;\;
|d^{\,J}_{M'M}\,(\beta )\,|^2
\end{equation}

\noindent and this quantity is symmetric in $M'$ and $M$.
Following the step illustrated in \cite{BiLo9} (Topic 9, Section 2)
we limit ourselves to analyse the case of an input quantum state characterised in the
original reference frame by a total angular momentum $\mathbf{J}$ maximally
oriented along the $z$-axis, namely $M=J$.
the probability that the angular momentum projection along the rotated $z'$-axis
has the value $M'=M$ is given by

\begin{equation}\label{probM}
P(M)\;=\;[d^{\,J}_{M'M}\,(\beta )\,]^2\;=\;
\binom{2J}{J-M}\,\left(\cos \frac{\beta}{2}\right)^{2(J+M)}\,
\left(\sin \frac{\beta}{2}\right)^{2(J-M)}.
\end{equation}

\noindent In the classical limit the most probable value of $M$
is distributed around the classical value $M_0 = J\cos \beta$ with
some fixed value of the classical probability $P(M_0)$. Then it
can be shown that the limiting value of \eqref{probM}
for $J\gg 1$ and $J\pm M_0 \gg (M-M_0)$ reads

\begin{equation}\label{classprM}
P(M)\;\;\sim\;P(M_0)\;\exp\,\left\{-\,\frac{1}{J}\;\left(\frac{M-M_0}{\sin \beta}
\right)^2\right\}.
\end{equation}

\noindent Thus the probability for $M$ is a Gaussian distribution around the 
classical value $M_0$ and the dispersion in the variable $(M-M_0)$, $\sqrt{J} \sin \beta$,
is of the order $\sqrt{J}$ by the assumption made above. The simplicity of this result is
due to the choice of $|JJ>$ as original states. More general types of classical limits
and asymptotics are discussed in \cite{BiLo9} (Topic 9, Section 10) and collected also in \cite{Russi}. 

Without entering into further technical details, the rationale underlying
our approach should have become clear: anyone of the computing classes of the
quantum simulator introduced in Section 4.2 as finite sequences
of elementary unitary $j$- and/or $M$-gates can be mapped onto a corresponding (classical)
non--deterministic circuit--type computing process based on combinations of \eqref{prob6j}
and/or \eqref{classprM}.
We argue that such a semiclassical model could be able to simulate physical systems made up by
many interacting constituents such as polyatomic molecules described by
pure states of some suitable angular momentum--type variables for high values of q--numbers.

\vspace{.5cm}

In order to complete our semiclassical picture we have to include the treatment of 
asymptotic limits of the transition amplitudes associated with the elementary gates
employed in the quantum model. As we shall see, the resulting setting is closely
related to $SU(2)$ {\em state sum models} introduced in a completely different context,
namely topological quantum field theories (TQFT) and Euclidean quantum gravity
defined on triangulated 3--dimensional manifolds (see \cite{Kau}, \cite{AmDuJo}
for extended reviews on such topics). 

The key point is the interpretation of the Ponzano--Regge asymptotic formula for the $6j$
symbol which reads \cite{PoRe}

\begin{equation}\label{PRasymt}
\begin{Bmatrix}
a & b & d\\
c & f & e
\end{Bmatrix}
\;\sim\;\; \frac{1}{\sqrt{24 \pi V}}\;
\exp\,\left\{i\,\left(\sum_{r=1}^{6}\,\ell_r \, \theta_r \,+\,\frac{\pi}{4}
\right)\right\}
\end{equation}

\noindent where the limit is taken for all entries $\gg 1$ (recall that $\hbar =1$)
and $\ell_r \equiv j_r +1/2$
with $\{j_r\}=\{a,b,c,d,e,f\}$. 
$V$ is the Euclidean volume of the tetrahedron with edges of 
lengths $\{\ell_r\}$ (note the shift $j \rightarrow j+ 1/2$ with respect
to the variables employed in calculating the volume in \eqref{6jsquare}) and
finally $\theta_r$ is the angle between the outer normals to the faces which
share the edge $\ell_r$.

\begin{itemize}
\item From a purely quantum mechanical point of view, the probability amplitude
\eqref{PRasymt} has the form of a semiclassical (wave) function since the factor 
$1/\sqrt{24 \pi V}$ is slowly varying with respect to the spin variables while
the exponential is a rapidly oscillating dynamical phase. Such behavior complies with
the fact that the square of the modulus of the asymptotic \eqref{PRasymt} reproduces Wigner's 
expression \eqref{6jsquare}. Moreover, according to Feynman path sum
interpretation of quantum mechanics, the argument of the exponential 
represents a classical action, and indeed it can be read as
$\sum p\,\dot{q}$ for pairs $(p,q)$ of canonical variables
(angular momenta and conjugate angles).
\item There exists another intriguing physical interpretation of \eqref{PRasymt}
if we recognise that the expression in the exponential represents the classical
Regge action \cite{Reg} -- namely the discretized version of Einstein--Hilbert
action of General Relativity  -- for the tetrahedron associated with the $6j$ symbol
in the asymptotic regime.\\ 
In Regge's approach the edge lengths of a triangulated spacetime
are taken as discrete counterparts of the metric tensor appearing in the usual action
for gravity and angular variables (deficit angles) are related to the scalar 
curvature obtained from the Riemann tensor. Strictly speaking, a $''$triangulated spacetime$\, ''$
is a piecewise linear (PL) manifold of dimension $D$ dissected into
simplices, namely triangles in $D=2$, tetrahedra in $D=3$, 4-simplices in $D=4$
and so on. Inside each simplex either an Euclidean or a Minkowskian metric
can be assigned: accordingly, spacetime manifolds obtained by gluing together
$D$--dimensional simplices acquire an overall $PL$ metric of Riemannian or Lorentzian
signature.\\ 
The Regge Calculus formalism became in the early  80's 
the starting point for a novel approach to quantization of General Relativity
known as Simplicial Quantum Gravity (see the review \cite{ReWi} and references therein). 
The quantization procedure
most commonly adopted is the Euclidean path sum approach, namely the discretized
version of Hartle--Hawking path integral describing
$D$--dimensional, locally Euclidean geometries undergoing $''$quantum fluctuations$\, ''$, 
possibly with the constraint of keeping some $(D-1)$--dimensional boundaries fixed. 

Coming back to the interpretation of \eqref{PRasymt},
we conclude that it represents the semiclassical functional -- to be intended
as the semiclassical limit of a sum over all quantum fluctuations -- associated with  
a very simple 3--dimensional $''$spacetime$\, ''$, the Euclidean tetrahedron.
\end{itemize}

On the basis of the remark above, we pass to describe in brief the Ponzano--Regge 
state sum model representing the (quantized) partition function of  
simplicial Euclidean 3--gravity. Denote by

\begin{equation}\label{jtriang}
\mathcal{T}^3\,(j)\;\;\rightarrow \;\;\mathcal{M}^3
\end{equation}

\noindent a particular triangulation of a closed 3--dimensional PL manifold 
$\mathcal{M}^3$ (of fixed topology) obtained by assigning $SU(2)$ $''$spin variables$\, ''$
$\{j\}$ to the edges of $\mathcal{T}^3$. The assignment must satisfy a number of conditions
which can be more easily illustrated if we introduce the {\em state functional}
associated with $\mathcal{T}^3 (j)$, namely
  
\begin{equation}\label{PRstfunct}
\mathbf{Z}[\mathcal{T}^3(j) \rightarrow \mathcal{M}^3; L]=
\Lambda(L)^{-N_0}\prod_{A=1}^{N_1} (-1)^{2j_A} \mathsf{w}_A\prod_{B=1}^{N_3}
\phi_B
\begin{Bmatrix}
j_1 & j_2 & j_3 \\
j_4 & j_5 & j_6
\end{Bmatrix}_B
\end{equation}

\noindent where $N_0, \, N_1,\, N_3$ denote the number of vertices, edges and tetrahedra 
in $\mathcal{T}^3(j)$, $\Lambda (L)=4L^3/3C$ ($C$ an arbitrary constant),
$\mathsf{w}_A \doteq$ $(2j_A+1)$ are the dimensions of 
irreducible representations of $SU(2)$ which weigh the edges,
$\phi_B =$ $(-1)^{\sum_{p=1}^6 j_p}$ and $\begin{Bmatrix}a&b&c\\d&e&f \end{Bmatrix}$ 
are $SU(2)$ $6j$ symbols to be associated with the tetrahedra of the triangulation. The Ponzano--Regge 
{\em state sum} is obtained by summing over triangulations
corresponding to all assignments of spin variables $\{j\}$ bounded by the cut--off $L$,
namely

\begin{equation}\label{PRstsum}
\mathbf{Z}_{PR}\,[\mathcal{M}^3]\;=\;
\lim_{L\rightarrow \infty}\:
\sum_{\{j\}\leq L}
\mathbf{Z}\; [\,\mathcal{T}^3(j) \rightarrow \mathcal{M}^3; L\,]\;,
\end{equation} 

\noindent where we formally remove the cut--off by taking the limit in
front of the sum. As already noted in \cite{PoRe}, the above state sum is a topological
invariant owing to the fact that its value is actually independent  of
the particular triangulation, namely does not change under
suitable topological transformations (the bistellar moves).  
These moves are expressed algebraically in terms of the Biedenharn-Elliott identity  
\eqref{BEid}
--representing the moves 
(2 {\em tetrahedra}) $\leftrightarrow$  (3 {\em tetrahedra})-- and of both the 
Biedenharn--Elliott identity and the orthogonality conditions 
\eqref{ort6j}
for $6j$ symbols, which represent the barycentric move together its inverse, namely 
(1 {\em tetrahedra}) $\leftrightarrow$  (4 {\em tetrahedra}). 

The state sum \eqref{PRstsum} (and, more generally, geometric partition functions of this type
built up in any dimension $D$ \cite{CaCaMa}) resembles the functional 
\eqref{pathsum1}
introduced in dealing with the $j$--computing class of the spin
network simulator (Section 4.2). This is due to the fact that the amplitude of the
$3nj$ symbol in \eqref{pathsum1}
can be factorized according to the general prescription 
\eqref{expect2}
into sums over intermediate angular momenta of products of $6j$ symbols weighted by suitable factors
and phases ({\em cfr.} 
\eqref{6j}, \eqref{exphase}). 
These two partition functions share the property of being combinatorially invariant
under topological moves expressed in terms of algebraic identities of the $6j$ symbols.
However, on the one hand, the Racah identity
\eqref{Rid}
does not appear in the Ponzano--Regge framework since it would correspond to a topological
transformation  (1 {\em tetrahedron}) $\leftrightarrow$  (2 {\em tetrahedra})
which is forbidden in the PL category. On the
other hand, in the spin network framework it is not required {\em a priori}
that the $6j$ symbols match together to give rise to a triangulation of a 
3--dimensional manifold. Moreover, if we fix $n$ to get a specific computational space modeled on
the graph $\mathfrak{G}_n$, we would not catch in $\mathfrak{G}_n$ all possible 
triangulations of a given PL 3--manifold. Although we may be tempted to claim that the spin network 
is able to simulate $SU(2)$--coloured 3--dimensional quantum gravity, we should bear in mind that 
we are actually dealing with a graphical device which encodes all types of $3nj$ symbols for 
any fixed $n$ \cite{Belgi99}. In this perspective it is interesting to recall that
Ponzano and Regge themselves \cite{PoRe} noted that the topology of a $9j$ symbol corresponds to
the real projective space $\mathbb{RP}^2$, in the same sense that the $6j$
has the topology of the 2--sphere bounding the tetrahedron. Indeed any particular type of
$3nj$ symbol may be associated with a closed, not necessarily oriented, surface
representing the boundary of a 3--dimensional polyhedron \cite{Aquila} obtained by 
duality from the graphical representations introduced in \cite{Lituani}.
In this sense the claim that the simulator can simulate some classes of extended triangulated 
objects is certainly true in dimension 2. \\

{\bf Remark 5.1.} In \cite{TuVi} a regularized version of \eqref{PRstsum} --based on representation
theory of a quantum deformation 
of the group $SU(2)$--  was proposed and shown to be a well--defined (finite)
{\em quantum invariant} for closed 3--manifolds. Its expression reads
 
\begin{equation}\label{TVstsum}
\mathbf{Z}_{\,TV}\,[\mathcal{M}^3;q]\,=\,\sum_{\{j\}}\;\mathbf{w}^{-N_0}\,
\prod_{A=1}^{N_1} \mathbf{w}_A
\,\prod_{B=1}^{N_3} \;
\begin{vmatrix}
j_1 & j_2 & j_3 \\
j_4 & j_5 & j_6
\end{vmatrix}_B \,,  
\end{equation}

\noindent 
where the summation is over all colourings $\{j\}$ labeling highest weight irreps of $SU(2)_q$ 
($q=\exp\{2\pi i /r\}$, with $\{j=0,1/2,1 \dots, r-1\}$), $\mathbf{w}_A\doteq$ 
$(-1)^{2j_A}[2j_A+1]_q$ 
where $[\,]_q$ denote a  quantum integer, 
$\mathbf{w}=2r/(q-q^{-1})^2$ and $|\,\;|_B$ represents the q--$6j$ symbol whose entries are the 
angular momenta $j_{\ell}\, , \, \ell =1,\dots ,6$ associated with tetrahedron $B$.
In \cite{Tur} the invariant \eqref{TVstsum} is shown to equal the 
square of the modulus of the Reshetikhin--Turaev invariant, which in turn represents
the Chern--Simons partition function written for a closed oriented manifold
$\mathcal{M}^3$ equipped with a surgery presentation. We may write
schematically

\begin{equation}\label{TVCS}
\mathbf{Z}_{\,TV}\,[\mathcal{M}^3;q\,]\,\longleftrightarrow\,
|\,\mathbf{Z}_{\,CS}\,[\mathcal{M}^3;k\,]\,|^2\,,
\end{equation}

\noindent where the level $k= 2(r-1)$ of the Chern--Simons functional is related to the
deformation parameter $q$ and also to the cosmological constant of the underlying
Euclidean gravity model ({\em cfr.} \cite{AmDuJo} (Ch 7) and \cite{Car} for reviews 
on Chern--Simons theory and its relations with 3--dimensional gravity) . 

As we shall see in the following section, functors derived from $SU(2)$ Chern--Simons
theory are the basic ingredients for implementing computation in the topological approach 
\cite{FrLaWa}. It will be shown that the spin network dynamics based
on $j$--gates can be mapped into the functorial approach in a way that resembles the
correspondence \eqref{TVCS}. 
$\blacktriangle$ \\

\section{Spin network and topological\\
 quantum computation}

We begin this section by introducing some basic ingredients of Chern--Simons--type
Topological Quantum Field Theories (CS TQFTs) in order to deal with the topological
approach to quantum computation. Our presentation will be necessarily sketchy, 
and we refer the reader to \cite{BiBlRa},
\cite{Ati}, \cite{Qui} for general reviews on TQFTs, while the 3--dimensional  CS case
is extensively addressed in \cite{Kau}, \cite{AmDuJo}, \cite{Car}. 

TQFTs are particular types of gauge theories, namely field theories quantized through the
(Euclidean) path integral presciption starting from a classical Yang--Mills action 
defined on a suitable $D$--dimensional space(time).
TQFTs are characterized by observables 
(correlation functions) which depend only on the
global features of the space on which these theories live, namely they
are independent of any metric which may be used to define the underlying classical theory.
The geometrical and topological generating functionals
and correlation functions of such theories are computable by standard techniques in quantum field
theory and provide novel representations of certain global invariants (for $D$-manifolds and/or 
for particular submanifols embedded in the ambient space) which are of prime interest
in mathematics.
In the 3--dimensional case, theories based on Chern--Simons--type actions (see below)
have been shown to incorporate significant generalizations of previously known 
invariants for both 3--manifolds (Witten--Reshetikhin--Turaev invariant) and knots/links 
(Jones polynomial). In particular, the Jones polynomials \cite{Jon} 
can be obtained as  correlation functions of Wilson line operators along closed loops
in the CS framework \cite{Wit}. While these mathematical advances are self--evident, CS theory
also provides a unifying 3--dimensional viewpoint for 2--dimensional Conformal Field Theory
as well as new results on 3--dimensional quantum gravity. \\
Since TQFTs are quite generally soluble, they could provide a testing ground for new 
approaches to the quantum theory of fields. It has been conjectured that TQFTs may represent different 
$''$phases$''$ --in which general covariance is unbroken-- of their more conventional counterparts.

\vspace{.5cm}

Denote by $\Sigma_1$ and $\Sigma_2$ a pair of 2--dimensional manifolds and by
$\mathcal{M}^3$ a generic 3--dimensional manifold with boundary $\partial \mathcal{M}^3$
$=\Sigma_1 \cup \Sigma_2$ (all manifolds here are compact, smooth and oriented). 
A unitary 3--dimensional quantum field theory corresponds to the assignment of \\
{\bf i)} finite dimensional Hilbert spaces (endowed with non--degenerate
bilinear forms) $\mathcal{H}_{\Sigma_1}$ and $\mathcal{H}_{\Sigma_2}$
to $\Sigma_1$ and $\Sigma_2$, respectively;\\
{\bf ii)} a map ($''$functor$\, ''$) connecting such Hilbert spaces

\begin{equation} \label{TopFunct}
\mathcal{H}_{\Sigma_1}\;
\xrightarrow{\mathbf{Z}\,[\mathcal{M}^3\,]}\;
\mathcal{H}_{\Sigma_2}
\end{equation}

\noindent where $\mathcal{M}^3$ is a manifold which interpolates between $\Sigma_1$
(incoming boundary) and  $\Sigma_2$ (outgoing boundary). 
Without entering into details concerning a few more axioms
(diffeomorphism invariance, factorisation {\em etc.}) we just recall that unitarity
implies that\\
{\bf iii)} if $\bar{\Sigma}$ denotes the surface $\Sigma$ with the opposite orientation, then
$\mathcal{H}_{\bar{\Sigma}}=$ 
$\mathcal{H}^{*}_{\Sigma}$, where $*$ stands for complex conjugation;\\ 
{\bf iv)} the functors \eqref{TopFunct} are unitary and 
$\mathbf{Z}[\bar{\mathcal{M}}^3]=$ 
$\mathbf{Z}^{*}[\mathcal{M}^3]$, where  $\bar{\mathcal{M}}^3$ denote the manifold
with the opposite orientation. 

In Chern--Simons theory the functor \eqref{TopFunct}
is the partition function $\mathbf{Z}_{CS}$ associated with the classical action

\begin{equation} \label{CSaction}
S_{CS}\,(A)\;=\;\int_{\mathcal{M}^3}\;\;
\left( A\,dA\,+\,\frac{2}{3}\,A \wedge A \wedge A \ \right) 
\end{equation}

\noindent written for simplicity for a closed manifold $\mathcal{M}^3$, $\partial
\mathcal{M}^3 = \emptyset$.
Here $A$ is an $SU(2)$--connection, namely a 1--form on the principal 
$SU(2)$--bundle over $\mathcal{M}^3$, $d$ is the exterior differential and $\wedge$
is the wedge product of differential forms. The partition function is obtained functionally
by integrating the exponential
of the classical action \eqref{CSaction} over the space of all $SU(2)$ connections. 
We formally write 

\begin{equation} \label{CSfunct}
\mathbf{Z}_{\,CS}\,[\mathcal{M}^3; k]\;=\;
\int [DA]\,\exp \left\{\frac{i\, k}{4 \pi}\,
S_{CS}\,(A)\,\right\}
\end{equation}

\noindent where the coupling constant $k$ (the level of the theory) must be
an integer. It can be shown that the partition function \eqref{CSfunct} actually
represents a topological invariant for closed 3--manifolds, related in turn 
to the Turaev--Viro invariant 
\eqref{TVstsum}
by the correspondence
\eqref{TVCS}
(where $|\mathbf{Z}_{CS}[\mathcal{M}^3]|^2$ stands for 
$\mathbf{Z}_{CS}[\mathcal{M}^3]$ $\mathbf{Z}^{*}_{CS}[\mathcal{M}^3]$).
 
The extension of \eqref{CSfunct} to the case  
$\partial \mathcal{M}^3 \neq \emptyset$
 requires modifications
of the classical action \eqref{CSaction} by suitable (Wess--Zumino--type) boundary terms \cite{Car}.
In view of applications in a computational context it is sufficient to note that for a boundary component $\Sigma$,
${\bf Z}_{CS} [{\cal M}^3 ](\Sigma )$ can be realized as restriction to invariant subspace $W \subseteq {\cal 
H}_{\Sigma }$ of transformations of the form $\displaystyle{{\prod_i g_i}}$ on the subspace 
of computational states of a quantum computer, where the $g_i$, interpreted as gates in a quantum 
circuit scheme, can be written as $''$words$\, ''$ in the standard
generators ({Dehn's twists}) of the {\em Mapping Class Group} of the surface $\Sigma$.
Also in the computational framework  the $''$observables$''$ of the theory turn out to be represented by 
 Wilson loops as discussed in \cite{PreLo} and in the following paragraph.
The explicit expression of Wilson loops operators ${\cal W}_k(K)$, 
namely holonomies
of the connection 1--form evaluated on closed curves $K$ in $\mathcal{M}^3$, read 

\begin{eqnarray}
{\cal W}_{k}(K) = \int_{A/{\cal G}} [D A]\, {\rm e}^{\frac{ik}{4\pi}S_{CS}(A)} {\rm Tr} \left ({\rm hol}
\, K \right ) \Big{/} \int_{A/{\cal G}} [D A]\, {\rm e}^{\frac{ik}{4\pi}S_{CS}(A)} 
\label{loop}
\end{eqnarray}

\noindent where ${\displaystyle{{\rm hol}\, K \doteq {\cal P}\exp \int_K A}}$, 
${\cal P}$ is the path ordering,  
and $A$ is now thought of as the connection over the ${\cal G}$--bundle of Lie algebra--valued 
1--forms tangent 
to ${\cal M}^3$ (${\cal G}$ being the gauge group). \\
It is worth to recall that the evaluation of Jones polynomials --
appearing in expressions like \eqref{loop} --
was shown to be computationally $\# {\bf P}$ \cite{JaVeWe} (namely 
essentially the enumerative equivalent of ${\bf NP}$--complete \cite{GaJo}).  \\

\subsection{Holonomic Quantum Computation}

Holonomic Quantum Computation (HQC) is an all--geometrical 
approach to quantum information processing. In the HQC 
strategy information is encoded in degenerate eigenspaces 
of a parametric family of Hamiltonians. The computational 
network of unitary quantum gates is realized by driving 
adiabatically the Hamiltonian parameters along loops in a 
control manifold. By properly designing such loops the 
non--trivial curvature of the underlying bundle geometry 
gives rise to unitary transformations, {\em i.e.} holonomies that 
implement the desired unitary transformations. Conditions 
necessary for universal QC are stated in terms of the 
curvature associated to the non--Abelian gauge potential 
 over the control manifold. In view of their 
geometrical nature the holonomic gates are robust against 
several kind of perturbations and imperfections. This,  
along with the adiabatic fashion in which gates are operated, 
makes in principle HQC an appealing way towards 
universal fault--tolerant QC. 

HQC as introduced in  
\cite{ZaRa}, \cite{PaZaRa}, is 
based on a novel gauge--theoretic framework in which one is 
supposed to be able to control a set of parameters $\lambda\in 
{\cal L}$, on which depends an iso--degenerate family ${\cal F}$ 
of quantum Hamiltonians $\{ H(\lambda )\}$. Information is 
encoded in a $\nu$-dimensional eigenspace ${\cal C}$ of a 
specific $H(\lambda_0) \in {\cal F}$. Universal QC \cite{DeBaEk} 
over ${\cal C}$ can be then obtained by adiabatically driving 
the control parameters along suitable loops $\gamma$ rooted at 
$\lambda_0$.
The key physical ingredient is provided by the appearance in such 
quantum evolutions of non--Abelian geometrical contributions \cite{WiZe} 
$U_{\gamma} \in U(\nu )$($\nu > 1$) given by holonomies associated with 
a gauge potential $A$ valued in the algebra of $U(\nu ) $\cite{WuYa}, 
\cite{Nak}. In other words quantum computation in the HQC approach is 
nothing but the parallel transport of states in ${\cal C}$ realized by 
the connection $A$. 
Therefore the computational power in the HQC approach relies on the 
non--triviality of the geometry of the bundle of eigenspaces of ${\cal 
F}$ over the manifold of control parameters, ${\cal L}$:
in this sense HQC is fully geometrical. It is worth observing that the 
computational subspace ${\cal C}$ can be thought of as the lowest--energy 
manifold of a highly symmetric quantum system; from this point of view 
HQC is a kind of ground--state computation. This last remark points out 
the potential existence of a fault--tolerant \cite{Pre} feature of HQC due 
to energy gaps and even spontaneous relaxation mechanisms. Further 
fault--tolerant characteristics of HQC are related to the fact that the 
holonomies $U_{\gamma}$ realizing quantum computations typically turn 
out to depend just on the areas of the surfaces that the generating 
loops $\gamma$ span on certain 2--dimensional submanifolds. When this 
area is given one can consider even very large, {\em i.e.} $''$far$\, ''$ from 
the identity deformations of $\gamma$, but as long as they are 
area--preserving no errors are induced. Moreover as far as the 
adiabaticity condition holds, $U_{\gamma}$ does not depend on the rate 
at which the control loops are driven. Hence, even with respect the 
issue of timing, HQC can be expected to be robust. 

The evolution of the quantum system is thought of as actively driven 
by the parameters $\lambda$, over which the experimenter is assumed to 
have direct access and control, being able to drive by a dynamical control 
process the parameter configuration $\lambda\in {\cal L}$ through a path 
$\gamma : [0, T] \to {\cal L}$. Hence, a one--parameter  
(time--dependent) family 

\begin{equation}
{\cal F}_{\gamma} \doteq  \bigl \{ H(t)\equiv H[\Phi \circ \gamma (t)] 
\big | t \in [0, T] \bigr \} \subset {\cal F} \; ,  
\end{equation}

\noindent is defined for all $\Phi : {\cal L} \mapsto U({\cal N})$ , ${\cal N}= {\rm dim} 
\, {\cal C}$, $\Phi$ being a smooth mapping, and ${\rm dim}\bigl ( U({\cal N}) \bigr ) 
= {\cal N}^2$. The quantum evolution associated to the family 
${\cal F}_{\gamma}$ is described by the time--dependent Schr\"odinger equation 
$i \partial_t |\psi (t)\rangle = H(t)|\psi (t)\rangle$ and hence it has the 
operator form 

\begin{equation}
U_{\gamma} \doteq {\bf T} \exp \left \{ -i \int_0^T {\rm d}t H(t) \right 
\} \in U({\cal N}) \; ,
\end{equation}

\noindent where ${\bf T}$ denotes chronological ordering. The above time--dependent 
quantum evolution, for a given map $\Phi$, depends in general on the path 
$\gamma$ and {\em not} just on the curve $\gamma ([0, T])$, namely the image 
of $\gamma$ in the control manifold. 
In other words the unitary transformation $U_{\gamma}$ contains a {\em 
dynamical} as well as a {\em geometrical} contribution, the former depends 
on the rate at which $\gamma ([0, T])$ is traveled along whereas the latter 
depends merely on the geometrical characteristics of the curve. 

Non--Abelian holonomies are a natural generalization of the Abelian Berry 
phases. The basic assumption is that ${\cal F}$ is an iso--degenerate 
Hamiltonian family, {\em i.e.} all the elements of ${\cal F}$ have the 
same degeneracy structure. When the system control parameters are driven 
adiabatically, slowly with respect to any time--scale associated 
to the system dynamics, along a loop $\gamma$ in ${\cal L}$ any initially 
prepared state $|\psi_{in}\rangle \in {\cal H}$ will be mapped after the 
period $T$ onto the state 

\begin{equation}
|\psi_{out}\rangle = U_{\gamma}\, |\psi_{in}\rangle \; ,\;  U_{\gamma} = 
\bigoplus_{\ell =1}^{R} {\rm e}^{i \phi_{\ell}} \Gamma_{A_{\ell}} (\gamma ) \; , 
\end{equation}

\noindent where $\displaystyle{\phi_{\ell} \doteq \int_0^T {\rm d}\tau \varepsilon_{\ell}
(\lambda_{\tau})}$ is the dynamical phase ($\varepsilon_{\ell}(\lambda )$ 
denoting the degenerate Hamiltonian eigenvalues) whereas the matrices 
$\displaystyle{\Gamma_{A_{\ell}} (\gamma )}$ represent the geometrical 
contributions. They are unitary mappings of ${\cal H}_{\ell}$ onto itself 
and they can be expressed by the following path ordered integrals 

\begin{equation}
\Gamma_{A_{\ell}} (\gamma ) \doteq {\cal P} \oint_{\gamma} A_{\ell} \; \in 
\; U(n_{\ell}) \; , \; \ell = 1,\dots ,R \; , 
\end{equation}

\noindent $n_{\ell}$ denoting the number of degenerate states of energy $\varepsilon_{\ell}$, 
and $R$ the number of degeneracies. 
These are the {\em holonomies} associated with the loop $\gamma$, and the {\em 
adiabatic connection forms} $A_{\ell}$. The latter have an explicit matrix form 
given by 

\begin{equation}
A_{\ell} = \sum_{\mu} A_{\ell ,\mu} {\rm d}\lambda_{\mu} \; , \; \bigl (A_{\ell ,
\mu}\bigr )^{\alpha ,\beta} \doteq \langle \psi_{\ell}^{\, \alpha} (\lambda ) | 
\frac{\partial}{\partial \lambda^{\mu}} | \psi_{\ell}^{\, \beta}(\lambda )\rangle 
\; , 
\end{equation}

\noindent with $\bigl \{ \lambda_{\mu} \bigr \}_{\mu =1}^d$ the local coordinates on ${\cal L}$, and ${\cal H}_{\ell} 
= {\rm span} \bigl \{ \psi_{\ell}^{\, \alpha} (\lambda ) \rangle \bigr \}_{\alpha =1}^{n_{\ell}}$, 
$\psi_{\ell}^{\, \alpha} (\lambda )\rangle$ denoting the eigenstates of $H(\lambda )$ corresponding to 
eigenvalue $\varepsilon_{\ell}(\lambda )$. The connection forms $A_{\ell}$ are nothing but the non--Abelian 
gauge potentials enabling the parallel transport over ${\cal L}$ of vectors of the fiber ${\cal H}_{\ell}$. 

The combinatorial setting of the spin network simulator, and in particular of its fiber space structure 
over ${\mathfrak{G}}_n$ (see Remark 4.1 at the end of Section 4.1 and \cite{MaRa2}), 
appears to provide the natural structure for 
a discrete--time implementation of HQC, at least when the Hamiltonian defined in 
(\ref{expect3}) exhibit the required degeneracy.

\subsection{Combinatorial setting of Topological\\ Quantum Computation}

The approach to quantum computation of Freedman and collaborators \cite{FrLaWa} is based on an
extension of Chern--Simons functor at level $k=3$ (CS3) to particular types of
2--dimensional boundary objects, namely closed disks with some marked points. 
Denote by ($D^2, 3$ pts) a closed disk --namely a set topologically
equivalent to the standard 2--disk $\{(x,y) \in \mathbb{R}^2$ $| x^2+y^2 \leq 1\}$--
with three points lying in the interior of $D^2$. These points, together with the
boundary $\partial D^2$ of the disk, are marked by four labels $\{a,b,c,d\}$
chosen in the set $\{0,1,2,\ldots k\}$, which reduces to
$\{0,1,2,3\}$ in the present case $k=3$.
Notice that the convention adopted by the authors of \cite{FrLaWa}
is different from ours: the level $k=3$ here
corresponds to their $k=5$.\\
  The disk considered so far and depicted in
Fig. 3 (left) is the support of the so--called
$''$topological qubit$\, ''$. More precisely, a topological qubit corresponds to the image of the usual
1--qubit space $\mathbb{C}^2$ into the Hilbert space $\mathcal{H}_{CS3\,} (D^2,3$ pts)
derived from CS theory. The inclusion

\begin{align}
\label{CSincl}
i\;:\;\mathbb{C}^2\,& \hookrightarrow\,\mathcal{H}_{CS3\,} (D^2,3 \,\text{pts})\nonumber\\
\text{with}\,i(\,\mathbb{C}^2\,)\,& \doteq\, V(D^2,3 \,\text{pts})
\end{align}

\noindent induces on the label set $\{a,b,c,d\}$ the identifications
$a\equiv b\equiv c\equiv1$ and $d\equiv 0$, namely a mapping onto the binary
digits $\{0,1\}$ (see Fig. 3, right).

\hspace{1cm}

\begin{center}
{\bf Figure 3}
\end{center}
  
\noindent In the present framework 1--qubit gates are implemented  by considering -- instead 
of the full Chern--Simons functor \eqref{TopFunct} -- a unitary action of the braid group
$B_3$ on the strands generated in the ambient space by $''$evolution$\, ''$ of the marked points.

\begin{center}
{\bf Figure 4}
\end{center}

\noindent If 

\begin{equation}\label{B3action1}
B_3\,\times \, (D^2,3 \,\text{pts})\;\rightarrow\,(D^2,3 \,\text{pts})
\end{equation}

\noindent denotes the action of the braid group $B_3$ on the topological support
(see Fig. 4) then we formally write the induced unitary functor between Hilbert spaces
as

\begin{equation}\label{B3action2}
V_{1\,}(D^2,3 \,\text{pts})\;
\xrightarrow{\mathbf{Z}_{\,B_3\,}}\;
V_{2\,}(D^2,3 \,\text{pts}).
\end{equation}

\noindent A configuration of $N$ qubits is supported by a 2--disk with $3N$ marked points
($D^2, 3N$ pts) and the associated Hilbert space is $V(D^2, 3N$ pts), namely the image of 
$(\mathbb{C}^2)^{\otimes N}$ in $\mathcal{H}_{CS5\,}(D^2, 3N$ pts), where the image is defined as 
in \eqref{CSincl}. An $N$--qubit gate is represented by a unitary map  

\begin{equation}\label{B3Naction}
V_{1\,}(D^2,3N \,\text{pts})\;
\xrightarrow{\mathbf{Z}_{\,B_{3N}\,}}\;
V_{2\,}(D^2,3N \,\text{pts}),
\end{equation}

\noindent where $B_{3N}$ is the braid group acting on the $3N$ strands generated by the marked points.

We may summarise the main results of the series of papers \cite{FrLaWa} as follows.
It is shown that their $CS5$--functor (and also any other $CSk$--functor, $k \geq 3, k \neq 4$) is universal
for quantum computation and in particular that 1 and 2--qubits topological gates are sufficient to 
reconstruct all other gates of the type \eqref{B3Naction}.
The resulting model is polynomially equivalent to the usual Boolean quantum circuit and,
conversely, it is actually shown that there exists (at least) a class of TQFTs which can 
be simulated on a quantum machine ({\em cfr.} the diagram at the end of Section 2). 
There are however some open problems in this approach, mainly due to the fact that
it is difficult to $''$localise$\, ''$ topological objects (such as the marked points on
disks) in order to provide local Hamiltonian operators.\\

{\bf Remark 6.1.}
It may be useful to recall some basic properties of the Artin braid group $B_n$ (see for instance
\cite{Kau}). $B_n$ has $n$ generators, denoted for the moment by $\{\sigma_1,\sigma_2,\ldots,
\sigma_n\}$, which satisfy the relations

\begin{align}\label {algYB1}
\sigma_i\,\sigma_j\,= &\,\sigma_j\,\sigma_i \;\;\;\;\text{if}\;\,\,|i-j| > 1 \nonumber \\
\sigma_i\,\sigma_{i+1}\,\sigma_i\,= &\,\sigma_{i=1}\,\sigma_{i}\,\sigma_{i+1}
 \;\;\;(\,i=1,2,\ldots,n).
\end{align}

\noindent This group acts naturally on topological sets of $n$ disjoint strands --
running downward and labeled from left to right --
in the sense that each generator $\sigma_i$ corresponds to a crossing of two contiguous 
strands labeled by $i$  and $(i+1)$, respectively (if $\sigma_i$ stands for the crossing
of the $i$--th strand over the $(i+1)$--th one,
then $\sigma_i^{-1}$ represents the inverse operation and 
$\sigma_i\,\sigma_i^{-1}$ $=\sigma_i^{-1}\sigma_i =$ Identity).
By slightly changing notations, denote by $R_{ij}$ the (over)crossing operation acting on
 two strands the endpoints of which are labeled by $i$ and $j$. Then the second relation in
\eqref{algYB1} can be recasted into the form

\begin{equation}\label{algYB2}
R_{12}\,R_{13}\,R_{23}\;=\;R_{23}\,
R_{13}\,R_{12}
\end{equation}

\noindent and represented pictorially as in Fig. 5, where operations are ordered downward.
Note that this picture can be viewed as a portion of an $n$--strands configuration  (and thus
$\{1,2,3\}$ may actually represent labels attached to any triad of contiguous strands) since the first
relation in \eqref{algYB1} ensures that other kinds of crossing are trivial.
The relation \eqref{algYB2} (or, alternatively, \eqref{algYB1}) is the algebraic Yang--Baxter equation
which characterises the algebraic structure of a number of models in statistical mechanics and field theory. 

\begin{center}
{\bf Figure 5}
\end{center}

\noindent The Artin braid group --which arises naturally in the CS--framework
and consequently in the topological setting for quantum computation described above--
plays a crucial role also in the approach recently proposed in 
\cite{KaLo}
where unitary representations of $B_4$ are shown to be universal gates for Boolean
quantum computation. We shall come back on this point at the end of
this section, after the analysis of the algebraic structure underlying the spin network
model. $\blacktriangle$\\

We come now to describe the algebraic content of the theory underlying the spin network simulator
by showing explicitly how binary coupling trees can be embedded into $''$combinatorial$\, ''$ 
2--disks with marked points
whose associated transformations satisfy the Racah and the Biedenharn--Elliott identities 
\eqref{BEid}
and \eqref{Rid}
(instead of the Yang--Baxter equation \eqref{algYB2} characterising the standard topological
approach).

Recall from Appendix A1 (see in particular Fig. 18) that the fundamental binary
trees on $(n+1)=3$ labeled leaves are of three types (twists are inessential). Choose for
instance the tree $\mathsf{T}_{12}$ corresponding to the bracketing scheme

\begin{equation}\label{fundtree1}
\mathsf{T}_{12}\;\longleftrightarrow\;
((j_1\,j_2\,)_{j_{\,12}}j\,_3)_{\,J}
\end{equation}

\noindent and pay attention to the fact that in the drawings we further simplify 
labels by setting
$j_1\,\equiv\,1;$ $\,j_2\,\equiv\,2;$ $\,j_{12}\,\equiv\,12;$
$\,j_3\,\equiv\,3$ while $J$ is left unchanged. Since now we have always 
placed the labels onto the nodes (following the notations of \cite{BiLo9} (Topic 12)) 
but here we switch to
the conventions of \cite{Lituani} by labeling edges. Thus the tree acquires a new edge 
springing from the vertex formerly labeled by $J$ and becomes
a 3--valent graph --denoted by $\mathsf{t}_{12}$-- as shown in Fig. 6.
We may interpret the vertices of such new graphs as $''$interaction vertices$\, ''$
for the pair of incoming spin variables.  

\begin{center}
{\bf Figure 6}
\end{center}

\noindent The procedure to pass from $\mathsf{t}_{12}$ to a decorated
2--disk with marked points is carried out in a few steps illustrated below.\\

{\bf Step 1.}   
A topological transformation can be performed on $\mathsf{t}_{12}$ (technically
it corresponds to the action of a thickening functor, see {\em e.g.} \cite{Qui})
which consists in $''$blowing inside$\, ''$ the graph and smoothing the corners.
The resulting topological 2--manifold, depicted in Fig. 7,
is a 2--sphere $S^2$ with boundaries represented by four circles (a $''$punctured$\, ''$
sphere). We label such boundary circles by arbitrary labels $\{a,b,c,d\}$ and
denote such a surface by $(S^2;\,a,b,c,d)$.

\begin{center}
{\bf Figure 7}
\end{center}

{\bf Step 2.} 
Fill up $(S^2;\,a,b,c,d)$ with an open set $\subset \mathbb{R}^3$: the resulting
3--space is (a portion of) a $''$precursor$\, ''$ of a handlebody $\mathcal{M}^3$ (a
handlebody is a closed oriented 3--manifold --the complement of a link in 
the 3--sphere $S^3$-- obtained by gluing tubular neighbourhoods of the link along the boundary
2--disks). Here we simply glue closed 2--disks along the boundary circles
of $(S^2;\,a,b,c,d)$ to get a 3--manifold with boundary --topologically equivalent to the
3--ball bounded by a 2--sphere-- denoted by $(D^3,S^2)_{\,(a,b,c,d)}$, where
$(a,b,c,d)$ may be thought of as labelings for 2--disks. The corresponding picture can be 
visualised by looking at the configuration  in Fig. 7 as representing a $''$solid$\, ''$ object 
with 2--dimensional disks placed over the circles $a,b,c,d$.\\

{\bf Step 3.} 
Consider an oriented embedding 
 
\begin{equation}\label{tembed}
\mathit{i}_{\,\mathsf{t}}\,:\,\mathsf{t}_{12}\;\hookrightarrow\;
(D^3,S^2)_{\,(a,b,c,d)}
\end{equation}

\noindent which can be realized as a smooth map by splitting the 3--valent vertices in 
$\mathsf{t}_{12}$ according to the rule illustrated in Fig. 8.

\begin{center}
{\bf Figure 8}
\end{center}

\noindent Then the image of $\mathsf{t}_{12}$ into 
$(D^3,S^2)_{\,(a,b,c,d)}$ is obtained by requiring that
$1\,\hookrightarrow\,a$, $2\,\hookrightarrow\,b$, $3\,\hookrightarrow\,c$,
$J\,\hookrightarrow\,d$ as shown in Fig. 9 (we agree to drop
out the auxiliary labels on the boundary 2--disks once the map $\mathit{i}_{\,\mathsf{t}}$
in \eqref{tembed} has been implemented). 

\begin{center}
{\bf Figure 9}
\end{center}

\noindent Having started from the fundamental binary tree of \eqref{fundtree1}
we end up with a set of three (embedded) disjoint strands with crossings. In TQFTs
configurations of this type represent $''$precursors$\, ''$ of knots/links, which
are the observables of the theory as explained in the introductory remarks of this section.

We may note that the procedure outlined so far depends on the choice of the
embedding map \eqref{tembed} and thus it is not uniquely defined. As discussed in \cite{Qui} (Ch. 7)
is always possible to establish a bijection between TQFTs belonging to the categories
$FCx^{0+1}$ (graphs) and $SDiff^{1+1}$ (smooth surfaces) but the construction is not
well defined since one should actually consider $''$categories of all possible choices of
embeddings and regular neighbourhoods, and get an induced TQFT by taking inverse limits over these
categories$\, ''$.\\  

{\bf Step 4.} 
By taking $''$time slicings$\, ''$ on the configuration  of embedded strands in Fig. 9 equipped
with the downward orientation we get the picture shown in Fig. 10, where points and boundaries of disks
inherit consistent labelings. 

\begin{center}
{\bf Figure 10}
\end{center}

\noindent The final configuration  (at the bottom of Fig. 10) represents a 2--disk
$(D^2;$  $j_1,j_2,j_3;$ $j_{12};J)$
with three marked points and one marked circle inside (binary marked 2--disk 
for short). Note that 
\begin{itemize}
\item $SU(2)$--labelings of marked points $(1,2,3)\equiv (j_1,j_2,j_3)$ can be freely chosen; 
\item the labelings of the internal circle ($12\equiv j_{12}$) and of the boundary of the disk
($J$) have suitable ranges ({\em cfr.} \eqref{j12}) and are induced by the original tree structure.
\end{itemize}

{\bf Step 5.}
If we take into account all the three fundamental binary coupling trees (each mapped into a
suitable binary marked 2--disk) we realize that the algebraic structure 
relating the associated Hilbert spaces is encoded into the Racah identity (see the explicit
expression \eqref{Rid})
which we write down schematically (apart from weights/phases) as

\begin{equation}\label{Racahtri}
\mathcal{R}\,(j_{12},\,j_{31})\;=\;\sum_{j_{23}}\;
\mathcal{R}\,(j_{12},\,j_{23})\;\,\mathcal{R}\,(j_{23},\,j_{31})
\end{equation}

\noindent The pictorial representation of \eqref{Racahtri} acting on (the Hilbert spaces of)
binary marked 2--disks is given by the triangular commutative diagram of Fig. 11
which has the same content of the triangular graph shown in Fig. 22 of 
Appendix A2 (where the edges are thought of 
as topological moves on binary trees).

\begin{center}
{\bf Figure 11}
\end{center}

{\bf Step 6.}
By taking into account the different binary coupling schemes of $(n+1)=4$
angular momenta we would get a combinatorial picture based on binary 2--disks with four
marked points and two marked circles, each associated with its own computational Hilbert 
space defined in \eqref{genba}.
Then the natural algebraic structure linking the five spaces (up to phases)
is provided by the Biedenharn--Elliott identity
\eqref{BEid} 
encoded in the pentagonal diagram depicted in Fig. 23 of Appendix A2.\\

{\bf Remark 6.2.}
As is well known from representation theory of simple Lie algebras 
(see {\em e.g.}
\cite{FuSc}), the multiple tensor products of irreducible modules
can be handled formally by employing intertwiner spaces. More precisely, if 
$\Lambda, \Lambda', \Lambda'',\ldots$ label highest weight representations
of the algebra, the isomorphisms $V_{\Lambda}\otimes V_{\Lambda'}$
$\cong V_{\Lambda'}\otimes V_{\Lambda}$ between the modules supporting (the
tensor product of) the irreps $\Lambda, \Lambda'$ is reflected into an isomorphism
of intertwiner spaces

\begin{equation}\label{inter1}
\digamma\;:\;\Upsilon_{\Lambda \Lambda'}^{\,\Lambda_i}\;\longrightarrow\;
\Upsilon_{\Lambda' \Lambda}^{\,\Lambda_i}
\end{equation}

\noindent
and the three--fold isomorphism
$(V_{\Lambda}\otimes V_{\Lambda'})\otimes V_{\Lambda''}$ $\cong$
$V_{\Lambda}\otimes (V_{\Lambda'}\otimes V_{\Lambda''})$
corresponds to an isomorphism

\begin{equation}\label{inter2}
\mathsf{R}\;:\;\sum_i \Upsilon_{\Lambda \Lambda'}^{\,\Lambda_i}
\otimes
\Upsilon_{\Lambda_i \Lambda''}^{\,\Lambda_j}
\;\longrightarrow\;
\sum_i\Upsilon_{\Lambda \Lambda_i}^{\,\Lambda_j}\otimes
\Upsilon_{\Lambda' \Lambda''}^{\,\Lambda_i}
\end{equation}

\noindent of suitably defined intertwiner spaces.
It can be shown that in fact the isomorphisms \eqref{inter1}
and \eqref{inter2} are all we need to treat arbitrary tensor products
of (a finite number of) irreducible modules provided that three compatibility
conditions are fulfilled, namely
a so--called {\em pentagon} and two {\em hexagon} identities.
In the case of the Lie algebra $\mathfrak{s}\mathfrak{l}(2)$ the pentagon relation
is the Biedenharn--Elliott identity \eqref{BEid}
while the two hexagon relations become identical and coincide with the
Racah identity \eqref{Rid}. Thus we recover the content of the 
Biedenharn--Louck Theorem of Section 4.1
stated on the basis of $SU(2)$ recoupling theory.
$\blacktriangle$

\vspace{.5cm}

Summing up, the algebraic structure underlying the kinematics of the spin
network simulator encodes automatically the pentagon and hexagon relations
without resorting to {\em ad hoc} hypotheses. We have also shown that
this combinatorial model for computation can be mapped --not uniquely--
into the topological approach \cite{FrLaWa} and is not affected by localisation
problems typical of any $''$purely topological$\, ''$ setting. Finally, all the gates
appearing in the spin network framework are unitary while unitary representations
of the braid group must be carefully picked up to fit with the usual
(Boolean) quantum circuit model (see \cite{KaLo}). For the convenience of the 
reader we collect below a concise dictionary of the basic objects, spaces and maps
employed in the two approaches.

We argue that the partition functions of the two models may be related to each 
other much in the same way as the (regularized) Ponzano--Regge functional 
corresponds to a double Chern--Simons  
({\em cfr.}
Remark 5.1 at the end of Section 5 and in particular
\eqref{TVCS}).  

\begin{table}[!ht]
\begin{center}
\begin{tabular}{c||c|c|}
 & {\bf combinatorial approach} &
{\bf topological approach}\\
\hline\hline
$\begin{array}{c}
\textbf{information}\\
\textbf{encoded into} 
\end{array}$ &
$\begin{array}{c}
\text{binary coupling trees}\\
\text{on 3 leaves} 
\end{array}$ &
$\begin{array}{c}
\text{topological qubits}\\
(D^2\;,3\,\text{pts}) 
\end{array}$\\
\hline
$\begin{array}{c}
\textbf{computational}\\
\textbf{Hilbert spaces} 
\end{array}$ &
$\mathcal{H}^J_n\,(\mathfrak{b})$ &
$\mathcal{H}_{CS3}(D^2\;,3\,\text{pts})$\\
\hline 
{\bf gates} &
phase and Racah transforms &
unitary actions of $B_{3N}$\\
\hline
$\begin{array}{c}
\textbf{compatibility}\\
\textbf{conditions} 
\end{array}$ &
$\begin{array}{c}
\text{B--E (pentagon) identity}\\
\text{Racah (hexagon) identity} 
\end{array}$ &
Yang--Baxter equation\\
\hline
\end{tabular}
\end{center}
\caption{A dictionary containing the basic ingredients of the combinatorial and
topological approaches.}
\label{TableTop}
\end{table}

\vfill
\newpage

\section*{Appendix A. Spin network combinatorics} 

The next two paragraphs are inspired 
by the basic reference \cite{BiLo9} (Topic 12) where binary couplings of $N$
$SU(2)$ angular momenta and unitary transformations between pairs of such
schemes (recoupling coefficients) are explored on the basis of their underlying
graphs combinatorics. The origin of such an approach based on graph theory dates back 
to the Russian school of nuclear physics (\cite{SmSh} and earlier references therein)
and gave rise to diagrammatical methods of vast applicability \cite{Lituani}, \cite{Russi}. 
Fack and collaborators have recently discussed some improvements concerning the
 efficiency of calculations for $3nj$ symbols \cite{Belgi99}, \cite{Belgi02}. 
The latter achievements are summarised
 in Appendix A3, together with other results exploited in Section 4.3 in connection
with spin network computational complexity.

\subsection*{A1. Binary coupling trees} 

A {\em rooted binary coupling tree} on $(n + 1)$ leaves
 (terminal nodes) is a tree $T$ --namely a connected graph with no cycles (closed loops)-- 
characterised as follows.

{\bf i)} There exists a special vertex, the root.$\blacktriangle$

 {\bf ii)} The tree is binary, namely each of its
 nodes has zero or two $''$siblings$\, ''$. More precisely, if we draw the tree with the root 
at the bottom and the leaves at the top as in all Figures of this section, the siblings
 of a particular node are the nodes lying in the nearest upper level which are connected 
to the given node by an edge. Thus the number of siblings (the out--degree) of both root
 and internal nodes is two, while each leaf has out--degree zero. The number of internal 
nodes of a rooted binary tree on $(n+1)$ leaves is $(n-1)$ and the total number of nodes 
is $2n+$ root $\equiv 2n+1$ which of course coincides with the cardinality of the 
tree as a graph.$\blacktriangle$

{\bf iii)} The leaves are decorated with $(n+1)$ distinct labels.
 
Generally speaking, we may 
attach to the leaves labelings $\{i_1,i_2,i_3,\ldots,$ $i_{n+1}\}$ thought of as a permutation 
of the integers $\{1,2,3,\ldots,n+1\}$. However, in the framework of the quantum
 theory of angular momenta, labels are to be interpreted as quantum numbers
 $\{j_{i_1}, j_{i_2}, j_{i_3},\ldots, j_{i_{n+1}}\}$ associated with a set of 
$(n+1)$ mutually commuting angular momentum operators 
$\mathbf{J}_1,$ $\mathbf{J}_2,$ $\mathbf{J}_3,$ $\ldots,\mathbf{J}_{n+1}$. Once assigned
 any such a labeling to the leaves, we induce a consistent decoration on the
 other nodes by associating with them --moving
downside along the tree-- the quantum numbers of the intermediate angular 
momentum operators arising from the pairwise couplings described by the tree.
 According to this rule the root acquires a label $J$ , the quantum number of the
 total angular momentum $\mathbf{J} =$ $\mathbf{J}_1 + \mathbf{J}_2 +$ 
$\cdots + \mathbf{J}_{n+1}$. 
Referring to \eqref{bcou} and \eqref{bbra} of
 Section 2 it follows that rooted binary coupling trees on $(n + 1)$ leaves represent 
in a faithful combinatorial way the structure of the computational Hilbert 
spaces $\mathcal{H}^J_n (\mathfrak{b})$ given explicitly in \eqref{genba}. 

As an example we sketch in Fig. 12  
the rooted binary coupling tree corresponding to the particular binary
bracketing structure of \eqref{seq} or \eqref{altbin2}.$\blacktriangle$

\begin{center}
{\bf Figure 12} 
\end{center}

{\bf Remark A.1.} As a matter of fact there exist other discrete structures 
in one-to-one correspondences with rooted binary trees and commonly 
used in theoretical computer science. Suppose you have a finite set of 
symbols $S$ ({\em e.g.} letters in an alphabet $A = \{x,y,z,\ldots \})$ endowed with 
a binary operation denoted by paired round parentheses

\begin{equation}\label{A2}
(x,y)\;\in \;(S\times S)\;\mapsto\,(xy)
\end{equation}

\noindent In the most general case the operation is neither commutative 
$[(xy)\neq (yx)]$ 
nor associative $[((xy)z) \neq (x(yz))]$. Note that we may enlarge the set $S$ to include 
also $''$($\, ''$ and $''$)$\, ''$, but then we should require that the admissible words possess 
an equal number of left and right parentheses. As a particular example consider the case where $S
= \{(, )\}$, namely no other symbol except the parentheses themselves. The objects 
we obtain -- with n $''$($\, ''$ and n $''$)$\, ''$ -- are called Dyck words of length $2n$ [{\em e.g.} 
(()()) 
and ((())) for $n=3$] and are enumerated by Catalan numbers. More generally, the
 structure induced by a binary operation on a finite set $S$ turns out to be faithfully 
encoded in binary trees. As shown in Fig. 13, we may represent any pairing of two 
symbols $(xy)$ by labeling with $x$ and $y$, from left to right, two nodes in the tree 
which meet in a third node, then pairing the resulting symbol with a $z \in S$ attached
 to another node, and so on (note that this procedure generates automatically a root
 in the tree).$\blacktriangle$

\begin{center}
{\bf Figure 13} 
\end{center}

For the convenience of the reader, we collect here some useful facts from the combinatorics of 
rooted binary trees ({\em crf.} \cite{BiLo9}, \cite{Belgi99}, \cite{Belgi02}
\cite{Sta}, \cite{Seq} and references therein).
Having in mind the definition of rooted binary tree given in {\bf i)-ii)} we are going 
to distinguish between $''$plane$\, ''$ and $''$not plane$\, ''$ unlabeled trees (here we use the 
adjective $''$plane$\, ''$ instead of $''$ordered$\, ''$ to avoid confusion with possible ordering 
of labels to be assigned to leaves: for the moment there is no labeling at all).
 A tree $T$ is {\em plane} if its nodes --except the root-- are put into an ordered partition 
of disjoint $\{T_1, T_2, T_3,\ldots, T_m\}$ in the Euclidean plane, where each $T_k$ is a plane
 tree. In other words, in plane trees we distinguish between left/right  and
 left/right subtrees. In Fig. 14 all plane unlabeled trees on 3 and 4 leaves are depicted.

\begin{center}
{\bf Figure 14} 
\end{center}

\noindent These trees (on $(n+1)$ leaves) are enumerated by Catalan numbers

\begin{equation}\label{A4}
C_n\, =\;\frac{1}{n+1}\, \binom{2n}{n}\,=\,\frac{(2n)!}{(n + 1)! n!}.
\end{equation}

\noindent The first few terms in the sequence \eqref{A4} are collected in Table 3 at 
the end of this paragraph, while more terms are listed in \cite{Seq}
 (ID Number A000108). According to Remark A.1, an identical counting holds true also for 
Dyck words and -- as illustrated in Exercise 6.19 of \cite{Sta} -- there is actually  a plenty 
of discrete structures from many branches of mathematics 
whose enumeration involves Catalan numbers.

{\em Not plane} trees are obtained as equivalence classes of plane trees of the same size 
under reflections with respect to vertical axes through each node which is not a leaf.
 The resulting structures are known as $''$types$\, ''$ (of not plane trees) and enumerated by
 Wedderburn--Etherington (W--E) numbers $B_{n+1}$ for which a closed form expression is not 
known. The first terms of this sequence are listed in Table 3 (see \cite{Seq}, ID Number 
A001190 for more terms). In Fig. 15 the trees of W--E--types on 3 and 4 leaves are shown.

\begin{center}
{\bf Figure 15} 
\end{center}

\noindent The classification for unlabeled trees considered so far is summarised in the 
upper row of Table 2 at the end of the section.

\vspace{.5cm}

We say that a tree is {\em labeled} when we assign some symbol to each 
of its leaves. Generally speaking, we may use either distinct labels (as in binary
coupling trees, see {\bf iii)} above) or a binary label $\{0, 1\}$ (as in search trees) 
or even a same label for each leaf.

{\bf Remark A.2.} To illustrate the subtleties arising from assignments of no label, a 
same label or distinct labels to leaves of trees, let us consider Catalan trees
 again. According to our previous definition they should be unlabeled, but it is 
easily recognised that the same combinatorics is shared by two more plane binary 
tree structures (see Fig. 16), namely
\begin{itemize}
\item Trees with all leaves labeled by a same $x$. In this case the trees encode
a non associative, partially commuting binary operation on the alphabet 
$\{x\}$. $\bigl [ \, ''$Partially commuting$\, ''$ means that inside each pairing $(xx)$ 
commutativity is trivially ensured, but for instance $((x^2)x) \neq (x(x^2)\, \bigr ]$. 
Note that the case of a non associative totally commuting binary operation corresponds 
to W--E--types labeled with a same $x$ 
since we would get in that case $((x^2)x) = (x(x^2))$ as should be clear by looking at Fig. 15, top).
\item Trees with a fixed sequence of distinct labels ({\em e.g.} ordered lexicographically 
from left to right) on their leaves.
\end{itemize}
\noindent In the following we shall refer to the above labeled structures as Catalan trees too. 
$\blacktriangle$

\begin{center}
{\bf Figure 16} 
\end{center}

From now on we are dealing with distinct labelings, chosen for simplicity in a
 string of $(n+1)$ Latin letters $\{a, b, c, d,\ldots\}$. 

Starting with the plane 
category, we have to decorate the leaves of Catalan trees with all possible 
permutations of labels. Thus the number of these objects for any $n$ is given 
by $(n + 1)!\, C_n$, where $C_n$ is the Catalan number defined in \eqref{A4}. This number,
 usually written as

\begin{equation}\label{A5}
\hat{C}_n\,=\;\frac{(2n)!}{n!}\;,
\end{equation}

\noindent is the quadruple factorial and in Fig. 17 all labeled plane 
trees on $n+1=3$ leaves are shown $(\hat{C}_2 = 3! C_2 = 12)$. The
 first terms of this integer sequence are listed in Table 3 
 and more terms can be found in \cite{Seq}, ID Number A001813.

\begin{center}
{\bf Figure 17} 
\end{center}

To enumerate labeled {\em not plane} trees on $(n+1)$ leaves we realize that  there exist 
$n$ axes through the $(n-1)$ internal nodes plus root. Thus, on the basis of
\eqref{A5} and looking also at Fig. 17, we have to drop out exactly $2^n$ configurations 
since the trees are binary (namely each of the former $n$ nodes has exactly two siblings).
The resulting counting reads

\begin{equation}\label{A6}
D_n\,\doteq\, \frac{\hat{C}_n}{2^n}\, = \,\frac{(2n)!}{n!2^n} = (2n - 1)!! , 
\end{equation}

\noindent where $(2n-1)!! \equiv 1 \cdot 3 \cdot 5 \cdot 7 \ldots$
is the double factorial (see Table 3 and \cite{Seq}, ID Number A001147). The combinatorics 
of this enumeration can be also understood by picking up each W--E type of unlabeled trees
 (see Fig. 15) and decorating consistently its leaves. Then

\begin{equation}\label{A7}
D_n\; =\;
\sum_{b=1}^{B_{n+1}}\;
p_{n+1} (b)
\end{equation}

\noindent where $p_{n+1}\, (b)$ represents the number of ways to decorate leaves of type 
$b$ trees with $(n + 1)$ distinct labels in such a way that $(xy) = (yx)$ for each binary 
parenthesization involving both nodes and subtrees. In Fig. 18 not plane labeled 
trees on 3 leaves are depicted ($D_2 = \hat{C}_2 /2^2 = 3$): they arise from the unique W--E 
type shown in Fig. 15, top.

\begin{center}
{\bf Figure 18} 
\end{center}

 In the following Table 2 the combinatorial enumerations for all trees 
considered so far are summarised, while in Table 3 the first few terms of the 
four integer sequences are written down to give the reader an idea of their rates of growth. 

\vfill\eject 

\begin{table}[!ht]
\begin{center}
\begin{tabular}{c||c|c|}
& {\bf plane} & {\bf not plane}\\
\hline \hline
{\bf unlabeled} & 
$\begin{array}{c}
\textit{Catalan number}\\
C_n=\frac{(2n)!}{n!(n+1)!}
\end{array}$ &
$\begin{array}{c}
 \textit{Wedderburn--Etherington number}\\
B_{n+1}
\end{array}$
\\
\hline
{\bf labeled} & 
$\begin{array}{c}
\textit{Quadruple factorial}\\
\hat{C}_n=\frac{(2n)!}{n!}\\
=\text{card}\,(\hat{\mathfrak{G}}_n (V,E)
\end{array}$ 
& 
$\begin{array}{c}
\textit{Double factorial}\\
D_n=\frac{(2n)!}{n!\,2^n}=(2n-1)!!\\
=\text{card}\,(\mathfrak{G}_n (V,E)
\end{array}$
\\
\hline
\end{tabular}
\end{center}
\caption{Enumerations of rooted binary trees on $(n+1)$
leaves according to {\bf plane/not plane} and {\bf unlabeled/labeled}
categories. In case of labeled trees the numbers represent the cardinalities of the
Twist--Rotation and Rotation graphs, respectively (see Appendix A2).}
\label{Table1}
\end{table}

\begin{table}[!ht]
\begin{center}
\begin{tabular}{l||lllllll}
{\bf n} & {\bf 1} & {\bf 2} & {\bf 3} & {\bf 4} & {\bf 5} & {\bf 6} & {\bf 7}\\
\hline\hline 
$B_{n+1}$ & 1 & 1 & 2 & 3 & 6 & 11 & 23\\
$C_n$ & 1 & 2 & 5 & 14 & 42 & 132 &  429\\
$D_n$ & 1 & 3 & 15 & 105 & 945 & 10395 & 135135\\
$\hat{C}_n$ & 2 & 12 & 120 & 1680 & 30240 & 665280 & 17297280\\
\end{tabular}
\end{center}
\caption{The first terms of the four integer sequences of Table 1 arranged according
to their rate of grow. $B_{n+1}$ are Wedderburn--Etherington numbers;
$C_n$ are Catalan numbers; $D_n$ are double factorial numbers; $\hat{C}_n$
are quadruple factorial numbers}
\label{Table2}
\end{table}

\subsection*{A2. Twist-Rotation and Rotation graphs} 

Here we explain --keeping on
 using extensively graph--theoretical tools as in \cite{BiLo9}, \cite{Belgi99}-- the 
construction underlying the discrete computational space of the spin 
network simulator discussed in Section 4.1. For simplicity we do not 
change the notation used there, although many definitions and results
 can be applied also to larger classes of graphs (the so--called $''$distance graphs$\, ''$ 
see \cite{BuHa}). 
 
Given a rooted binary tree $T$ on $(n+1)$ labeled leaves ($n \geq 2$), 
two kinds of topological operations (moves) can be considered, namely rotations and 
twists ($''$rotation$\, ''$ is not to be confused with rotation matrices or operators used in 
Sections 3.2, 4.1 and Appendices B). These moves represent alterations in the shape of 
the tree generated around either an internal node (rotation) or a node which is not 
a leaf (twist) and all trees in a given class can be reached on applying admissible
 moves to an arbitrary tree chosen in that class (roughly speaking the moves are $''$ergodic$\, ''$).

{\em Rotations} around non--root internal nodes consist in swapping subtrees 
or nodes as shown in Fig. 19. In the upper region there appears on the left a generic 
(portion of a) tree $T$ with subtrees $A, B, C$ and a fourth subtree $R$ containing the root: 
the rotation around the node $x$ transforms $T$ into $T'$ and, conversely, the rotation 
around $x' \in T'$
 changes back $T'$ into $T$ . In the lower region a rotation on a particular 
tree is shown, together with its inverse operation.

\begin{center}
{\bf Figure 19} 
\end{center}

\noindent Since there are $(n-1)$ internal nodes, $(n-1)$ different rotations can be
 performed on any tree (with fixed labelings on its leaves). Note that a rotation alterates 
the shape of the tree near the corresponding node but leaves the rest of its structure intact.

{\em Twists} around non terminal nodes consist in exchanging left and right subtrees (or nodes). 
We draw in Fig. 20 both a general twist around a node $x \in T$ and a particular twist on a specific tree.

\begin{center}
{\bf Figure 20} 
\end{center}

\noindent There are $n$ possible twists on a rooted labeled tree on $(n+1)$
leaves and these transformations may alterate the global shape of the tree.

Coming back to combinatorics of labeled rooted binary trees (see Table 2), we realize 
that every category may support rotations but only $''$plane$\, ''$ trees (enumerated by the quadruple 
factorial $\hat{C}_n$) 
admit twists. For what concerns labeled Catalan trees in Fig. 16 (bottom) we see that they are actually connected 
by the rotation represented in Fig. 19, but no twist is allowed since the sequence of labels must be kept fixed. 
Looking at Fig. 17, where plane trees labeled in all possible ways are shown, we see clearly
 that both rotations 
and twists appear as admissible moves connecting pairs. Finally, not plane trees (enumerated 
by the double factorial $D_n$ and shown in Fig. 18) undergo only rotations since they do not
 distinguish between left and right at each non terminal node. The further step consists 
in building up new graphs associated with each of the above categories (these structures 
are called $''$distance graphs$\, ''$ and the reason for such terminology will become clear in the
 next paragraph, where distances will be introduced). Denote generally such a graph by $\mathbf{G}_n 
(V, E)$ (as in Section 4.1), where $V$ and $E$ are the vertex and the edge sets, respectively. 
The vertices are in one--to--one correspondence with rooted binary
trees on $(n+1)$ labeled leaves, namely

\begin{equation}\label{A9}
T^{(n)}\;\longleftrightarrow\; v\,\in V, 
\end{equation}
 
\noindent where $T^{(n)}$ stands for a tree in a given 
class. The edge set $E$ is generated by linking pairs of vertices 
$v, v' \in V$ by an undirected edge $e(v, v')$ if, and only if, the
 corresponding trees are related to each other either by a rotation or by a twist. Formally

\begin{equation}\label{A10}
e(v, v')\,\in E   \Longleftrightarrow 
 T^{(n)}_v\,\leftrightarrow \,  T^{(n)}_{\,v'}\;,
\end{equation}

\noindent  where $\leftrightarrow$ stands for one of the 
topological moves described above. According to the remarks on admissible
 moves acting in different categories of trees we specialise $\mathbf{G}_n(V, E)$ as follows.

$\bullet$ {\bf Labeled plane trees} on $(n+1)$ leaves represent the vertex set of the 
Twist-Rotation graph $\mathbf{G}^{TR}_n (V, E)$ denoted as in \eqref{TR1}, namely

\begin{equation*}
\mathbf{G}^{TR}_n (V, E) \;\doteq\; \hat{\mathfrak{G}}_n (V, E). 
\end{equation*}

\noindent Its cardinality is the quadruple factorial
 number given in \eqref{A5}

\begin{equation}\label{A11}
\text{card} ( \hat{\mathfrak{G}}_n (V,E))\;\doteq\;|V |\,=\, \hat{C}_n 
\end{equation}

\noindent and each edge represents either a rotation or a twist ({\em cfr.} also \eqref{TR2}).
 
For any $n \geq 2$ $\hat{\mathfrak{G}}_n (V, E)$ is an undirected regular 
cubic graph (namely there is no ordering on its vertices and each vertex 
has valence three). It is also a planar graph, {\em i.e.} it can be drawn onto the
2--dimensional sphere without crossings. For $n = 2$ card$(\hat{\mathfrak{G}}_2) = 12$ 
({\em cfr.} the trees in Fig. 17) and $\hat{{\mathfrak{G}}}_2$ 
has the shape of a truncated tetrahedron made up of 
triangular and hexagonal faces. $\hat{\mathfrak{G}}_3$ has cardinality 120 and represents the graph
 of a 3--valent polyhedron made up of pentagons and hexagons. A portion of this 
graph is shown in Fig. 21 (reprinted from \cite{AqCo}) and its shape looks quite
 familiar since the discover of fullerene.

\begin{center}
{\bf Figure 21} 
\end{center}

$\bullet$ {\bf Labeled not plane trees} on $(n+1)$ leaves represent the vertex set of 
the Rotation graph 
$\mathbf{G}^{R}_n (V, E)$ denoted as in \eqref{R1}, namely

\begin{equation*}
\mathbf{G}^{R}_n (V, E) \;\doteq\; \mathfrak{G}_n (V, E). 
\end{equation*}

\noindent Its cardinality is the quadruple factorial number given in \eqref{A6}

\begin{equation}\label{A12}
\text{card} (\mathfrak{G}_n (V,E))\;\doteq\;|V |\,=\, D_n 
\end{equation}

 \noindent and each edge represents a rotation ({\em cfr.} also \eqref{R2}).
 For any $n \geq 2$, $\mathfrak{G}_n (V,E)$ is an undirected regular graph 
of valence $2(n-1)$ 
which turns out to be not planar for $n\geq 3$. 
 
The graph $\mathfrak{G}_2$ 
with cardinality 3 shown in Fig. 22 is a trivial example from the combinatorial point of view 
but -- by exploiting the encoding map \eqref{A8}
-- we see that it shares the same 
content as the $''$triangular$\,''$ Racah algebraic identity \eqref{Rid} introduced in Section 4.1.

\begin{center}
{\bf Figure 22} 
\end{center}

\noindent The graph $\mathfrak{G}_3$ (whose vertices are trees on 4 leaves) is 
depicted in Fig. 1 of Section 4.1: 
from a topological point of view we see that some vertices have been doubled 
to avoid crossings; when we identify by an antipodal mapping opposite vertices 
and edges we realize that this graph lies in the real projective space $\mathbb{R}
\mathbb{P}^2$ (namely it is not planar). Other pictures of $\mathfrak{G}_3$ 
can be found in \cite{Belgi99} and \cite{Belgi02}. 
In Fig. 23 one of the pentagons belonging to both $\mathfrak{G}_3$ in Fig. 1
and $\hat{\mathfrak{G}}_3$ in Fig. 21 is depicted: edges represent rotations from a topological 
point of view and at the same time the figure is the diagram encoding through \eqref{A8} 
the Biedenharn--Elliot (pentagon) identity \eqref{BEid}.

\begin{center}
{\bf Figure 23} 
\end{center}

$\bullet$ {\bf Labeled Catalan trees} on $(n+1)$ leaves represent
 the vertex set of the Rotation graph denoted by

\begin{equation}\label{A13}
\mathbf{G}^R_n (V, E)\;\doteq\;\mathfrak{g}_n (V, E) 
\end{equation}

\noindent to distinguish it from the previously defined $\mathfrak{G}_n$. 
$\mathfrak{g}_n (V, E)$ is a regular graph of 
valence $(n-1)$ and its cardinality is given by the Catalan number \eqref{A4}, namely

\begin{equation}\label{A14}
\text{card} (\mathfrak{g}_n (V, E))\;\doteq\; |V | \,=\, C_n.
\end{equation}

\noindent Although this kind of 
Rotation graph is not suitable to model the computational space 
of the quantum simulator (Catalan coupling trees are not sufficiently
 general) we meet these structures in connection with issues
 on combinatorial complexity discussed in Section 4.3 and in the next paragraph.

\subsection*{A3. Combinatorial complexity} 

All graphs described in Section A2 are actually
$''$distance graphs$\, ''$ \cite{BuHa} since they encode in their edge sets operations
 on the basic objects (trees) associated with their vertex sets. To evaluate 
quantitatively how far away pairs of rooted labeled binary trees are --and using
 standard terminology in discrete mathematics-- we introduce explicitly 
Twist--Rotation and Rotation distances. 

Given two binary trees $T^{(n)}_1$, $T^{(n)}_2$ --thought
of as vertices in $\hat{\mathfrak{G}}_n$, $\mathfrak{G}_n$, $\mathfrak{g}_n$, respectively-- 
their distance is the length of the shortest
path joining them, namely the minimum number of topological operations needed to transform
one tree into the other. With an obvious meaning of symbols we set

\begin{align}\label{A15}
d^{TR}\,(T^{(n)}_1 ,T^{(n)}_2)\;\doteq & \;\text{min}\,\{\text{lengths of paths}\;
T^{(n)}_1 \leftrightarrow T^{(n)}_2 \,\text{in} \,\hat{\mathfrak{G}}_n\}\nonumber\\
d^{R}\,(T^{(n)}_1 ,T^{(n)}_2)\;\doteq &\; \text{min}\,\{\text{lengths of paths}\;
T^{(n)}_1 \leftrightarrow T^{(n)}_2 \,\text{in} \,\mathfrak{G}_n\}\nonumber\\
d^{r}\,(T^{(n)}_1 ,T^{(n)}_2)\;\doteq & \;\text{min}\,\{\text{lengths of paths}
T^{(n)}_1 \leftrightarrow T^{(n)}_2 \,\text{in}\, \mathfrak{g}_n\}.
\end{align}

\noindent The diameters of the above graphs are naturally defined in 
terms of distances according to

\begin{align}\label{A16}
Diam\,(\hat{\mathfrak{G}}_n) \,=\;&\,\text{max}\; \{d^{TR}\,(T^{(n)}_1,T^{(n)}_2)\,|\, 
T^{(n)}_1,T^{(n)}_2 \in\,\hat{\mathfrak{G}}_n\}\nonumber\\
Diam\,(\mathfrak{G}_n) \,=\;&\,\text{max}\; \{d^{R}\,(T^{(n)}_1,T^{(n)}_2)\,|\, 
T^{(n)}_1,T^{(n)}_2 \in\,\mathfrak{G}_n\}\nonumber\\
Diam\,(\mathfrak{g}_n) \,=\;&\,\text{max}\; \{d^{r}\,(T^{(n)}_1,T^{(n)}_2)\,|\, 
T^{(n)}_1,T^{(n)}_2 \in\,\mathfrak{g}_n\}.
\end{align}

\noindent In addressing combinatorial complexity we are primarily interested in 
computing or estimating such distance functions and diameters since in Section 4.3 
we relate combinatorial complexity to computational complexity of the spin network 
simulator on the basis of the quantum encoding maps introduced in  
\eqref{A1} and \eqref{A8}. 
In this respect it may be useful to collect here some known results from graph theory 
(addressed in a classical computational complexity framework).

$\bullet$ For what concerns Catalan trees equipped with rotation distance $d^r$, 
a major breakthrough was achieved in \cite{Thu}, where the authors proved
the existence of a tight bound on the diameter of $\mathfrak{g}_n$ given by

\begin{equation}\label{A17}
Diam\, (\mathfrak{g}_n)\, <\, 2n-8
\end{equation}

\noindent for trees on $(n+1)$ terminal nodes. Their elegant proof relies
 on the translation of the combinatorial problem for trees into
 an equivalent geometrical one, namely enumerating triangulations 
of a polygon with $(n+1)$ edges and finding the maximum number of diagonal
 flips needed to convert one triangulation into another (diagonal flips are
 topological moves on 2--dimensional triangulations which transform a 
quadrilateral dissected into two triangles into the configuration  generated
 by cutting along the other diagonal). The argument to get the final result
 involves the construction of 3--dimensional hyperbolic polyhedra and calls into play volume 
estimates in hyperbolic geometry.

$\bullet$ In \cite{Belgi02}, purely combinatorial counting tools used previously in \cite{Bio96}
are improved to establish lower and upper bounds for the diameter of
the Rotation graph $\mathfrak{G}_n$. The explicit form of the upper bound is 
$(\lg = log_2)$

\begin{equation}\label{A18}
Diam\,(\mathfrak{G}_n)\, <\, n \lg(n) + n - 2\lg(n) + 1  
\end{equation}

\noindent which is compatible with older estimates
(see {\em e.g.} \cite{RotDist}). 
Therefore we may conclude
that the diameter of
$\mathfrak{G}_n$ grows no faster than polynomially in $n$, namely

\begin{equation}\label{A19}
Diam \,(\mathfrak{G}_n) \,\lesssim\, n \lg(n) + O (n). 
\end{equation}

$\bullet$ A crucial open question remains the complexity status of computing the rotation distance
 even for the simplest case of $d^r$ on the graph $\mathfrak{g}_n$ made up by unlabeled
 Catalan trees (binary search trees in computer science language). 
In particular ({\em cfr.} the recent papers 
\cite{Pal00}, \cite{Pal03}) it is not known
\begin{itemize}
\item whether the problem is {\bf NP}--complete;
\item whether $d^r$ can be determined in time polynomial in $n$ (that is to say,
whether there exists an efficient algorithm to compute it exactly).
\end{itemize}
\noindent There are however (classical) polynomial time algorithms which estimate
 this distance (or its lower/upper bounds) (\cite{Pal00}, \cite{Rog}). 

$\bullet$ The two remarks above hold for the rotation distance $d^R$ too, since $\mathfrak{G}_n$ is 
much bigger than $\mathfrak{g}_n$: more precisely, if we think about $''$labeled$\, ''$ Catalan 
trees (as pointed out in Remark A.2, Appendix A1) then $\mathfrak{g}_n$ is actually
 a subgraph of $\mathfrak{G}_n$. To our knowledge there is no algorithm for computing
$d^R$ in $\mathfrak{G}_n$, although we may consider lower and upper bounds
on $Diam\, (\mathfrak{G}_n)$ \cite{Belgi02} (which are both of order $n \lg(n)$) as 
an estimate of the number of elementary operations --Racah 
transforms-- appearing in an $''$optimal$\, ''$ expansion of a $3nj$ 
symbol. (Note however that such optimal
expansion does exist only if we could actually computing
$d^R(T^{(n)}_1 ,T^{(n)}_2)$ for arbitrary $T^{(n)}_1 ,T^{(n)}_2$ $\in \mathfrak{G}_n$)

$\bullet$ In order to try to overcome the difficulties outlined above, 
a basic strategy emerges in current literature, namely the 
idea of introducing in $\mathfrak{g}_n$  some more specific notion of distance function. 
In \cite{Cle} a $''$restricted$\, ''$ rotation distance has been 
considered, and linear lower and upper bounds are estimated in terms of the
 number of interior nodes of trees. In \cite{Pal03} a $''$right--arm$\, ''$ rotation
 distance is shown to be computable by an efficient algorithm (in polynomial
 time $O(n^2)$ for trees on $(n+1)$ leaves).

$\bullet$ In the field of molecular biology, phylogenetic trees (or dendograms) 
for groups of species turn out to be powerful tools to address the study of 
similarities and dissimilarities appearing in biological evolution (see {\em e.g.} 
\cite{Bio00} and references therein). Theoretical biologists use different kinds
 of definitions with respect to our previous classifications from discrete mathematics. 
In particular, their rooted binary phylogenetic trees may have labels only on leaves
 (the different species) while the role of internal nodes is secondary. The distance
 function which is commonly used in this context is the $''$nearest neighbour interchange$\, ''$
(nni) distance which is surely more restrictive than the rotation distances introduced
 in \eqref{A15}, although it is not so easy to establish connections with the restricted
 rotation distances considered in our previous remark. The authors of \cite{Bio00}
claim to have proved the fact that computing the nni distance is {\bf NP}--complete (both in 
the unlabeled and in the labeled cases) by a reduction from Exact Cover by 3--sets) (X3C) 
which is known to be an {\bf NP}--complete problem \cite{GaJo}.

\vfill
\eject

\section*{Appendix B}
\subsection*{B1. Composition of W--rotation matrices}

In this Appendix, following the standard reference \cite{Russi}, we illustrate the derivation 
of the symbolic expression given in \eqref{NWcsser} which represents --in the language of quantum theory 
of angular momenta-- 
the generalised Clebsch--Gordan expansion involving a finite number of
W--rotation matrices with the same arguments. 

Denote as usual by

\begin{align}\label{incom}
{\bf J}_1,& \,{\bf J}_2\,,\,{\bf J}_3\,,\ldots,{\bf J}_{N};\nonumber\\
j_1,& \,j_2\,,\,j_3\,,\,\ldots,j_N,\nonumber\\
m_1,& \,m_2\,,\,m_3\,,\,\ldots,m_N,\nonumber\\
m'_1,& \,m'_2\,,\,m'_3\,,\,\ldots,m'_N
\end{align}

\noindent an set of $N$ commuting angular momentum operators and the
corresponding sets of quantum numbers. 

Consider the operators

\begin{equation}\label{keys}
\boldsymbol{\mathcal{K}}_i\;\doteq\;{\bf J}_1\,+\,{\bf J}_2\,+\,{\bf J}_3\,+\dotsb +{\bf J}_i
\end{equation}

\noindent which are defined, for each $1\leq i \leq N$, by taking 
any kind of binary coupling consistent with this vector addition rules.
We use here $\boldsymbol{\mathcal{K}}$'s
instead of ${\bf K}$'s used in Section 2 to stress the fact that we can choose {\em anyone}
of the schemes considered there (note that in \eqref{keys}
there appear also $\boldsymbol{\mathcal{K}}_1\;\equiv\;{\bf J}_1$ and 
$\boldsymbol{\mathcal{K}}_N\;\equiv\;{\bf J}$, so that the counting of intermediate
operators is not in contradiction with previous statements).
The quantum numbers of these operators are denoted by

\begin{equation}\label{jqnumK}
\kappa_1\,,\,\kappa_2\,,\,\kappa_3\,,\,\ldots\kappa_N,
\end{equation}

\noindent while (two possible sets of) magnetic quantum numbers are given by

\begin{align}\label{mqnumK}
M_i & =\;m_1\,+\,m_2\,+\,m_3\,+\,\dotsb +\,m_i\, \nonumber\\
M'_i & =\;m'_1\,+\,m'_2\,+\,m'_3\,+\,\dotsb +\,m'_i\,.
\end{align}

Coming to the composition of W--matrices, we start by considering the Clebsch--Gordan series, 
namely the expansion of the product of two W--matrices (with the same arguments) 
labeled by $j_1$ and $j_2$. It reads

\begin{equation*} 
D^{\,j_1}_{\,m_1\,m'_1}\;(\alpha \beta \gamma)\;
D^{\,j_2}_{\,m_2\,m'_2}\;(\alpha \beta \gamma)\;=
\end{equation*} 
\begin{equation}\label{Wcsser1} 
\sum_{J=|j_1-j_2|}^{j_1+j_2}
\;\sum_{M\,M'}\;
C^{\,J\,M}_{j_{1}m_{1}\,j_2\,m_2}\;
D^{\,J}_{\,M\,M'}\;(\alpha \beta \gamma)\;
C^{\,J\,M}_{j_{1}m'_{1}\,j_2\,m'_2},
\end{equation} 

\noindent where we denote simply by ${\bf J}$ the sum ${\bf J}_1 +{\bf J}_2$
and by $J,M, M'$ the corresponding quantum numbers (leaving aside for the moment the notations
introduced in \eqref{keys}, \eqref{jqnumK} and \eqref{mqnumK}). By using the orthogonality 
condition of the Clebsch--Gordan coefficients $C^{\bullet \bullet}_{\bullet \bullet \;\bullet \bullet}$
we can invert \eqref{Wcsser1} to get 

\begin{equation*} 
D^{\,J}_{\,M\,M'}\;(\alpha \beta \gamma)\;=
\end{equation*} 
\begin{equation}\label{Wcsser2} 
\sum_{\substack{m_1\, m_2\\
m'_1\, m'_2}}
C^{\,J\,M}_{j_{1}\,m_{1}\;j_2\,m_2}\;
D^{\,j_1}_{\,m_1\,m'_1}\;(\alpha \beta \gamma)\;
D^{\,j_2}_{\,m_2\,m'_2}\;(\alpha \beta \gamma)\;
C^{\,J\,M}_{j_{1}\,m'_{1}\;j_2\,m'_2}\;\;,
\end{equation} 

\noindent where the $j$'s entries of each Clebsch--Gordan coefficient must fulfill 
the triangular inequality (here in particular we have just to require 
$|j_1 -j_2| \leq J \leq j_1+j_2$).\\ 
The decomposition established above represents the starting point to
address the more general case of $N > 2$ incoming angular momenta. 
By fully restoring the notations explained at the beginning of this section and applying
successively \eqref{Wcsser2} together with the orthogonality conditions whenever necessary, we get
the {\em generalised} C--G expansion involving the product of $N$ W--rotation matrices. 
Its explicit expression reads

\begin{equation*} 
D^{\,\kappa_N}_{\,M_N\,M'_N}\;(\alpha \beta \gamma)\;=
\end{equation*} 
\begin{equation}\label{NWcsser1} 
\sum_{\substack{m_1 \ldots m_N\\
m'_1\ldots m'_N}}\;\;
\prod_{i=1}^{N}\;
C^{\,\kappa_i\,M_i}_{\kappa_{i-1}\,M_{i-1}\;j_i\, m_i}\;\,
D^{\,j_i}_{\,m_i\,m'_i}\;(\alpha \beta \gamma)\;\,
C^{\,\kappa_i\,M'_i}_{\kappa_{i-1}\,M'_{i-1}\;\,j_i\, m'_i}\;,
\end{equation} 

\noindent where it has been assumed $\kappa_0$= $M_0$ = $M'_0\,=0$ and all the triads
$\{j_i\,,\kappa_{i-1}\,,\kappa_i\}$ satisfy triangular inequalities.\\
We may recast the previous expansion in a symbolic form ({\em cfr.} \eqref{NWcsser} 
in Section 3.2) by keeping only the dimensions of representations 
both in $D$'s and $C$'s (the typographical changes are introduced to remind that
these objects are matrices)

\begin{equation*} 
\boldsymbol{D}^{\,J}\;(\alpha \beta \gamma)\;=
\end{equation*} 
\begin{equation}\label{NWcsser2} 
\sum_{\{m,\,m'\}}\;\,
\prod_{i=1}^{N}\;\left(\,
\boldsymbol{C}^{\,\kappa_i}_{\kappa_{i-1}\;j_i}\;\,
\boldsymbol{D}^{\,j_i}\;(\alpha \beta \gamma)\;\,
\boldsymbol{C}^{\,\kappa_i\,}_{\kappa_{i-1}\;j_i}\;\right).
\end{equation} 

\noindent Here we set

\begin{align}\label{totkappa}
{\bf J}\;\doteq\; &
\boldsymbol{\mathcal{K}}_N\;\equiv\;{\bf J}_1\,+\,{\bf J}_2\,+\,{\bf J}_3\,+\dotsb +{\bf J}_N;\nonumber\\
J\;\doteq\; & \kappa_N 
\end{align}

\noindent to comply with the notation used in the main text and in the particular case \eqref{Wcsser1}
(consequently the matrix indices
of $\boldsymbol{D}^{\,J}$ $(\alpha \beta \gamma)$ are intended to be 
$M=m_1+m_2+m_3+ \dotsb + m_N$;
$M'=m'_1\,+\,m'_2\,+\,m'_3\,+\,\dotsb +\,m'_N$).\\
The summation over the magnetic numbers of the incoming ${\bf J}$'s which appears in 
front of \eqref{NWcsser1} 
(or \eqref{NWcsser2}) is a partial trace on the $N$--product of triples contained in round parenthesis:
any such triple combination can be directly evaluated by substituting the numerical values of
the $''$elementary$\, ''$ W--matrices and of the suitable pair of C--G coefficients. 

The structure displayed in \eqref{NWcsser1} is drastically simplified if we look at two particular cases.

\vspace{0.5cm}

$\bullet$ {\bf Fermionic case} ($N$  spins $=\frac{1}{2}$ in the symmetric multiplet) 

For

\begin{equation*}
j_1\,=\,j_2\,=\,j_3\,=\dotsc=\,j_N\,=\,1/2
\end{equation*}

\noindent and
 
\begin{equation*}
\kappa_{i+1}\,=\,\kappa_i \,+\,1/2 \;\;\;\;\;\; (\Rightarrow J\equiv \kappa_N\,=\,N/2)
\end{equation*}

\noindent the expansion \eqref{NWcsser1} becomes

\begin{equation}\label{sp12ser} 
\Delta\;\cdot D^{\,J}_{\,M\,M'}\;(\alpha \beta \gamma)\;=
\sum_{\substack{m_1+ \ldots +m_N=M\\
m'_1+\ldots +m'_N=M'}}\;\;\;
\prod_{i=1}^{N}\;\,
D^{\frac{1}{2}}_{m_i\;m'_i}\;(\alpha\,\beta\,\gamma)
\end{equation}

\noindent where the weight $\Delta$ is given by
 
\begin{equation}\label{comdelta}
\sqrt{(J+M)!\,(J-M)!\,(J+M')!(J-M')!}\,/\,(2J)! 
\end{equation}

\noindent and each factor $D^{\frac{1}{2}}_{m\,m'}$ represents
the W--matrix in the fundamental representation written in terms of 
the Euler angles $\alpha\,\beta\,\gamma$, namely

\begin{equation}\label{D12}
D^{\frac{1}{2}}_{m\,m'}\;(\alpha\,\beta\,\gamma)\;=\;
\begin{pmatrix}
e^{-i\alpha/2}\cos (\beta/2)\;e^{-i\gamma/2} &  -e^{-i\alpha/2}\sin (\beta/2)\;e^{i\gamma/2}\\
e^{i\alpha/2}\sin (\beta/2)\;e^{-i\gamma/2} &  e^{i\alpha/2}\cos (\beta/2)\;e^{i\gamma/2}
\end{pmatrix}\,.
\end{equation}

\vspace{0.5cm}

$\bullet$ {\bf Bosonic case} ($N$ spins $=1$ in the symmetric multiplet) 

For

\begin{equation*}
j_1\,=\,j_2\,=\,j_3\,=\dotsc=\,j_N\,=\,1
\end{equation*}

\noindent and
 
\begin{equation*}
\kappa_{i+1}\,=\,\kappa_i \,+\,1 \;\;\;\;\;\; (\Rightarrow J\equiv \kappa_N\,=\,N)
\end{equation*}

\noindent the expansion \eqref{NWcsser1} becomes

\begin{equation*}
\Delta\;\cdot D^{\,J}_{\,M\,M'}\;(\alpha \beta \gamma)\;=
\end{equation*}
\begin{equation}\label{sp1ser} 
\sum_{\substack{m_1+ \ldots +m_N=M\\
m'_1+\ldots +m'_N=M'}}\;\;\;
\prod_{i=1}^{N}\;\,\,
\sqrt{(1+\delta_{m_{i}\,0})\,(1+\delta_{m'_{i}\,0})\;}\;
D^{1\,}_{m_i\;m'_i}\;(\alpha\,\beta\,\gamma)\,,
\end{equation}

\noindent where $\Delta$ is given in \eqref{comdelta}
and each factor $D^{1\,}_{m\,m'}$ represents
the W--matrix in the $j=1$ representation written in terms of 
the Euler angles $\alpha\,\beta\,\gamma$, namely

\begin{equation}\label{wigj1}
D^{\,1}_{\,M\,M'}\;(\alpha \beta \gamma)\;=\;e^{-iM\alpha}\;d^{\,1}_{\,M\,M'}\;(\beta)\;
e^{-iM'\gamma}
\end{equation}

\noindent with

\begin{equation}\label{rewj1}
d^{\,1}_{\,M\,M'}\;(\beta)\;=\;
\begin{pmatrix}
\frac{1+\cos \beta}{2} & -\frac{\sin \beta}{\sqrt{2}} & \frac{1-\cos \beta}{2} \\
\frac{\sin \beta}{\sqrt{2}} & \cos \beta & -\frac{\sin \beta}{\sqrt{2}}\\
\frac{1-\cos \beta}{2} & \frac{\sin \beta}{\sqrt{2}} & \frac{1+\cos \beta}{2} 
\end{pmatrix}\,.
\end{equation}

{\bf Remark B.1}. Equations \eqref{sp12ser} and \eqref{sp1ser} turn out to be useful for evaluating
in practice matrix elements of W--rotations acting on irreducible representation spaces. 
Notice in particular that matrix elements
\eqref{sp1ser} can be actually obtained by considering the symmetric 
combination of two spins $=1/2$, namely by exploiting \eqref{sp12ser} with $N=2$. More generally,
any $(2j_i+1)\times$ $(2j_i+1)$ W--matrix (as those appearing in \eqref{NWcsser1}) acts
on a representation space which is irreducible under the action of $SU(2)$ and (according to 
Majorana splitting \cite{Maj})
such eigenspace may be always viewed as a completely symmetric multiplet generated
by  $2j_i$ $\frac{1}{2}$ spin kinematically independent $''$particles$\, ''$. By taking advantage of this remark
we may factorize  each $D^{j_i}_{m_i\,m'_i}$ in terms of \eqref{D12}, as discussed in relation with
the estimate given in \eqref{couM1}. 
$\blacktriangle$

\subsection*{B2. $U$--rotation matrices}

In this appendix we collect from \cite{Russi} some useful definitions and examples concerning 
rotation matrices expressed in terms of the direction of the rotation axis and of the 
rotation angles, namely alternative forms for the $M$--gates introduced in Section 3.2. 

Let $\mathbf{n}(\Theta, \Phi)$ a unit vector in the direction of the rotation axis and
$\omega$ the rotation angle. An element of an $U$--matrix is defined according to

\begin {equation}\label{B1}
U^J_{M M'\,}(\omega; \Theta, \Phi)\;\doteq\;\langle JM\,|\,\exp\{-i\omega \mathbf{n}\cdot
\mathbf{J}\}\;|\,JM'\,\rangle 
\end{equation}  

\noindent in a representation space labeled by the quantum numbers $J,M$.
The $W$--rotation matrices introduced in \eqref{genW}
are related to \eqref{B1} by

\begin{equation}\label{B2}
U^J_{M M'\,}(\omega; \Theta, \Phi)\;=\;\sum_{M''}D^J_{M M''}(\Phi, \Theta, -\Phi)\,e^{-i\,M''\omega}\,
D^J_{M'' M'}(\Phi, -\Theta, -\Phi). 
\end{equation}

\noindent For rotations around coordinate axes we find the expressions
(the reduced $W$--matrices $d^J_{MM'}$ were introduced in \eqref{redD}
and given explicitly in \eqref{D12} and \eqref{rewj1} 
of Appendix B1 for $J=\frac{1}{2},1$). 

$\bullet$ ($x \;\text{axis}; \Theta =\frac{\pi}{2};\Phi =0$):

\begin{equation}\label{B3}
U^J_{M M'}(\omega; \frac{\pi}{2},0)\;=\;D^J_{M M'}(\frac{\pi}{2},\omega,-\frac{\pi}{2})\,=\,
(-i)^{\,M-M'}\,d^J_{MM'}\,(\omega)
\end{equation}

$\bullet$ ($y \;\text{axis}; \Theta =\frac{\pi}{2};\Phi =\frac{\pi}{2}$):

\begin{equation}\label{B4}
U^J_{M M'}(\omega; \frac{\pi}{2},\frac{\pi}{2})\;=\;
D^J_{M M'}(0,\omega,0)\,=\,
d^J_{MM'}\,(\omega)
\end{equation}

$\bullet$ ($z \;\text{axis}; \Theta =0$):

\begin{equation}\label{B5}
U^J_{M M'}(\omega; 0,\Phi)\;=\;
D^J_{M M'}(0,\omega,0)\,=\,
\delta_{MM'}\,e^{-i\,M\,\omega}
\end{equation}

\section*{Acknowledgements}

We are pleased to thank Vincenzo Aquilanti, Mauro Carfora, Silvano Garnerone and Tullio Regge
for interesting discussions. We are in debt with M. Carfora also for his unexhaustible 
enthusiasm in helping us with the preparation of all figures, and with V. Aquilanti and C. 
Coletti for the permission to reproduce Fig. 21 from \cite{AqCo}. 
\vfill
\eject

\end{document}



\begin{figure}
\begin{center}
\includegraphics[bb= 0 0 600 350, scale=0.5]{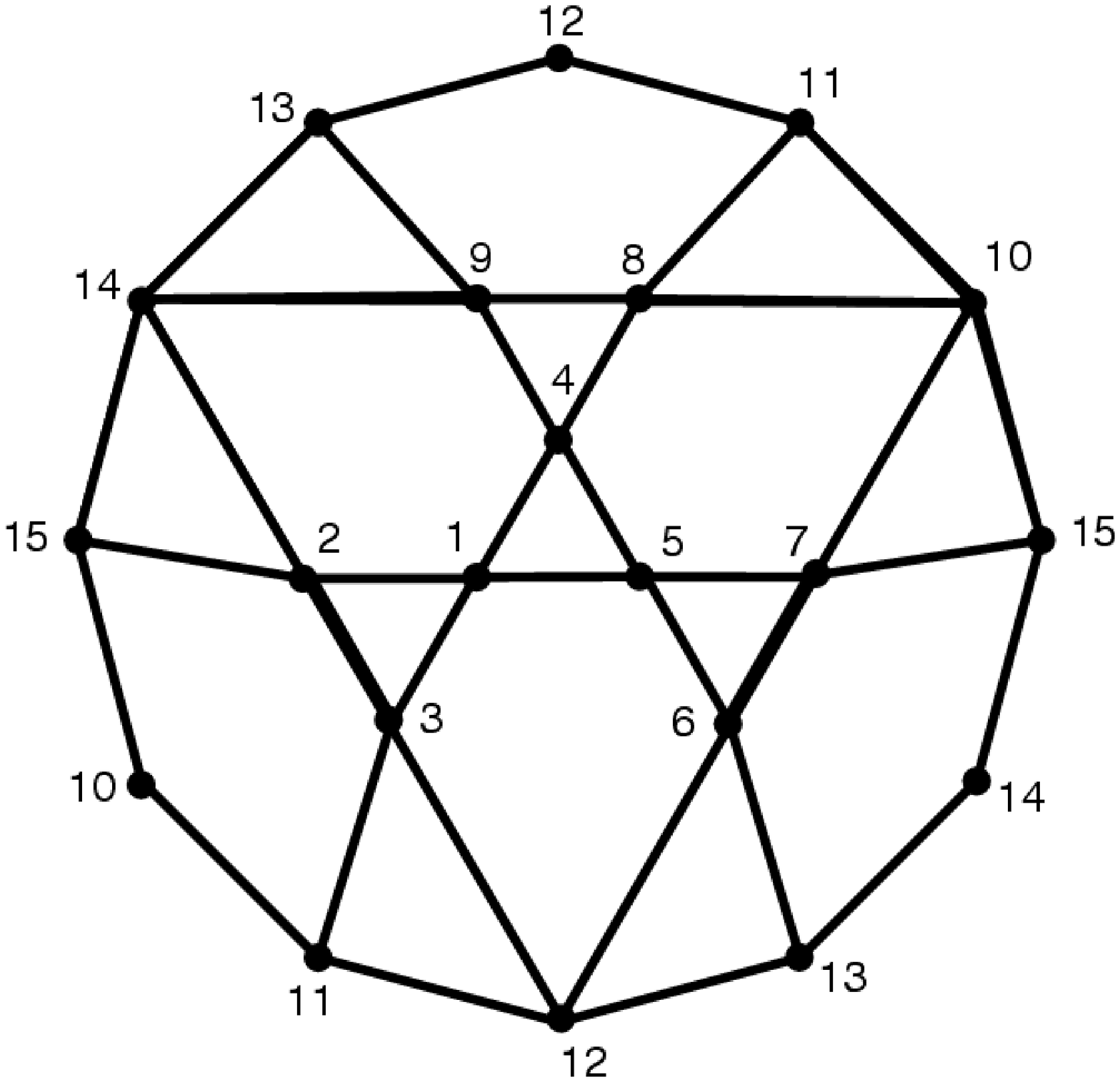}
\end{center}
\caption{The rotation graph $\mathfrak{G}_3$ where each vertex represents a binary coupling scheme
of $(n+1)=4$ angular momenta. Vertices on the perimeter have to be identified through
the antipodal map, showing that $\mathfrak{G}_3$ is not planar. Using the notations of 
Appendix A1, the $1:1$ correspondences between vertices and binary parenthesization 
on four letters, say $\{a,b,c,d\}$, read:
$1 \leftrightarrow (d(b(ac)));$ 
$2 \leftrightarrow (b(d(ca)));$ 
$3 \leftrightarrow ((ac)(bd));$ 
$4 \leftrightarrow (d(a(bc)));$ 
$5 \leftrightarrow (d(c(ab)));$ 
$6 \leftrightarrow (c(d(ab)));$ 
$7 \leftrightarrow ((ab)(cd));$ 
$8 \leftrightarrow (a(d(bc)));$ 
$9 \leftrightarrow ((ad)(bc));$ 
$10 \leftrightarrow (a(b(cd)));$ 
$11 \leftrightarrow (a(c(bd)));$ 
$12 \leftrightarrow (c(a(bd)));$ 
$13 \leftrightarrow (c(b(ad)));$ 
$14 \leftrightarrow (b(c(da)));$ 
$15 \leftrightarrow (b(a(cd)))$.}
\label{fig.4.1}
\end{figure}


\begin{figure}
\begin{center}
\includegraphics[bb= 0 0 600 350, scale=0.7]{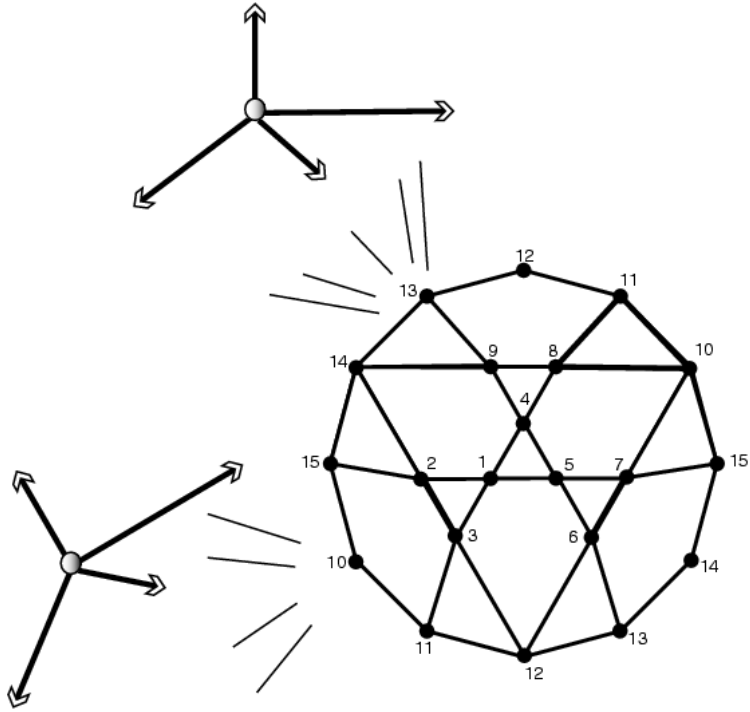}
\end{center}
\caption{The computational setting based on the
combinatorics of the
Rotation graph $\mathfrak{G}_3 (V,E)$ can be 
pictorially represented by blowing out each vertex 
into the corresponding computational Hilbert space
and by looking at the gates as elementary operations on the base space
and/or on the fiber ({\em cfr.} the remark at the end of this section
and the analysis carried out in the next section).}

\label{fig.4.2}
\end{figure}



\begin{figure}
\begin{center}
\includegraphics[bb= 0 0 600 350, scale=0.7]{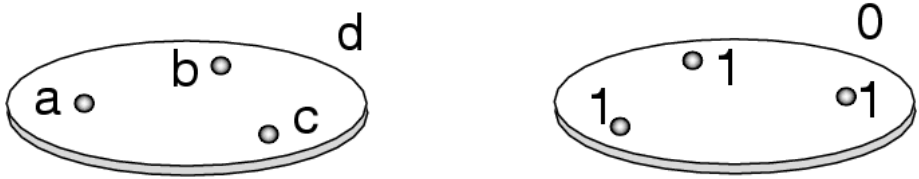}
\end{center}
\caption{Left: a marked $2$--disk corresponding to the Hilbert space $\mathcal{H}_{CS3}$
$(D^2, 3\, \text{pts})$ with $\{a,b,c,d\}$ $\in \{0,1,2,3\}$. Right: the $2$--disk supporting 
the topological qubit $V(D^2, 3 \,\text{pts})$.}

\label{fig.6.1}
\end{figure}


\begin{figure}
\begin{center}
\includegraphics[bb= 0 0 600 360, scale=0.7]{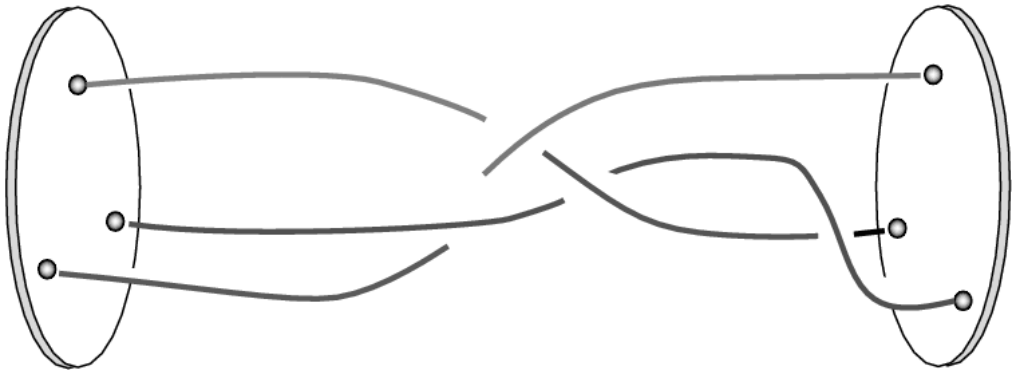}
\end{center}
\caption{Action of the braid group $B_3$ on the topological support of a single qubit.}

\label{fig.6.2}
\end{figure}


\begin{figure}
\begin{center}
\includegraphics[bb= 0 0 600 350, scale=0.7]{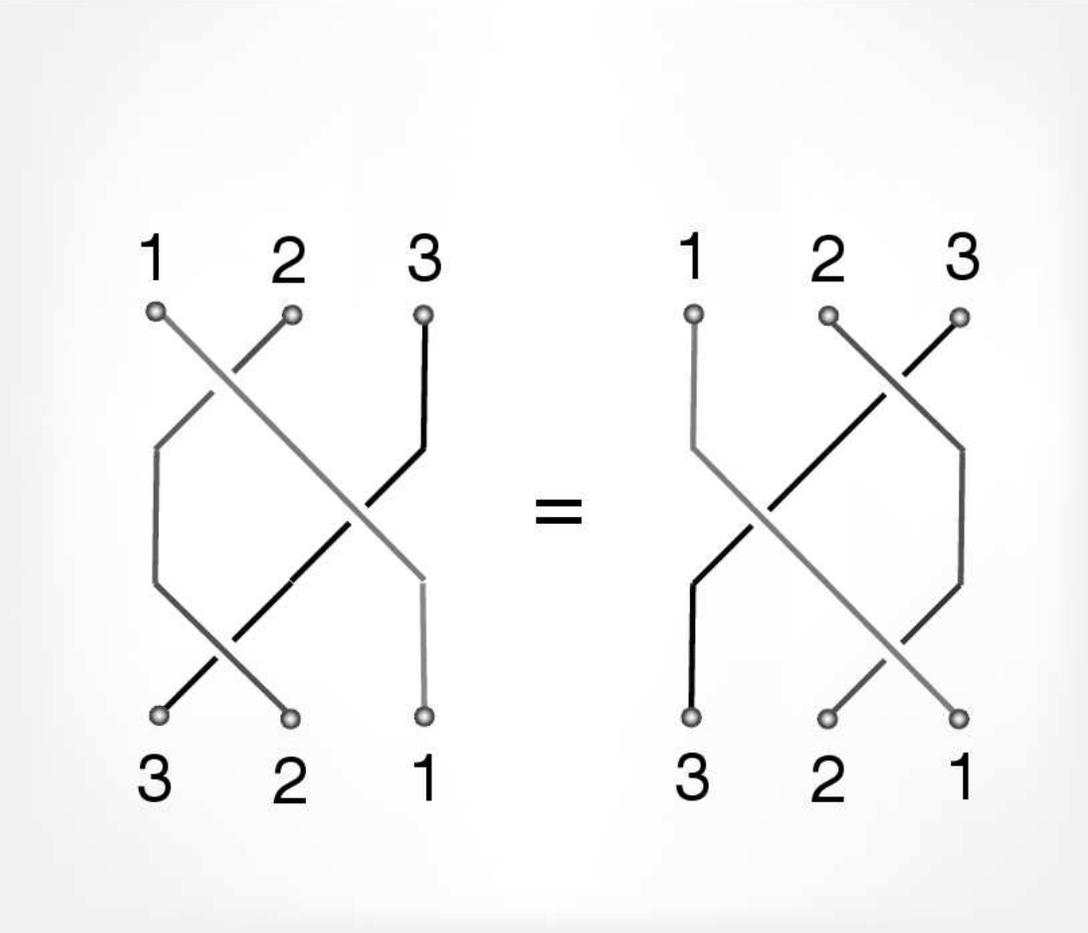}
\end{center}
\caption{The Yang--Baxter equation $R_{12}R_{13}R_{23}=$
$R_{23}R_{13}R_{12}$.}

\label{fig.6.3}
\end{figure}


\begin{figure}
\begin{center}
\includegraphics[bb= 0 0 600 350, scale=0.7]{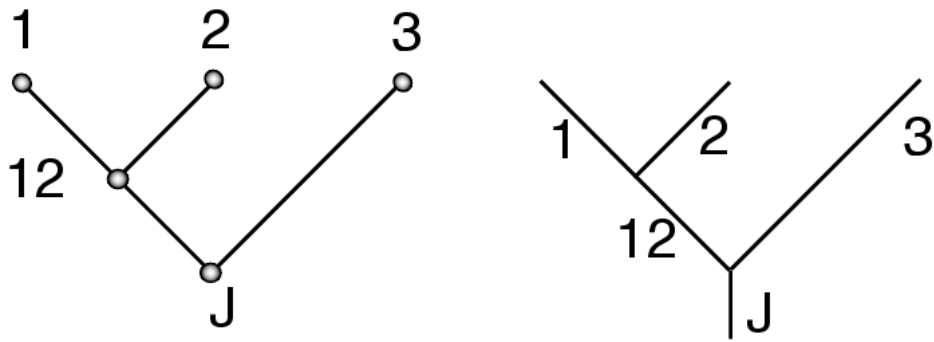}
\end{center}
\caption{The fundamental tree $\mathsf{T}_{12}$ with labels on nodes (left)
and the associated 3--valent graph $\mathsf{t}_{12}$ with labels on edges (right).}

\label{fig.6.4}
\end{figure}


\begin{figure}
\begin{center}
\includegraphics[bb= 0 0 600 400, scale=0.5]{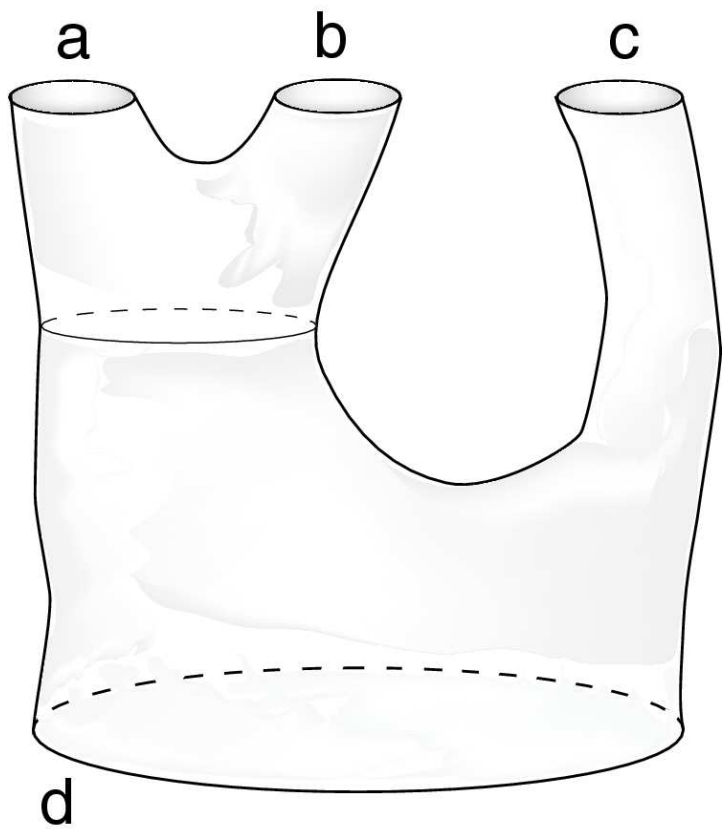}
\end{center}
\caption{The punctured 2--sphere $(S^2;\, a,b,c,d)$ with labels on the
boundary circles.}

\label{fig.6.5}
\end{figure}


\begin{figure}
\begin{center}
\includegraphics[bb= 0 0 600 380, scale=0.5]{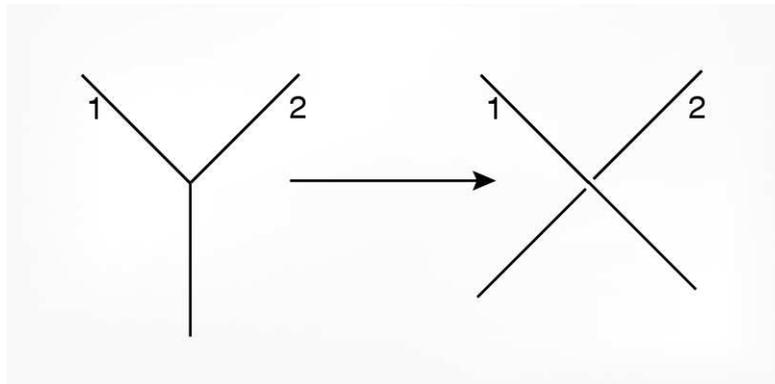}
\end{center}
\caption{An interacting 3--valent vertex with two incoming edges is splitted by
requiring that the first entry in the binary coupling overcrosses the second entry (and viceversa).}

\label{fig.6.6}
\end{figure}


\begin{figure}
\begin{center}
\includegraphics[bb= 0 0 600 350, scale=0.7]{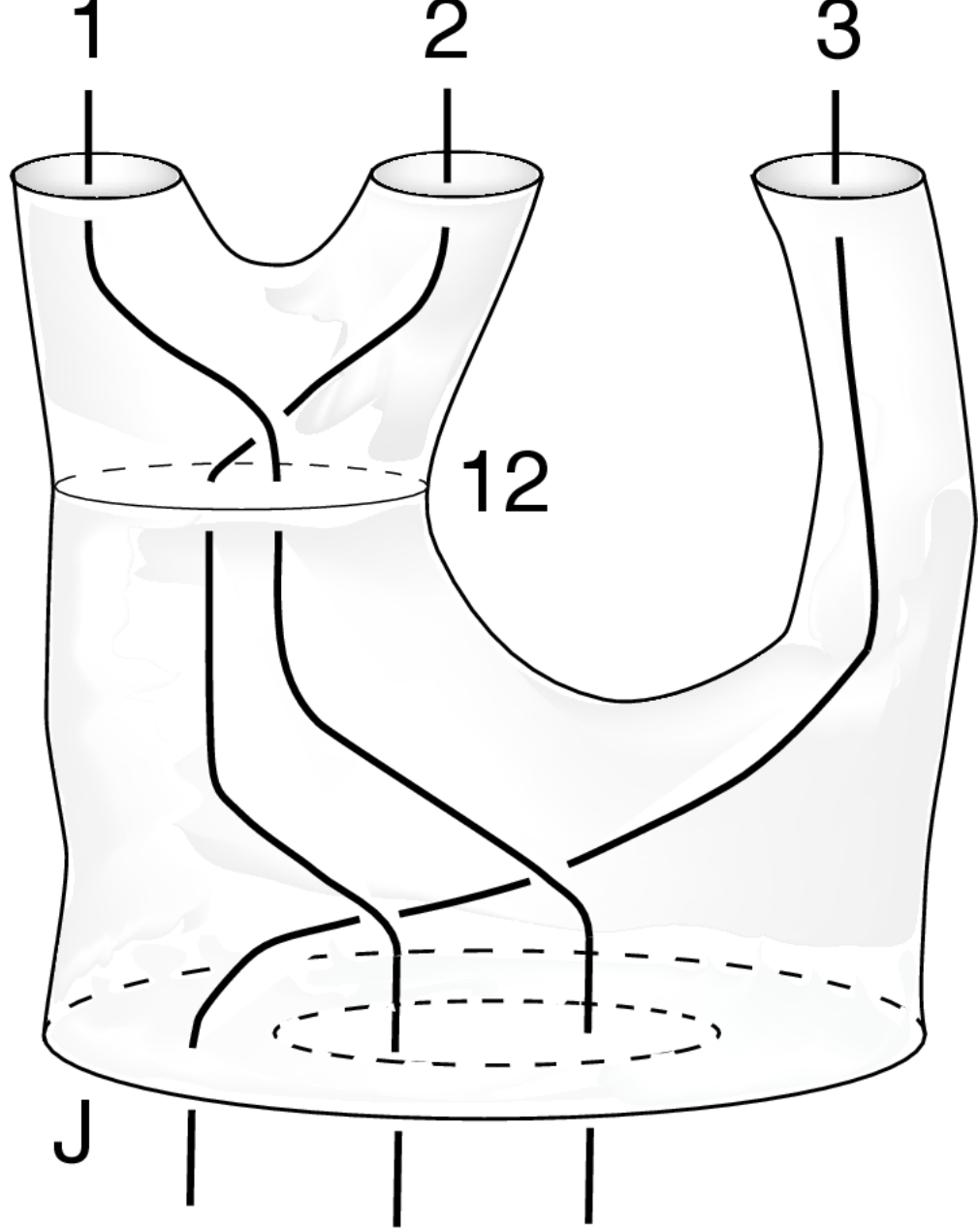}
\end{center}
\caption{An embedding of the fundamental graph $\mathsf{t}_{12}$ into $(D^3,\,S^2)_{(a,b,c,d)}$
where we drop out the labels $(a,b,c,d)$ on the boundary circles which appeared in Fig.7.}

\label{fig.6.7}
\end{figure}


\begin{figure}
\begin{center}
\includegraphics[bb= 0 0 600 380, scale=0.6]{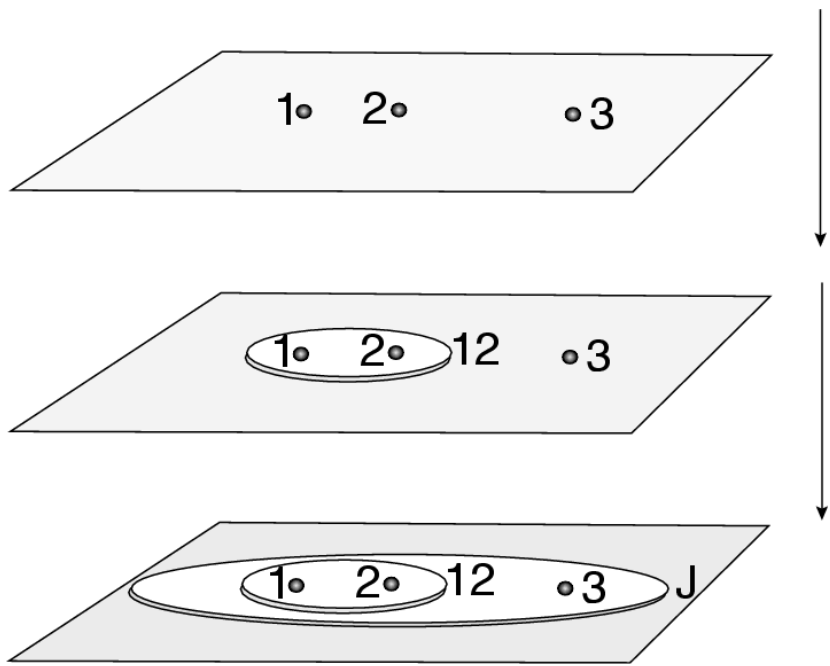}
\end{center}
\caption{The level set (``time slicing") of the configuration of Fig.9.}

\label{fig.6.8}
\end{figure}


\begin{figure}
\begin{center}
\includegraphics[bb= 0 0 600 350, scale=0.7]{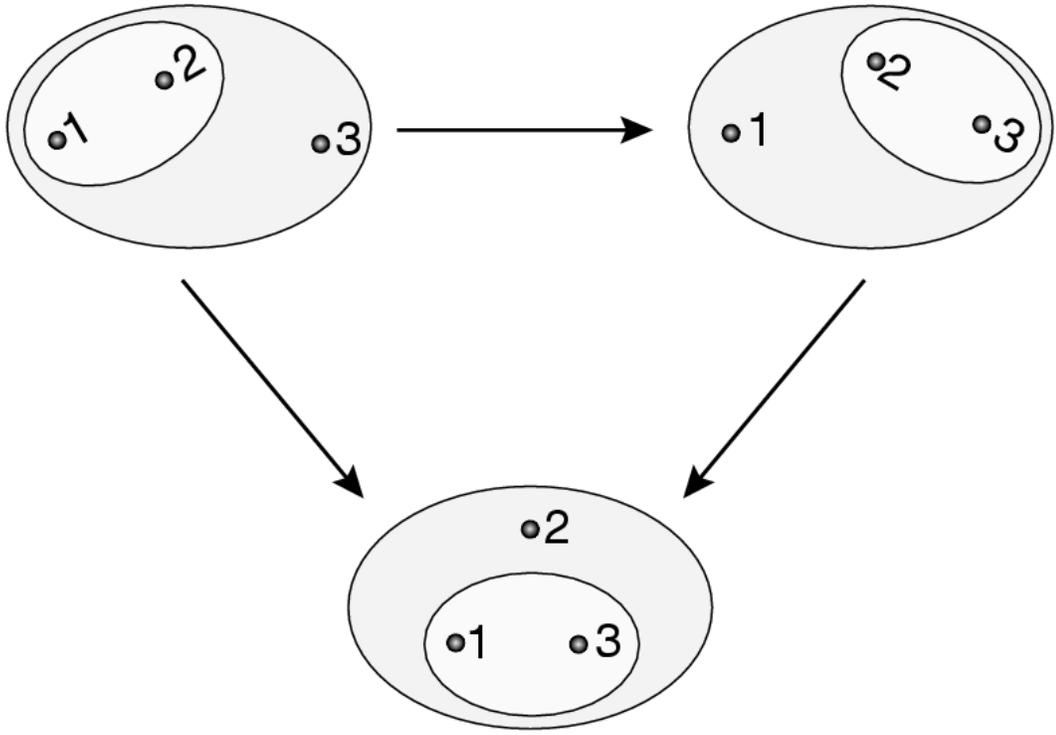}
\end{center}
\caption{The triangular commutative diagram representing the Racah identity.}

\label{fig.6.9}
\end{figure}



\begin{figure}
\begin{center}
\includegraphics[bb= 0 0 600 380, scale=0.5]{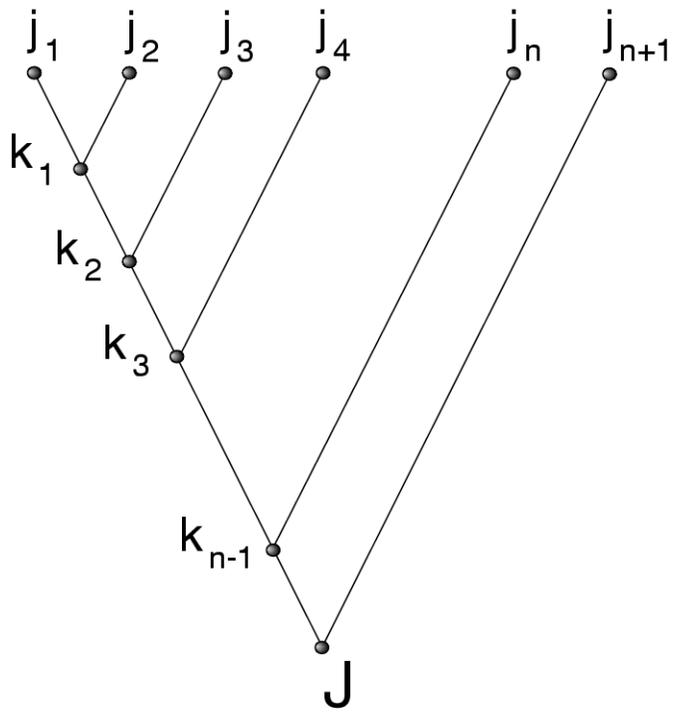}
\end{center}
\caption{A particular rooted binary coupling tree on $(n+1)$ leaves where incoming angular
momenta are coupled sequentially.}

\label{fig.A0}
\end{figure}


\begin{figure}
\begin{center}
\includegraphics[bb= 0 0 600 480, scale=0.5]{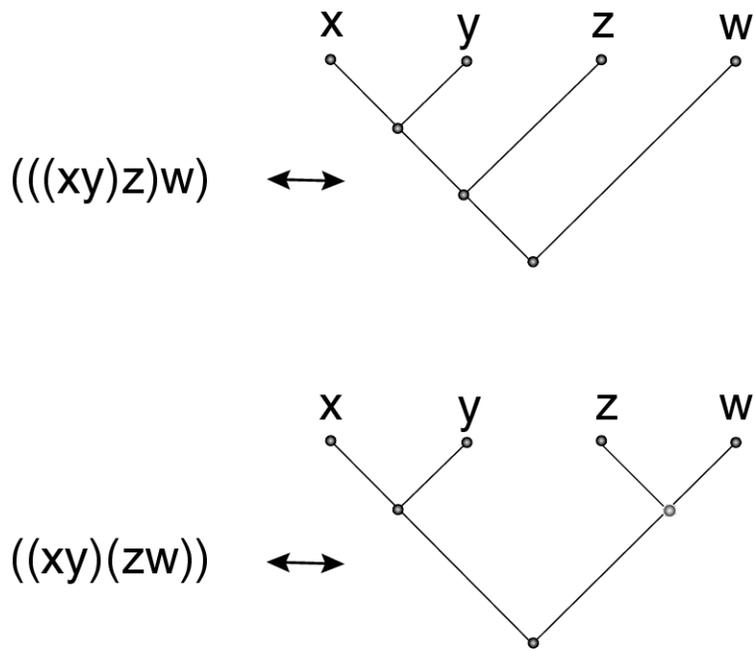}
\end{center}
\caption{Explicit bijections between binary parenthesizations over an alphabet of 4 letters and rooted
binary trees on 4 leaves. Here the ordering of $\{x,y,z,w\}$ is fixed and only two correspondences
are depicted [the other ones arise from $(x(y(zw)))$; $((x(yz))w)$; $(x((yz)w))$].}

\label{fig.A1}
\end{figure}


\begin{figure}
\begin{center}
\includegraphics[bb= 0 0 600 460, scale=0.5]{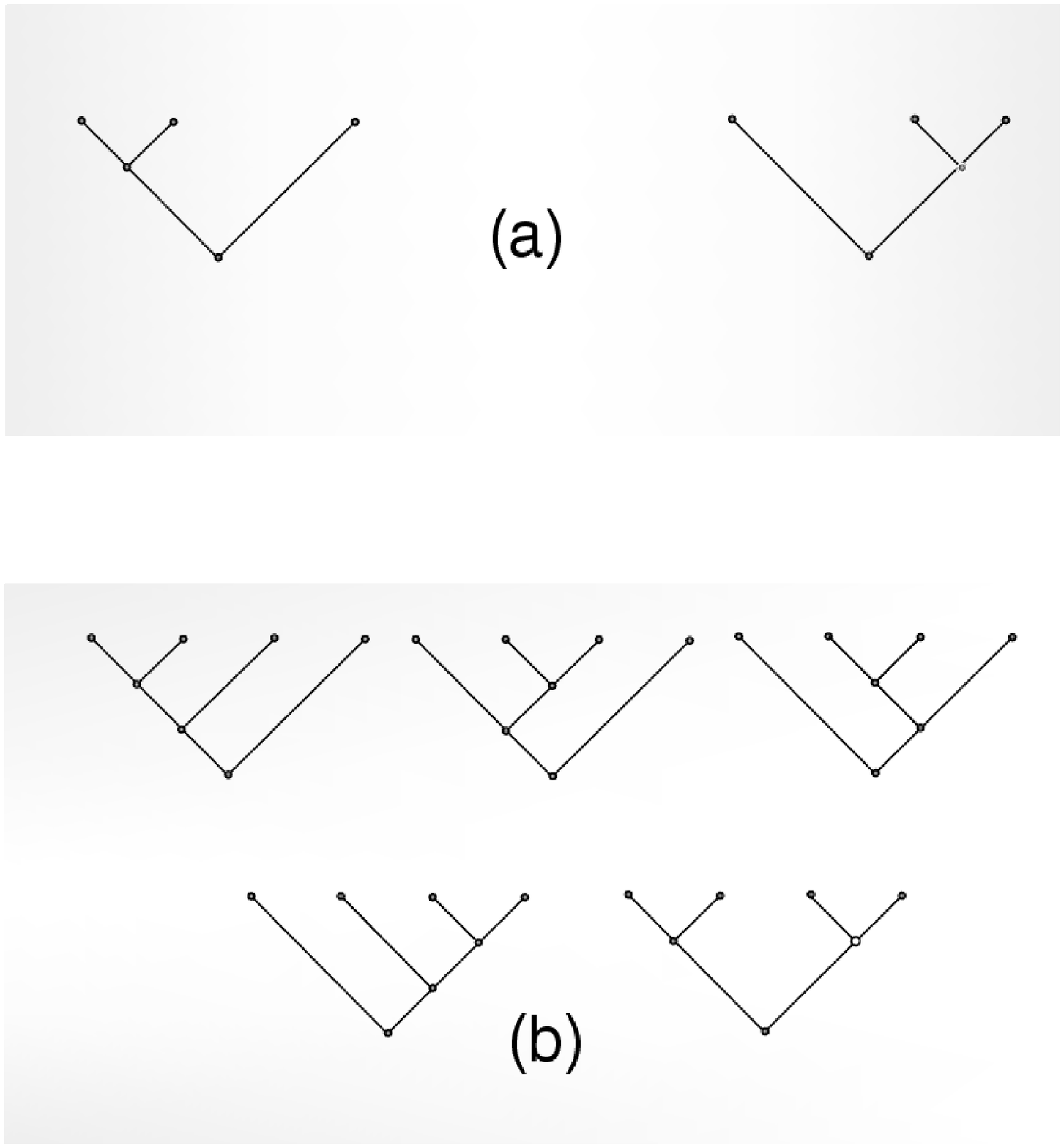}
\end{center}
\caption{Plane unlabeled rooted binary trees on 3 leaves (a) and 4 leaves (b).}

\label{fig.A2}
\end{figure}


\begin{figure}
\begin{center}
\includegraphics[bb= 0 0 600 450, scale=0.7]{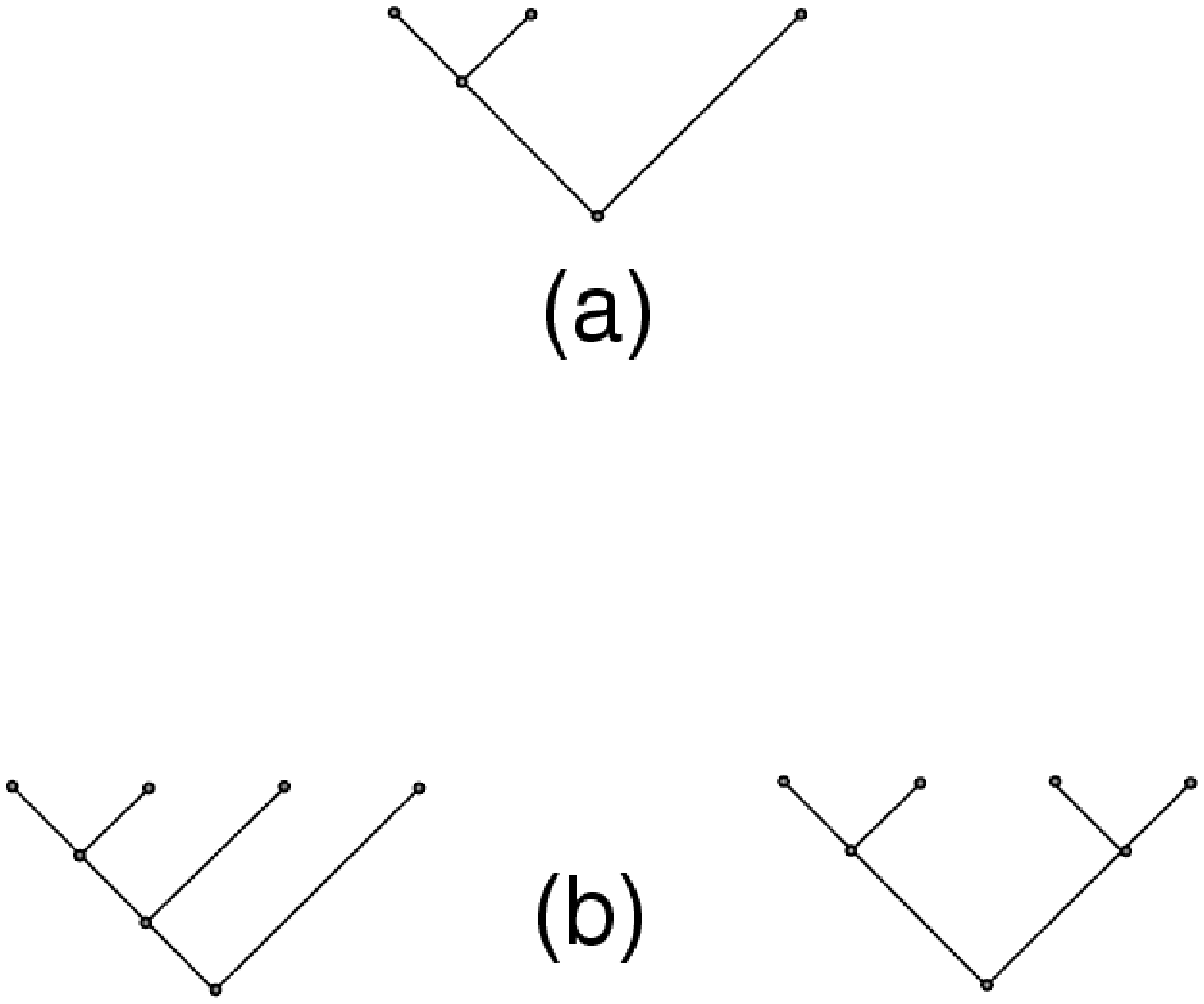}
\end{center}
\caption{Not plane and unlabeled rooted binary trees on 3 leaves (a) and 4 leaves (b).}

\label{fig.A3}
\end{figure}


\begin{figure}
\begin{center}
\includegraphics[bb= 0 0 600 300, scale=0.7]{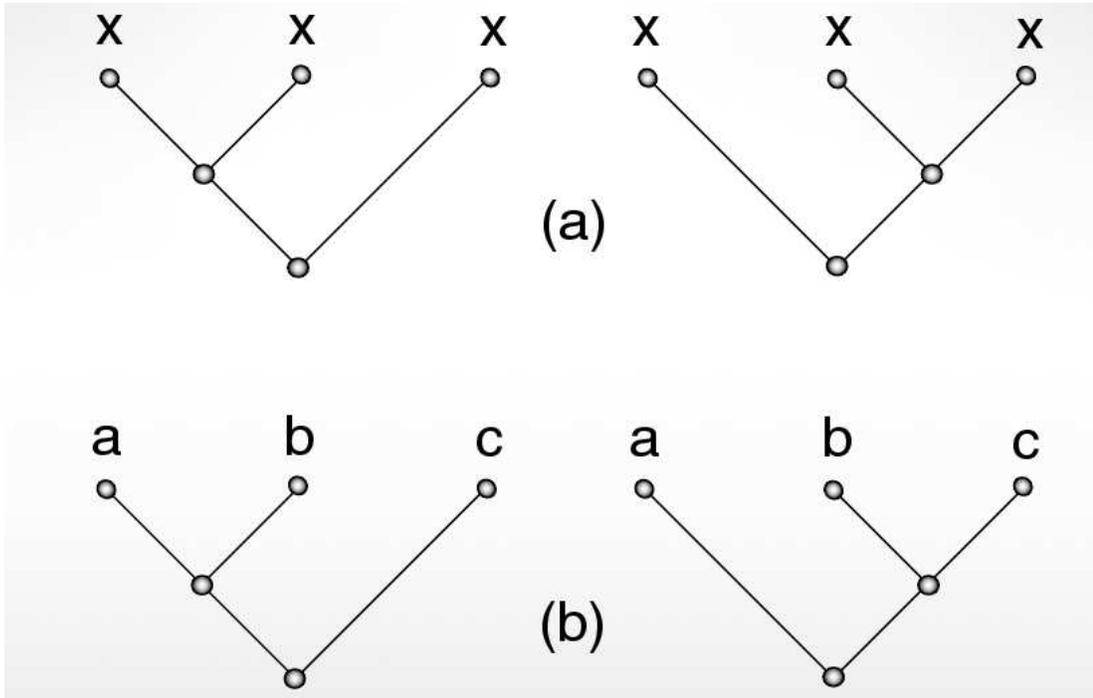}
\end{center}
\caption{Presentation of Catalan trees on 3 leaves as labeled trees: a same label assigned to each leave
(a); a fixed sequence of distinct labels assigned to leaves (b).}

\label{fig.A4}
\end{figure}


\begin{figure}
\begin{center}
\includegraphics[bb= 0 0 600 380, scale=0.7]{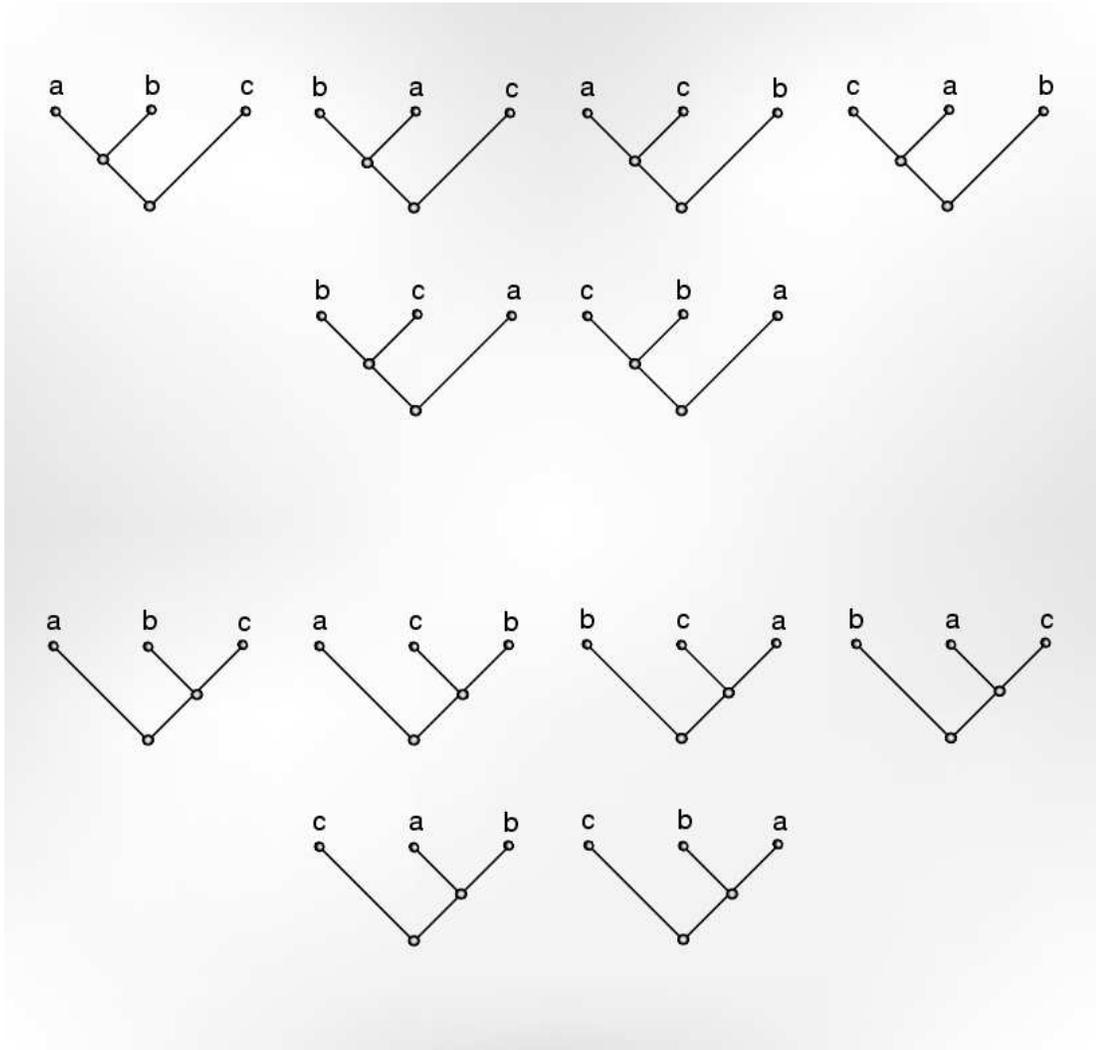}
\end{center}
\caption{Plane rooted binary trees on 3 leaves labeled in all possible ways with distinct symbols.}

\label{fig.A5}
\end{figure}
\vspace*{\fill}
\newpage

\begin{figure}
\begin{center}
\includegraphics[bb= 0 0 600 350, scale=0.7]{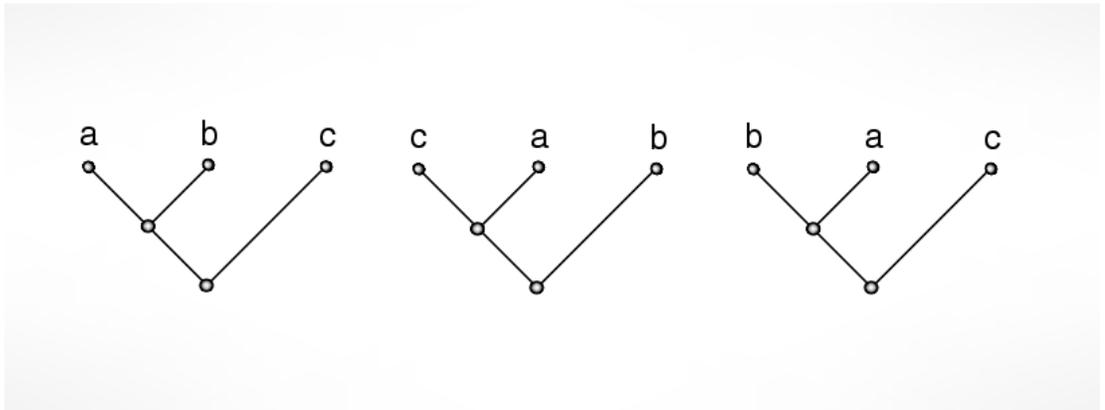}
\end{center}
\caption{Not plane rooted binary trees on 3 leaves labeled with no distinction between left and right.}

\label{fig.A6}
\end{figure}


\begin{figure}
\begin{center}
\includegraphics[bb= 0 0 600 600, scale=0.7]{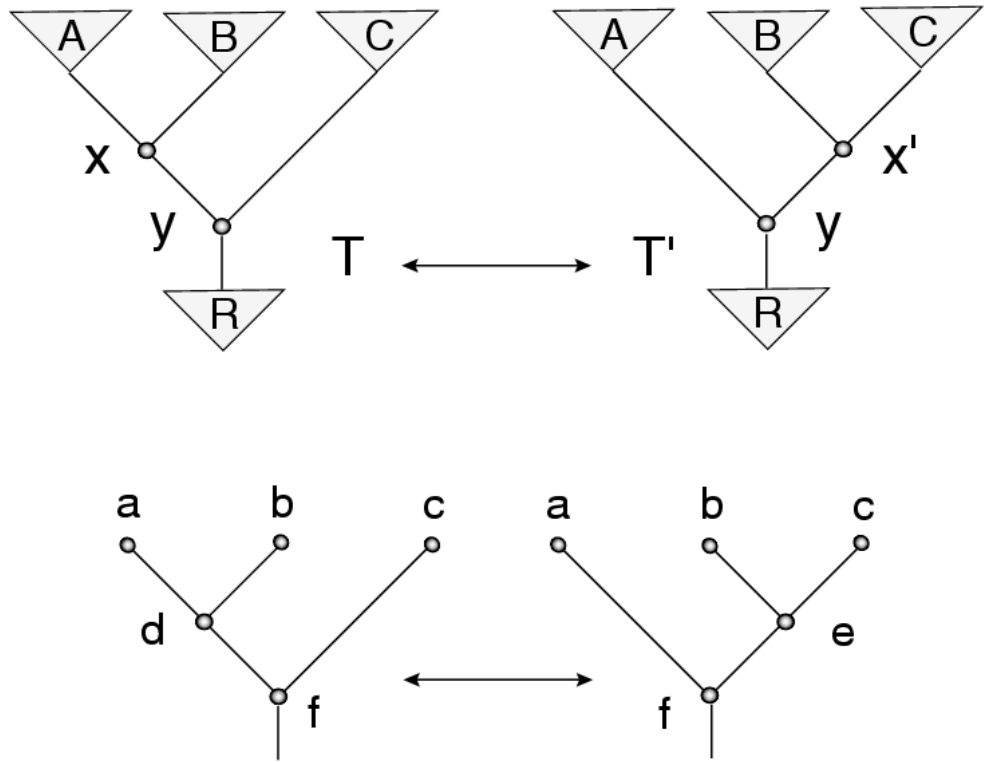}
\end{center}
\caption{A general rotation (top) and a particular rotation (bottom) on rooted binary trees. A,B,C
are generic subtrees, and R is a subtree containing the root.}

\label{fig.A7}
\end{figure}


\begin{figure}
\begin{center}
\includegraphics[bb= 0 0 600 300, scale=0.7]{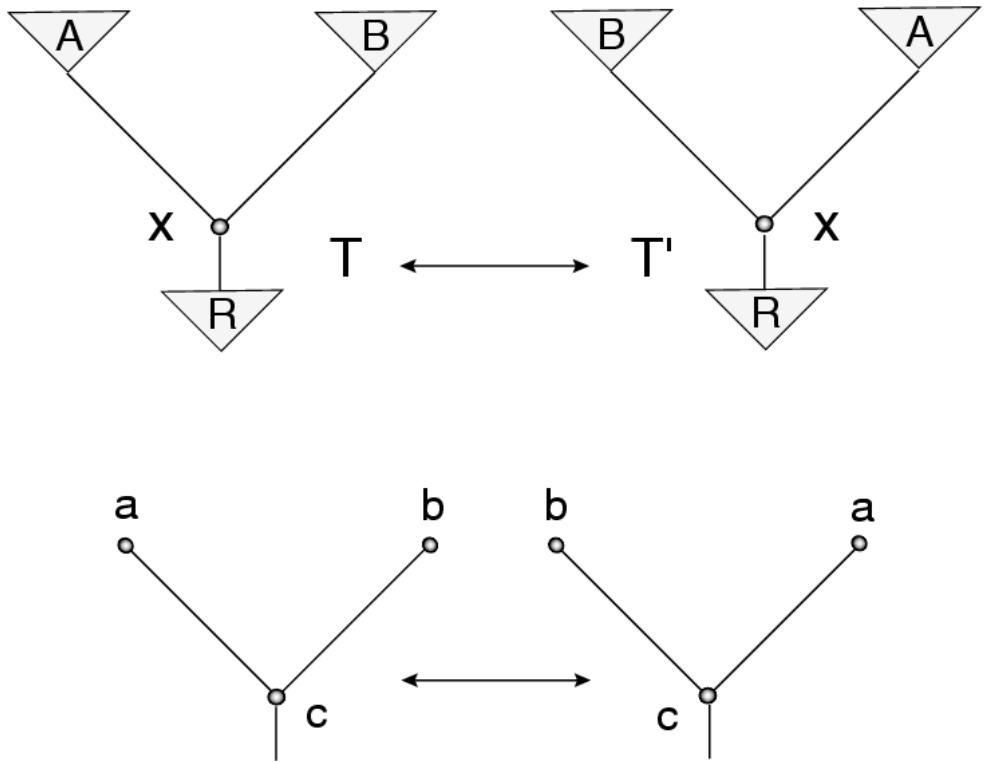}
\end{center}
\caption{A twist on a generic tree (top) and a particular twist (bottom).}

\label{fig.A8}
\end{figure}

\vspace*{\fill}
\newpage

\begin{figure}
\begin{center}
\includegraphics[bb= 0 0 600 300,scale=0.6]{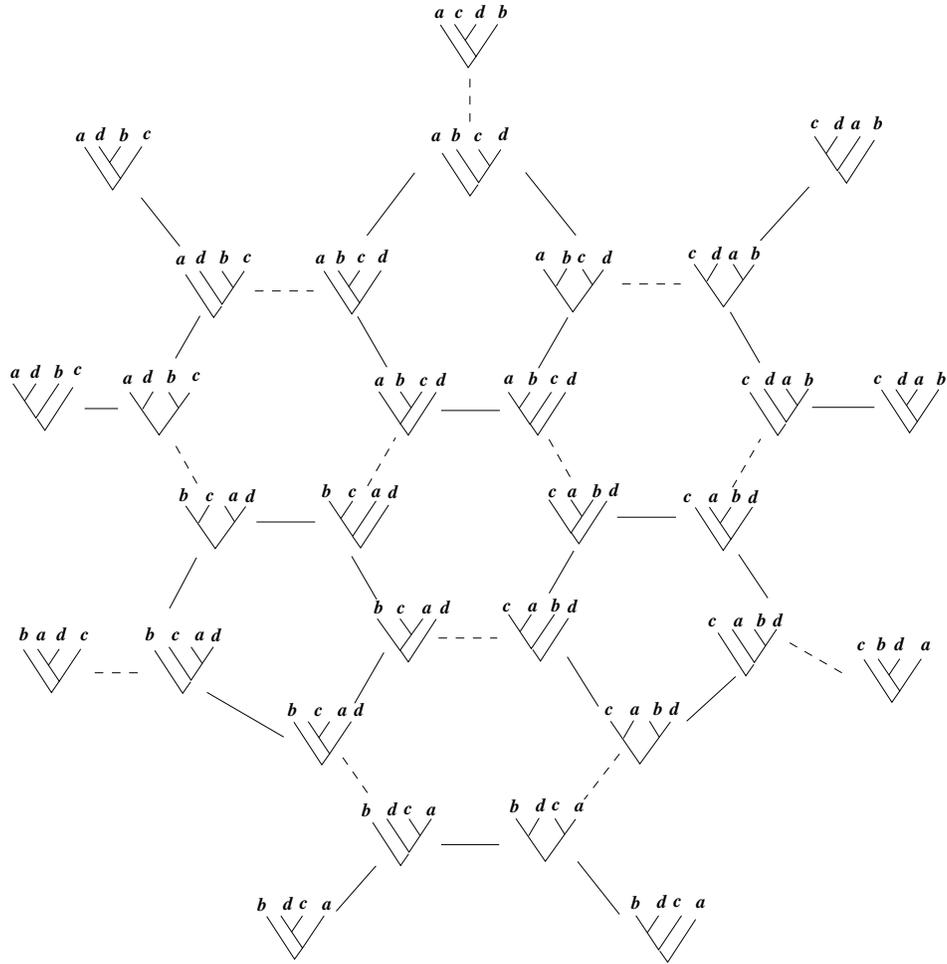}
\end{center}
\caption{A portion of the Twist--Rotation graph $\hat{\mathfrak{G}}_3$ representing the vertex--edge
set of a truncated icosahedron. Only 30 of the 60 vertices are shown, and the picture can be completed by
taking the mirror image of each tree at the antipodal vertex. The remaining 60 vertices of
$\hat{\mathfrak{G}}_3$ are arranged into an isomorphic graph which is obtained from the former one
by exchanging a pair of labels, {\em e.g.} $(a,b) \rightarrow (b,a)$.}

\label{fig.A9}
\end{figure}


\begin{figure}
\begin{center}
\includegraphics[bb= 0 0 600 300, scale=0.7]{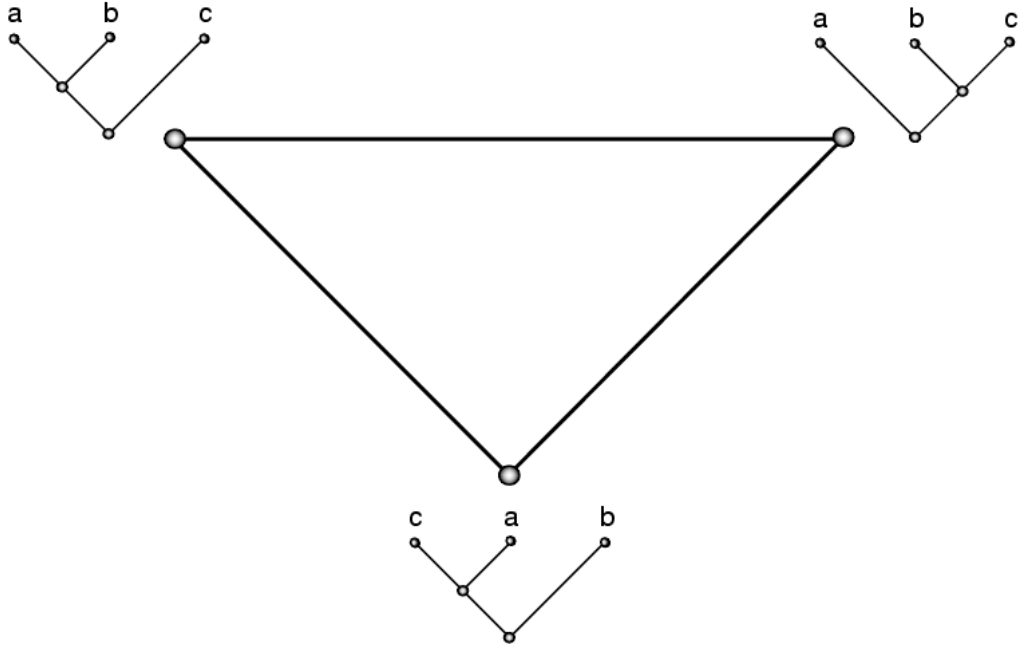}
\end{center}
\caption{The Rotation graph $\mathfrak{G}_2$ is depicted as a triangle with binary trees attached 
to its vertices. Edges correspond to
rotations (topologically) but -- when dealing with binary coupling trees -- they
represent three Racah transforms related by the Racah identity 
(see also Section 4.1, Section 6 and Fig.11). Similar configurations are associated to 
triangular plaquettes in the Rotation graph $\mathfrak{G}_3$ of Fig.1.}

\label{fig.A10}
\end{figure}

\vspace*{\fill}
\newpage

\begin{figure}
\begin{center}
\includegraphics[bb= 0 0 600 350, scale=0.7]{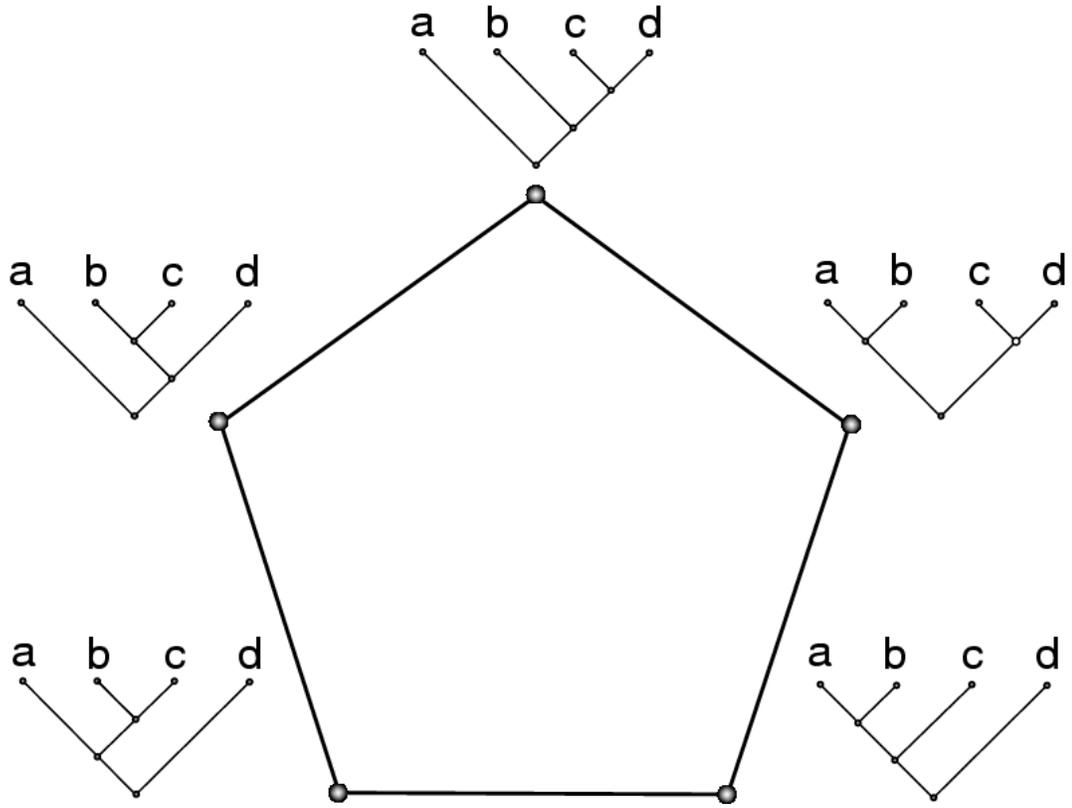}
\end{center}
\caption{A pentagonal plaquette  in the Twist--Rotation graph $\hat{\mathfrak{G}}_3$ of Fig.21 or
in the Rotation graph $\mathfrak{G}_3$ of Fig.1. When binary coupling trees are considered, edges represent Racah 
transforms related by the pentagon (Biedenharn--Elliott)  
identity, see Section 4.1.}

\label{fig.A11}
\end{figure}